%% file: main.tex
 \documentclass[final,3p,times,twocolumn]{elsarticle}

\usepackage{graphicx}
\usepackage{amssymb}

\usepackage{amsmath}

\usepackage{longtable}

\usepackage{lineno}

\journal{Nuclear Data Sheets}

\begin{document}

\begin{frontmatter}


\title{Measurement of the $^{239}$Pu(n,f)/$^{235}$U(n,f) Cross-Section Ratio with the NIFFTE fission Time Projection Chamber}



\cortext[corresponding]{Corresponding author}

\author[LLNL]{L.~Snyder}\corref{corresponding}
\ead{snyder35@llnl.gov}

\author[LLNL]{M.~Anastasiou}
\author[LLNL]{N.S.~Bowden}
\author[CSM]{J.~Bundgaard}
\author[LLNL]{R.J.~Casperson}
\author[UC]{D.A.~Cebra}
\author[LLNL]{T.~Classen}
\author[LLNL]{D.H. Dongwi}
\author[LANL]{N.~Fotiades}
\author[UC,LLNL]{J.~Gearhart}
\author[LANL]{V.~Geppert-Kleinrath}
\author[CSM]{U.~Greife}
\author[LLNL]{C.~Hagmann}
\author[LLNL]{M.~Heffner}
\author[CSM]{D.~Hensle}
\author[CSM,LANL]{D.~Higgins}
\author[ACU]{L.D.~Isenhower}
\author[LLNL]{K.~Kazkaz}
\author[CAL]{A.~Kemnitz}
\author[OSU]{J.~King}
\author[CAL]{J.L.~Klay}
\author[CSM]{J.~Latta}
\author[LANL]{E.~Leal-Cidoncha}
\author[OSU]{W.~Loveland}
\author[LLNL]{J.A.~Magee\fnref{AA}}
\author[LANL]{B.~Manning}
\author[LLNL]{M.P.~Mendenhall}
\author[LLNL]{M.~Monterial}
\author[LANL]{S.~Mosby}
\author[LANL]{D.~Neudecker}
\author[LANL]{C.~Prokop}
\author[LLNL]{S.~Sangiorgio}
\author[LANL]{K.T.~Schmitt\fnref{ORNL}}
\author[LLNL]{B.~Seilhan}
\author[LANL]{F.~Tovesson\fnref{ANL}}
\author[ACU]{R.S.~Towell}
\author[LLNL]{N.~Walsh}
\author[ACU]{T.S.~Watson}
\author[OSU]{L.~Yao}
\author[LLNL]{W.~Younes}
\author[]{\protect\\(The NIFFTE Collaboration)}

\address[LLNL]{Lawrence Livermore National Laboratory, Livermore, CA 94550, United States}
\address[LANL]{Los Alamos National Laboratory, Los Alamos, NM 87545, United States}
\address[ACU]{Abilene Christian University, Abilene, TX 79699, United States}
\address[CAL]{California Polytechnic State University, San Luis Obispo, CA 93407, United States}
\address[CSM]{Colorado School of Mines, Golden, CO 80401, United States}
\address[OSU]{Oregon State University, Corvallis, OR 97331, United States}
\address[UC]{University of California, Davis, CA 95616, United States}

\fntext[ANL]{Current Address: Argonne National Laboratory, Lemont, IL 60439, United States}
\fntext[ORNL]{Current Address: Oak Ridge National Laboratory, Oak Ridge, TN 37830, United States}
\fntext[AA]{Current Address: Avrio Analytics LLC, Knoxville, TN, 37917 United States}

\date{\today}
	
\begin{abstract}
The $^{239}$Pu(n,f)/$^{235}$U(n,f) cross-section ratio has been measured with the fission Time Projection Chamber (fissionTPC) from 100~keV to 100~MeV.  The fissionTPC provides three-dimensional reconstruction of fission-fragment ionization profiles, allowing for a precise quantification of measurement uncertainties.  The measurement was performed at the Los Alamos Neutron Science Center which provides a pulsed white source of neutrons.  The data are recommended to be used as a cross-section ratio shape.  A discussion of the status of the absolute normalization and comparisons to ENDF evaluations and previous measurements is included.
\end{abstract}

\end{frontmatter}

\tableofcontents{}


\input{preamble}
\newpage
\section{Introduction}
\input{Introduction}

\section{Cross-section Measurement Formulation}
\input{XS_Calc}

\section{Experiment Description}
\input{Short_Experiment_Description}

\section{Energy Dependent Cross-section Ratio Terms}
\label{sec:shape}
In this section we describe how the various energy dependent terms of Eq.~\eqref{eqn:xsCalc} are measured or calculated.  The uncertainties for each of the various terms will also be presented, with additional discussion of our overall uncertainty quantification methodology in Sec.~\ref{sec:results}. 
\subsection{Neutron Time-of-Flight and Energy Determination}
\input{Shape/Neutron_ToF}

\subsection{Fission Fragment Selection Cuts, $C_{ff}$}
\input{Shape/Fission_Count}

\subsection{Fission Fragment Selection Efficiency, $\epsilon_{ff}$}
\input{Shape/Efficiency}

\subsection{Beam and Target Spatial Distribution, $\Phi$, $\sum_{XY}(\phi_{XY}\cdot n_{XY})$}
\input{Shape/Overlap}

\subsection{Beam-Correlated Background, $C_r$, $C_w$, and $C_b$}
\input{Shape/Beam_Correlated}

\subsection{Beam-Uncorrelated Background, $C_{\alpha}$}
\input{Shape/Beam_Uncorrelated}

\subsection{Detector Livetime, $\omega$}
\input{Shape/Livetime_L2}

\subsection{Beam Attenuation and Scattering, $\kappa$}
\input{Shape/Attenuation}

\section{Cross-section Ratio Results and Uncertainties}
\input{Results}

\section{Measurement Validations}
\input{Validations}

\section{Cross-section Ratio Comparisons And Discussion}
\input{Comparisons}

\section{Acknowledgments}
This work performed under the auspices of the U.S. Department of Energy by Lawrence Livermore National Laboratory under Contract No.~DE-AC52-07NA27344. The neutron beam for this work was provided by LANSCE, which is funded by the U.S. Department of Energy and operated by Los Alamos National Security, LLC, under Contract No.~89233218CNA000001.
University collaborators acknowledge support for this work from the DOE-NNSA Stewardship Science Academic Alliances Program, under Grant No.~DE-NA0002921, and through sub-contracts from LLNL and LANL.  The authors would like the thank R.~Williams (LLNL) for providing mass spectrometry results.



\twocolumn
\bibliographystyle{elsarticle-num}
\bibliography{references.bib}








\newpage
\appendix

\onecolumn
\section{Measured Cross-section Ratio Table}
\input{Results_Table}

\section{Cross-section Ratio Partial Uncertainties Table}
\input{Uncertainty_Table}

\end{document}

%% file: preamble.tex
\newcommand{\etal}{\emph{et al.}}
\newcommand{\ie}{\emph{i.e.}}
\newcommand{\eg}{\emph{e.g.}}
\newcommand{\etc}{\emph{etc.}}
\newcommand{\daq}{DAQ}
\newcommand{\pu}[1][239]{$\mathrm{^{#1}}$Pu}
\renewcommand{\u}[1][235]{$\mathrm{^{#1}}$U}
\newcommand{\hyd}[1][1]{$\mathrm{^{#1}}$H}  
\newcommand{\talpha}{$\alpha$}
\newcommand{\talphas}{$\alpha$s}
\newcommand{\ftpc}{fissionTPC}
\newcommand{\mev}{MeV}
\newcommand{\mus}{~$\mu$s}
\newcommand{\massunits}{\ensuremath{\mathrm{\mu g /cm^{2}}}}

%% file: Introduction.tex

\label{sec:introduction}

A wide variety of nuclear data parameters have been systematically measured since the advent of the nuclear age in the 1940s.  Neutron-induced fission cross sections have played a prominent role due to their importance in nuclear security and energy production.  Along with advances in computing power, modeling, and theory, there has been a renewed interest in better understanding and quantification of nuclear data uncertainties that feed into calculations and inform theoretical models.  Many experimental data sets used for nuclear-data evaluation over the preceding 80 years lack the level of detail regarding uncertainty quantification that is desired for modern data evaluation techniques \cite{Smith2012}.  Different experiments measuring the same reaction channels often have considerable systematic differences that are outside of the reported uncertainties.  In such cases, the age of the reports and the lack of detail can render it impossible to determine which, if any, of the data sets should be considered to be reliable.  Recent efforts have been made by the evaluation community to quantify ``Unrecognized Sources of Uncertainties" (USU) as detailed by Capote \etal~\cite{Capote2020}.  One approach is to use the spread in various data sets to infer the magnitude of underestimated or unreported uncertainties in the experimental data.

\begin{figure}[ht]
	\centering
    \includegraphics[width=1.\linewidth]{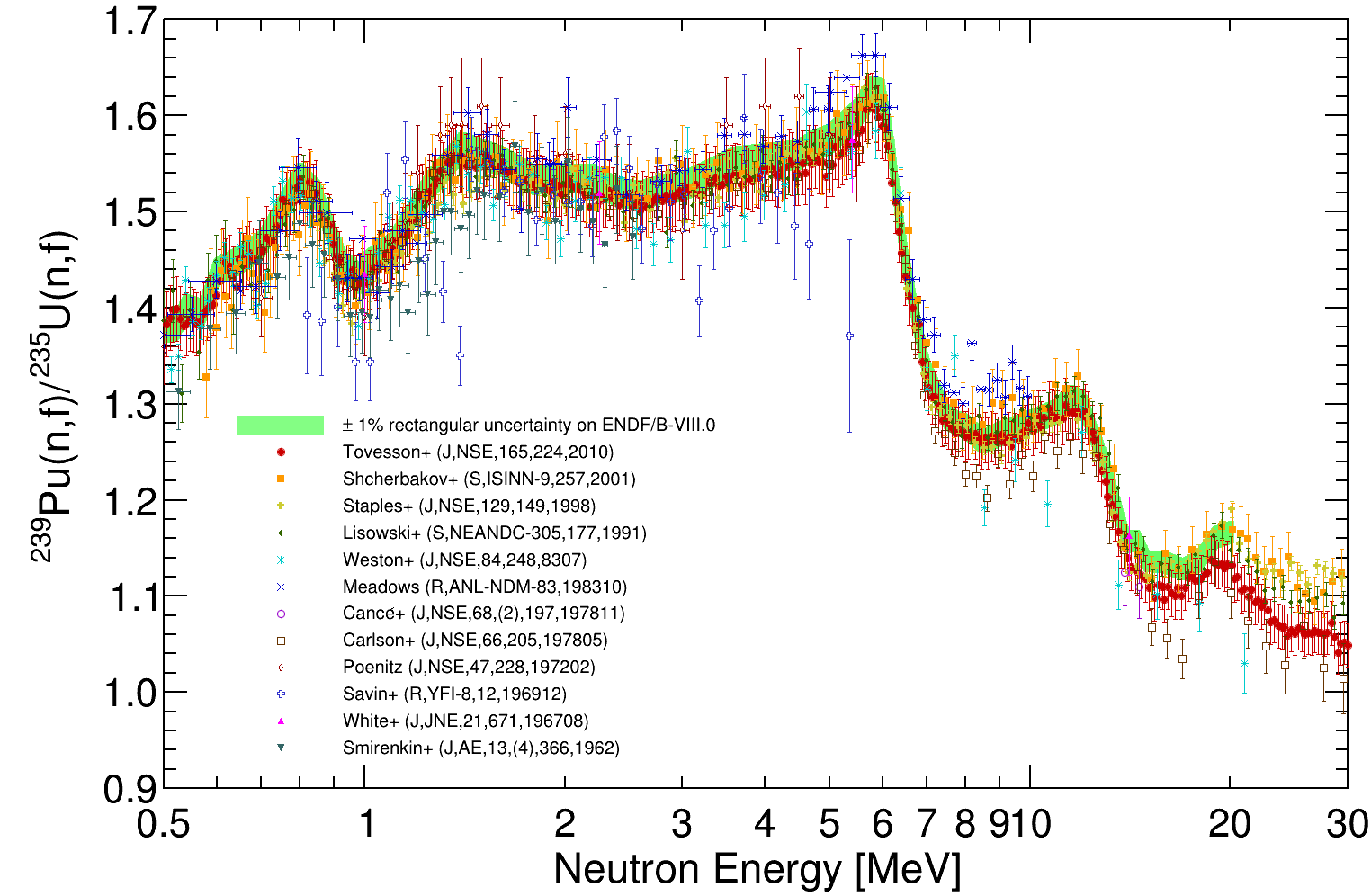}
    \caption
    {\label{fig:exfor_data}Previous measurements of the \pu(n,f)/\u(n,f) cross-section ratio; the solid-line is ENDF/B-VIII.0, while the green band shows approximately a 1\% uncertainty range to guide the eye. Note that although each datum claims high precision, the spread in the data is rather large.}
\end{figure}

As the \pu(n,f)/\u(n,f) cross-section ratio data in Fig.~\ref{fig:exfor_data} illustrate, the spread between measurements is largely inconsistent with the uncertainty of any particular measurement, suggesting potential USU.  Accordingly, the ENDF/B-VIII.0 evaluation incorporated USU into the \pu{}(n,f) evaluation \cite{ENDF8}.  

Neudecker \etal~\cite{Neudecker2020} systematically reviewed previous measurements of the \pu{}(n,f) cross section and developed a template to estimate unreported uncertainties for (n,f) measurements and to provide a guide for future measurements to follow.  Tovesson \cite{Tovesson2015} detailed the uncertainty quantification needed for fission cross-section measurements taken specifically at the Los Alamos Neutron Science Center (LANSCE), where the measurements presented in this paper were made.   Informed by these works, the NIFFTE collaboration has measured the \pu(n,f)/\u(n,f) fission cross-section ratio for incident-neutron energies between 0.2 MeV and 100 MeV. The total uncertainty for the cross-section ratio {\em shape} as a function of neutron energy has been carefully quantified and is below 1\%.

To reach sub-percent precision, the NIFFTE collaboration approached the measurement by designing a time projection chamber, optimized for measuring fission -- the \ftpc.
The \ftpc{} enables direct measurement of several relevant quantities, such as in-situ determination of neutron beam and target deposit uniformity and detector efficiency, thus eliminating a number of assumptions in experimental design and analysis. 

In addition, the data for every event delivers full 3-dimensional reconstruction of each fission, thereby yielding information beyond the traditional energy and time-of-flight quantities recorded by fission chambers. In short, the \ftpc{}  provides an independent, high-precision measurement with distinct systematic uncertainties that can be used to confirm or improve previous evaluations. The \ftpc{} and its capabilities are presented in full in Ref.~\cite{Heffner2014}.

Given the novelty of the experimental apparatus and approach, the collaboration has recognized the importance of careful validation of the detector response and analytical methods. 
This validation program is documented in a number of publications, which include the  $^{238}$U(n,f)/$^{235}$U(n,f) cross-section ratio shape as a function of neutron energy \cite{Casperson2018}, an instrumentation paper on the behavior of the \ftpc{} charge amplification in a neutron beam \cite{Snyder2018}, a publication on the fission fragment anisotropy in \u(n,f) \cite{Geppert-Kleinrath2019}, and a publication on the fission fragment anisotropy and linear momentum transfer \cite{hensle2020}.


In this article we begin with an explanation of how the cross-section ratio is formulated in Sec.~\ref{sec:ratioCalc}.
The \ftpc{} and the various measurement data-sets and targets are  presented in Sec.~\ref{sec:experiment_description}.  The cross-section shape is discussed in Sec.~\ref{sec:shape}. The results and uncertainty calculations are addressed in Sec.~\ref{sec:results}, and a series of validation studies are described in Sec.~\ref{sec:validations}.  Finally the results are compared to previous measurements and the ENDF/B-VIII.0 evaluation in Sec.~\ref{sec:compare}.  Also discussed in Sec.~\ref{sec:compare} is the status of the absolute normalization of the \ftpc{} cross-section ratio.  Details of the absolute normalization measurement are presented in Ref.~\cite{Monterial2021}.

%% file: XS_Calc.tex
\label{sec:ratioCalc}

The cross-section ratio measured is defined by Eq.~\eqref{eqn:xsCalc}.  In this formulation $x$ denotes the unknown actinide, in this case \pu{}, and $s$ the reference standard actinide, in this case \u{},

\begin{align}
\label{eqn:xsCalc}
\frac{\sigma_x}{\sigma_s} =& \frac{N_s}{N_x}\frac{\omega_s}{\omega_x}\frac{\kappa_s}{\kappa_x}\frac{\epsilon_{ff}^s}{\epsilon_{ff}^x} \frac{\Phi_s}{\Phi_x} \frac{\sum_{XY}(\phi_{s,XY}\cdot n_{s,XY})}{\sum_{XY}(\phi_{x,XY}\cdot n_{x,XY})} \nonumber \\ 
~& \times \left(\frac{\left[(C_{ff}^x - C_r^x - C_\alpha^x) - C_{w}^x\right] \cdot C_b^x}{\left[(C_{ff}^s - C_r^s - C_\alpha^s) - C_{w}^s\right] \cdot C_b^s} \right).
\end{align}
In this formulation $N$ denotes the number of target atoms, $\omega$ is the detector livetime, $\kappa$ accounts for the down-scatter and attenuation of neutrons from transport through the detector material and target backing, $\epsilon_{ff}$ denotes the fission fragment detection efficiency, $\Phi$ represents the neutron flux and $\sum_{XY}(\phi_{XY}\cdot n_{XY})$ is the spatial beam and target overlap term, which accounts for non-uniformity of the beam and target shape.  The $C$ terms are detector counts, with $C_{ff}$ being the number of fission fragment candidate events which are then corrected for various background classes. $C_\alpha$ is the beam-uncorrelated, pile-up $\alpha$-decays misidentified as fission. $C_r$ is the beam-correlated background, or neutron-induced events misidentified as fission.  $C_w$ is the wraparound correction, which are fission events assigned an incorrect neutron energy.  Finally $C_b$ is the correction for contamination of the targets with other fissile isotopes.  

Traditionally, the ratio formulation in Eq.~\eqref{eqn:xsCalc} is used to cancel terms that are the same for both actinides, such as $\epsilon$ and $\Phi$.  However, making such cancellations involves many assumptions.  In the case of the \ftpc{}, with its high fidelity data, these terms can be explored more directly.  The approach taken to determine each term in Eq.~\eqref{eqn:xsCalc} is summarized below, while Table \ref{tab:sections} provides a guide to the relevant sections of this article where details of the measurements and their uncertainties are discussed. 

The ratio of the livetimes of the two volumes of the detector $\omega_s/\omega_x$ was shown to be unity. The correction for beam scattering and attenuation in the detector and target backing, $\kappa$, was determined with a simulation using the MCNP package\cite{MCNP}.  Detector efficiency, $\epsilon_{ff}$, is determined by measuring the energy loss of fission fragments as a function of the exit angle from the surface of the target, $\theta$, in combination with a Monte Carlo simulation.  The neutron flux ratio $\Phi_s/\Phi_x$ is constructed to be 1 by the arrangement of back-to-back targets while the overlap term for non-uniformity was corrected for by measuring the relative distribution of target material with $\alpha$-decay vertex tracking, and the relative distribution of beam flux with fission fragment vertex tracking.  Fission fragment identification $C_{ff}$ was achieved with information from the specific ionization of the tracks in the \ftpc{}. The beam-correlated background $C_r$ was found to be negligible by measuring a blank Al target in beam. The beam wraparound contribution $C_w$ was corrected for by fitting the pattern of fragment counts in the known beam timing structure. Beam-uncorrelated background $C_\alpha$ was found to be negligible by measuring $\alpha$-decays during beam-off periods.  
The $C_b$ term accounts for contaminant isotopes, whose contribution is determined with a combination of mass spectroscopy data and published fission cross-section evaluations of the contaminant isotopes.  The $C_b$ term scales with the total fission rate and is therefore a multiplicative ($<$ 1) correction.
Finally, the ratio $N_s/N_x$ is measured using a silicon detector system separate from the \ftpc{}~\cite{Monterial2021}.  

\begin{table}[ht]
    \centering
    \begin{tabular}{c|l|l}
    \hline\hline
        Term in Eq.~\eqref{eqn:xsCalc} & Description & Section \\
        \hline
        $\omega$ & detector livetime & \ref{sec:livetime} \\
        $\kappa$ & beam attenuation & \ref{sec:attenuation} \\
        $\epsilon_{ff}$ & detection efficiency & \ref{sec:efficiency} \\
        $\Phi$ & neutron flux & \ref{sec:overlap} \\
        $\sum_{XY}(\phi_{XY}\cdot n_{XY})$ & beam-target overlap & \ref{sec:overlap} \\
        $C_{ff}$ & fission counts & \ref{sec:frag_selection}\\
        $C_{r}$ & beam background & \ref{sec:nuclear_recoils} \\
        $C_{\alpha}$ & $\alpha$-decay background & \ref{sec:beam_uncorrelated}\\
        $C_{w}$ & beam wraparound & \ref{sec:wraparound} \\
        $C_{b}$ & isotopic contaminant & \ref{sec:contamination}\\
        $N$ & target atom number & Ref.~\cite{Monterial2021} \\
        \hline\hline
    \end{tabular}
    \caption{Summary of terms in Eq.\eqref{eqn:xsCalc} with section where each is discussed.}
    \label{tab:sections}
\end{table}

%% file: Short_Experiment_Description.tex
\label{sec:experiment_description}
In this section we provide a brief description of the \ftpc{}. Further details on the design and operation of the \ftpc{} can be found in Ref.~\cite{Heffner2014}.  The \ftpc{} readout hardware consists of custom data acquisition electronics with an ethernet interface (EtherDAQ).  A complete description of the electronics can be found in Ref.~\cite{Heffner2013}.
Details on tracking data reconstruction for a fission cross-section ratio analysis are described in Ref.~\cite{Casperson2018}. 

The experiment was conducted at the Los Alamos Neutron Science Center (LANSCE) Weapons Neutron Science (WNR) facility on the 90L flight path \cite{LANSCE}.  The LANSCE facility provides a pulsed white neutron source.  For this experiment the pulses were delivered at 100 Hz.  Each pulse of approximately 625 $\mu$s length consisted of micro-pulses spaced by 1.8 $\mu$s.  A neutron time-of-flight measurement was employed to determine incident-neutron energies.

\subsection{\ftpc{} Design}
\label{sec:ftpc_design}
\begin{figure}[ht]
\centering\includegraphics[width=1.\linewidth]{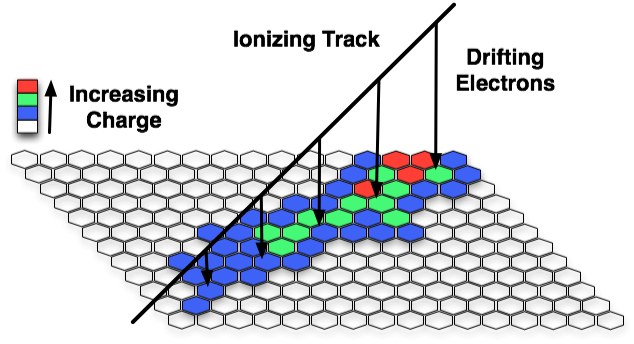}
\caption{\label{fig:hex_drift} A schematic representation of the \ftpc{} data reconstruction.  Each hexagonal segment of the anode is instrumented and the signal size and time are recorded. The positions of the segments provide 2-D information on the charge cloud shape while the relative arrival time of the signals can be used to reconstruct the third dimension if the constant electron drift velocity is known.}    
\end{figure}

A TPC \cite{Nygren} works on the basic principle that electrons will drift through a gas at a fixed average velocity when in a uniform electric field.  By segmenting the anode and instrumenting each segment with a separate channel, a 2-D image of the charge cloud generated by an ionizing particle (\eg{} a fission fragment) projected on the anode can be reconstructed.  By monitoring the relative arrival time of each signal the 3rd dimension of the track can then be reconstructed.  Fig.~\ref{fig:hex_drift} shows a schematic representation of the \ftpc{} data reconstruction process.

\begin{figure}[ht]
\centering\includegraphics[width=1.\linewidth]{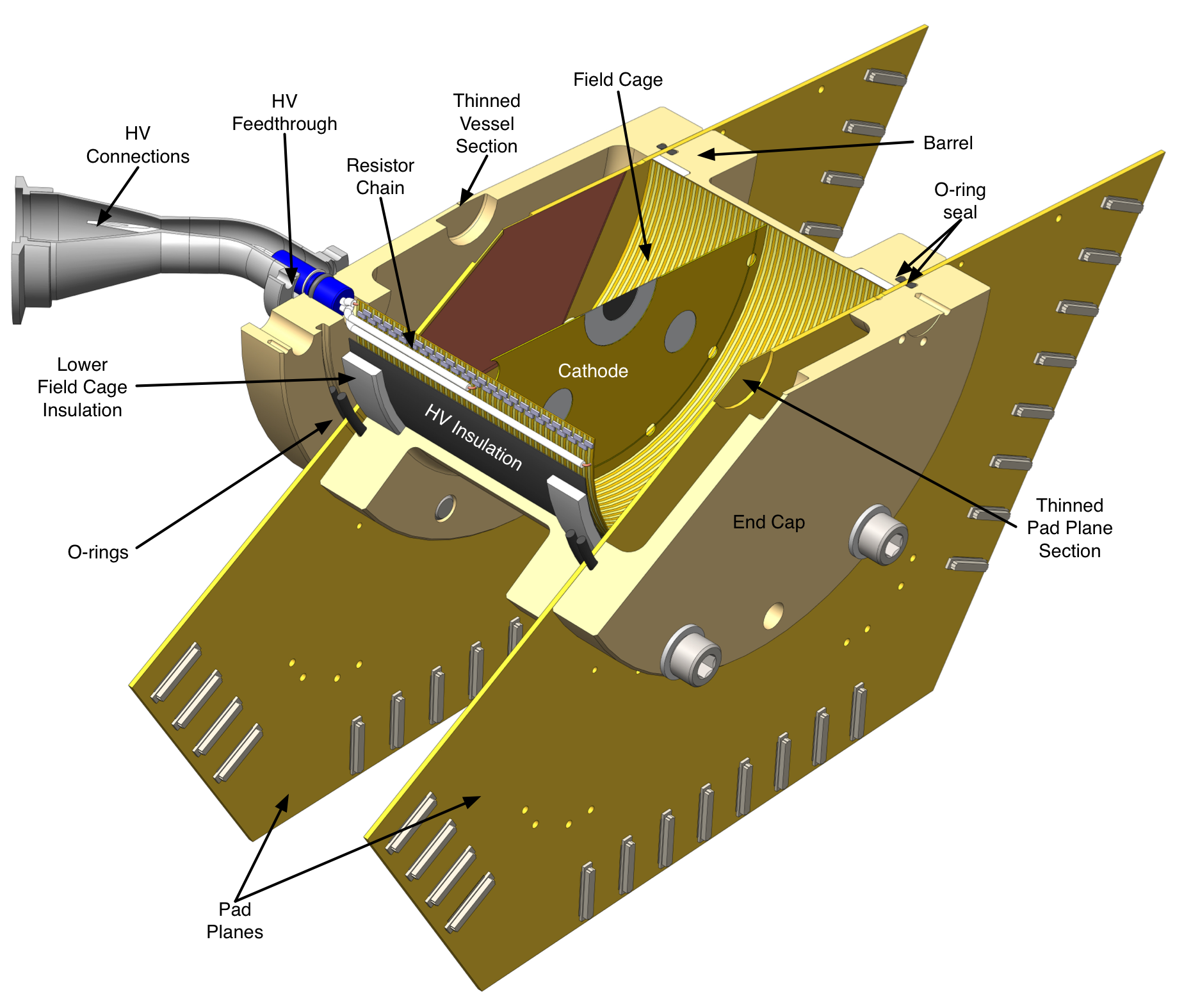}
\caption{\label{fig:cutaway} A cutaway image of the \ftpc{}.  The neutron beam passes through the thinned sections of the vessel and pad plane.  The actinide target is mounted in the center of the cathode.  Taken from Fig.~2 of Ref.~\cite{Heffner2014}.}    
\end{figure}

The \ftpc{} consists of two gas-filled  detection volumes viewing a single central target (see Fig. \ref{fig:cutaway}).  The two cylindrical volumes are filled with a common drift gas, in this case a mixture of argon with 5\% isobutane at a pressure of 550 torr.  The target is mounted on the central plane, which acts as a cathode.  

The actinide target for this measurement consisted of a 0.25 mm thick aluminum backing with a 90 $\mu$g/cm$^2$ vacuum volatilized uranium deposit on one side and a 118 $\mu$g/cm$^2$ molecular plated plutonium deposit on the other.  The actinide deposits are circular with 1 cm radius.  The targets were produced at Oregon State University and the process is described in Ref.~\cite{Loveland2016}.  The target isotopic composition is discussed in Sec.~\ref{sec:contamination}.

The cathode is connected to a current pre-amplifier through a high voltage decoupling capacitor.  When a charge cloud begins to drift away from the cathode an image charge is generated on the cathode.  The pre-amplifier records this charge and provides a high speed signal that is used to determine the neutron time-of-flight (see Sec.~\ref{sec:neutronEnergy}).  

Electron multiplication on the anode is achieved with a Micro MEsh GAseous Structure (MICROMEGAS) \cite{Giomataris}, which is typically operated with a gain of around 50.  The detector is sensitive to ions ranging from protons to fission fragments. 

\subsection{Particle Identification with the \ftpc{}}
\label{sec:particle_id}
One of the main advantages of the \ftpc{} is its superb particle identification (PID) capability, which comes primarily from track reconstruction. The value of
the specific ionization, $\mathrm{d}E/\mathrm{d}x$, along the length
of a track (the ``ionization profile'') is a near unique discriminant for the
particle types of primary interest: protons, \talpha-particles, and fission
fragments. Four parameters derived from reconstructed tracks can be used for
particle identification:

\begin{enumerate}
  \item Energy, $E$: Reconstructed track energy
  \item Length, $L$: Length of the reconstructed track
  \item Bragg Value, $BV$: The maximum $\mathrm{d}E/\mathrm{d}x$ of the
    ionization profile
  \item Bragg Position, $BP$: The position of the maximum $\mathrm{d}E/\mathrm{d}x$ value, relative to the track length (\ie, in the range [0,1]).
\end{enumerate}

\begin{figure}[ht]
\centering
\includegraphics[width=1.\linewidth]{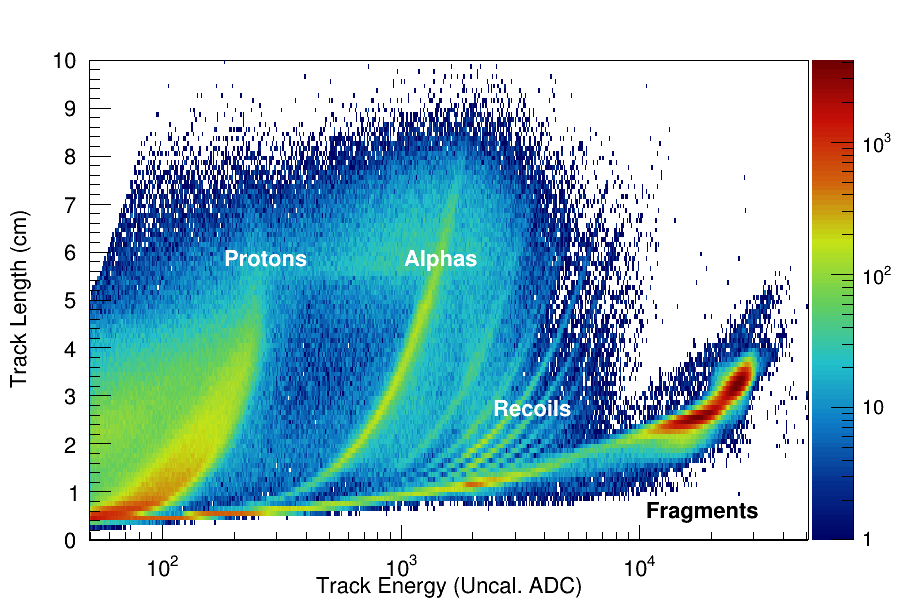}
\caption{\label{fig:pid_example} Track length vs.~particle energy during data collection, with various interactions labeled. Note the rich dynamic range: both the energy and intensity (color) axes are logarithmic.}
\end{figure}

Examination of this four parameter phase space reveals regions populated by the different particle types of interest. In practice, reconstructed track length and energy provide sufficient information to separate the various particle interactions of interest (see Fig.~\ref{fig:pid_example}).

Recoil protons from p(n,n) interactions, primarily with the isobutane component of the detector gas, that deposit all of their energy in the \ftpc{} active volume (``fully contained'') form a clear band at the left of Fig.~\ref{fig:pid_example}. At low proton energies, this band merges with recoils from heavier nuclei or degraded fission fragments.  The smearing of the band to lower energies is the result of partial proton tracks that were not fully contained within the active area of the detector.
The \talpha{}-band from spontaneous decay and (n, \talpha{}) reactions appears to the right of the proton recoils. The lightly populated bands at energies above the \talpha{}-band are recoils from heavier nuclei (such as $^{12}$C or $^{16}$O) or break-up reactions from high energy neutron interactions in the detector or gas. Removal of potential \talpha{} backgrounds in the fission count is discussed in Sec.~\ref{sec:beam_uncorrelated}.

The band at the highest energies in Fig.~\ref{fig:pid_example} is the fission fragment distribution. The structure exhibits the familiar double peak distribution of the light and heavy fragments. The low-energy tail comes from straggling and energy-loss in the target deposit, and eventually merges with the low-energy tails of the other distributions. In practice, we implement an energy cut at 7000 ADC to remove the low energy contaminants. Particle identification cuts for fission fragments and corrections for the particle selection efficiency are discussed in Sec.~\ref{sec:frag_selection} and Sec.~\ref{sec:efficiency} respectively.

%% file: Shape/Neutron_ToF.tex
\label{sec:neutronEnergy}
 
A measurement of the neutron time-of-flight (nToF) is used to determine the incident-neutron energy.  A signal from the LANSCE accelerator provides a reference for the start time, while detection of a fission event with the \ftpc{} fast cathode amplifier provides a stop time.  The construction of the cathode is described in Sec.~\ref{sec:experiment_description} and Ref.~\cite{Heffner2014}. Photons ($\gamma$-rays) produced in  the tungsten spallation target arrive at the detector simultaneously, producing a photo-fission peak in the timing spectrum. This allows for the determination of any time delay between the accelerator pulse signal and the fast cathode signal and provides a measurement of the timing resolution of the system.
 
The flight path length is determined by placing a carbon filter in the beamline between the neutron source and the \ftpc{} for a small subset of the data. $^{12}$C has strong neutron scattering resonances at well known energies \cite{Chadwick2011}.  The carbon filter produces a ``notch" in the measured nToF distribution at 2.08 MeV.  The notch in combination with the photo-fission peak allows us to determine the distance between the tungsten spallation target and the actinide targets in the \ftpc{}.  
The measured target distance of 8.059$\pm$0.003 m is thus used.  The primary source of uncertainty in this value comes from event statistics. 

\begin{figure}[ht]
\centering
\includegraphics[width=1.\linewidth]{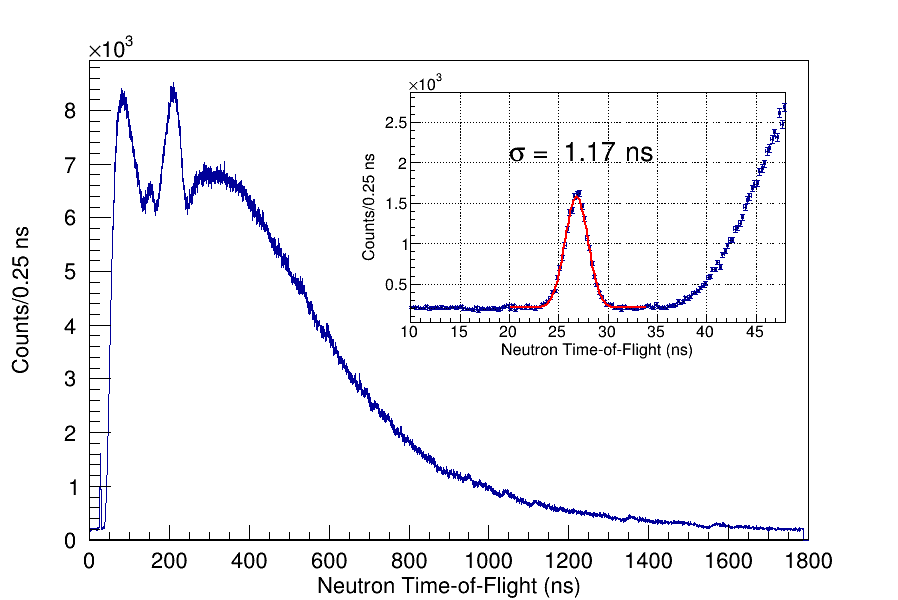}
\caption{Neutron time-of-flight distribution of the combined \pu[239] and \u[235] targets. The inset shows a Gaussian fit to the photo-fission distribution, yielding a timing resolution of $\sigma = 1.17$ ns.}
\label{fig:ToF}
\end{figure}

The timing of the cathode signal is extracted by applying a digital moving-average filter \cite{BOGOVAC20092073} and interpolating the rising edge back to the zero-crossing. The resulting timing measurement is found to be correlated with the cathode signal amplitude, which is corrected for in  post-processing with a quadratic function determined by fitting the distribution of cathode timing and amplitude for events from photo-fission.   The nToF resolution of 2.75$\pm$0.02 ns FWHM is determined by fitting the photo-fission peak with a Gaussian distribution on a flat background (Fig.~\ref{fig:ToF}).

Every anode track that falls within a given time window from each cathode signal is assigned the same corresponding nToF. The time window equals 1.5 times the maximum drift time of the electrons in one \ftpc{} volume.  This simple approach prioritizes nToF assignment to fragment tracks, disregarding whether other types of tracks receive a correct nToF.  Under this track-nToF correlation algorithm, fission fragments resulting from separate neutron interactions within the same event might be assigned the same nToF instead of two separate ones. The low fission rate in the \ftpc{} ($\sim$ 1 interaction per every 5 macropulses) reduces this pile-up likelihood. Indeed, only 0.12\% of fragment tracks with energy $>10000$ ADC appear with another fragment in the same event and receive the same nToF.   

The cathode fission fragment detection efficiency relative to the anode for the two actinide deposits is found to be 99.8\% for events that pass the fission selection cuts (see Sec.~\ref{sec:frag_selection}). No correction is needed as the efficiency cancels in the ratio with an uncertainty of $<0.1$\%. 

%% file: Shape/Fission_Count.tex
\label{sec:frag_selection}

\begin{figure}[ht]
	\centering
   \includegraphics[width= 1.\linewidth]{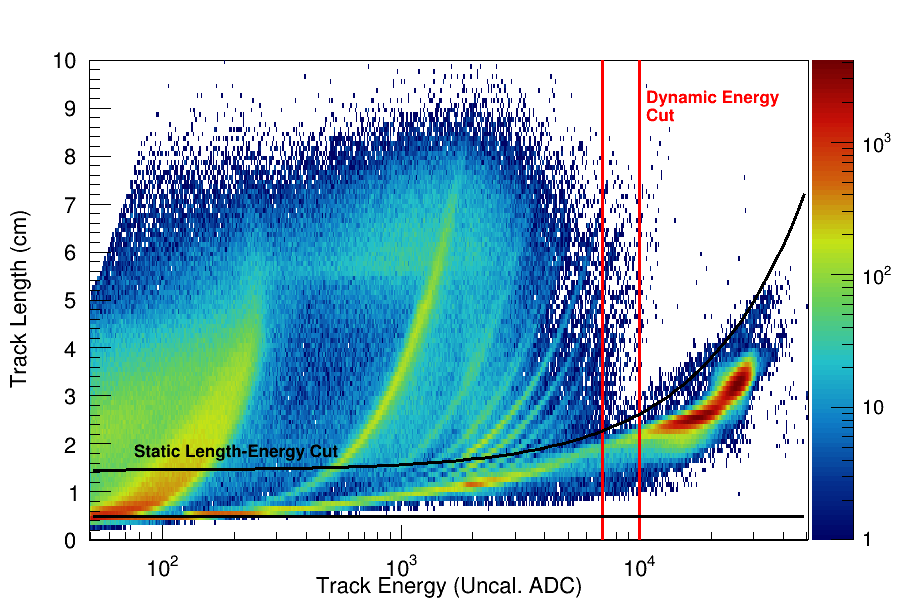}
   \caption{Selection cuts applied to the length vs.~track energy distribution of selected particles. Two static cuts remove background contributions, while one dynamic cut is varied within the range demonstrated. The dynamic cut is used to determine residual uncertainties from the efficiency correction.}
   \label{fig:lvadc_labelled}
\end{figure}

As described in Sec.~\ref{sec:particle_id}, excellent particle identification is a central capability of the \ftpc{}, which derives
primarily from track reconstruction. The primary parameter space used for particle identification is track length (L) plotted against track energy (E) (Fig.~\ref{fig:lvadc_labelled}).

To select fission fragments we apply cuts beyond a simple minimum energy threshold:

\begin{itemize}
    \item Require track start location to reside within the target area,
    \item Require a minimum reconstructed track length of 0.5 cm to remove the low-energy tail,
     \item Require an energy-dependent reconstructed track length maximum, which removes overlapping fragment tracks above the main distribution,
    \item Require a reconstructed track polar angle of cos($\theta$) $>$ 0.2 to remove tracks that have the greatest amount of straggling in the target material, 
    \item Accept only events with a valid neutron time-of-flight,
    \item Apply cuts requiring the fragment track to have occurred during the incident-neutron pulse.
\end{itemize}


Visual representations of the fission fragment selection cuts applied to the $L$~vs.~$E$ distribution are displayed in Fig.~\ref{fig:lvadc_labelled}. 
Two static particle identification cuts remove non-fragment background, while a dynamic cut is used to estimate residual uncertainties in the fission fragment selection efficiency.  The effects of the dynamic energy and cos($\theta$) cuts on the fragment selection efficiency is described in Sec.~\ref{sec:efficiency}.  The uncertainty calculation using the dynamic cuts is discussed in Sec.~\ref{sec:results}.

%% file: Shape/Efficiency.tex
\label{sec:efficiency}
The efficiency with which the experimental configuration detects fission fragments, $\epsilon_{ff}$, is of central importance to the \ftpc{} cross-section ratio measurement. A thorough description of the efficiency model can be found in Ref.~\cite{Casperson2018}, the publication of the $^{238}$U/\u{} cross-section ratio.  For completeness, some of that discussion is revisited here. Additionally, some improvements to the model that have been made since the previous publication are described.

The \ftpc{} detection efficiency method corrects for lost fission fragments using a Monte Carlo-based physics simulation. The microscopic structure of the target and uncertainties in the model inputs make a full, physically accurate simulation intractable, so the model instead is fit to the data using a number of parameters.  The efficiency fitting procedure simply fits the detector effects and target interactions parametrically rather than attempting to correct the data using a model with unknowns.  

The detailed tracking information recorded in the \ftpc{} data can be used to capture a variety of fission fragment transport and loss effects.  In addition the model captures neutron-energy dependent angular distributions and fission fragment mass yields, and accounts for analysis data selections described in Sec.~\ref{sec:frag_selection}.

Parameters required to accurately represent the \ftpc{} data include fission product yields (FPY), kinematic boost, fission anisotropy, electronics saturation, fission fragment stopping power, target thickness, composition, and surface roughness. 
Monte Carlo simulations of these effects are included in the efficiency model, with the parameters determined by fitting observable distributions to the \ftpc{} data. 

The fission fragment detection efficiency changes as fragment energy cuts are applied, primarily as a result of variable energy loss in the target as a function of the fragment emission angle with respect to the target surface, $\cos(\theta)$.  A fragment traveling perpendicular to the target plane, $\cos(\theta)=1$, will have minimal energy loss, while one traveling parallel to the plane can have significant enough energy loss as to prevent it from being detected, by either falling below the minimum energy cut or stopping completely in the target.

\begin{figure}[ht]
\centering
\includegraphics[width= 1.\linewidth]{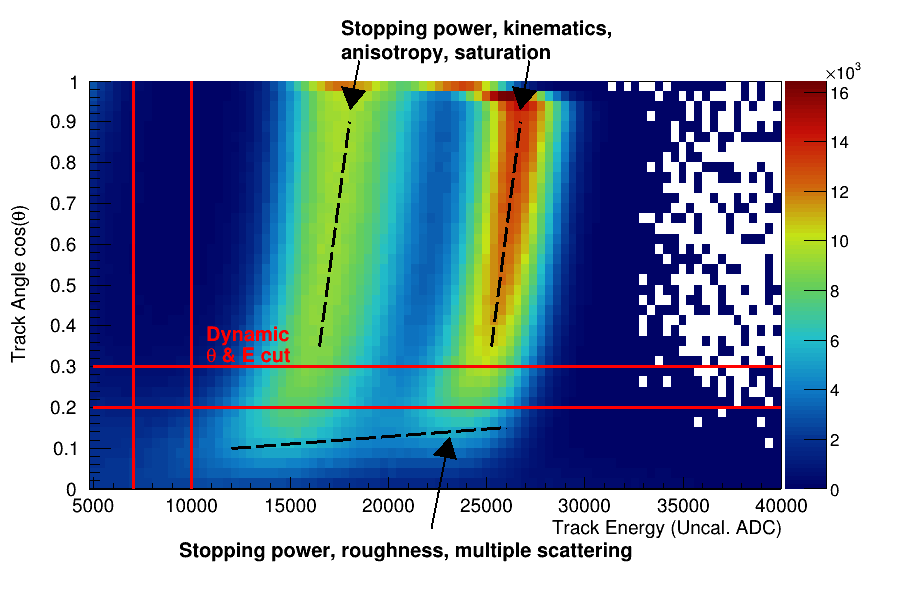}
\caption{\label{fig:Angle_vs_Energy_Explain} Fission fragment energy as a function of angle with respect to the target plane, where $\cos(\theta)=1$ is perpendicular to the plane.    
The two bands represent the light and heavy fission fragments. Labels are included to show the physical effects that control the shape of the fragment bands, in particular how they vary as a function of $\cos(\theta)$. A selection cut is applied to lower $\cos(\theta)$ and ADC values as the myriad effects that control the shape in this region cannot be accurately modeled at this time. The dynamic cuts are used in a variational analysis to estimate residual uncertainties and validate the method.}
\end{figure}

The energy angle relationship can be observed in Fig.~\ref{fig:Angle_vs_Energy_Explain}, where the distribution bends towards lower energies as the angle approaches a value of $\cos(\theta)=0$.  The angle vs.~energy distribution is the primary feature of the \ftpc{} data that is used to constrain the efficiency model.  

Annotations on Fig.~\ref{fig:Angle_vs_Energy_Explain} explain which physical processes affect different parts of the angle vs.~energy distribution for fission fragments. 
At forward emission angles close to the perpendicular, the primary processes that affect the shape of these distributions are:
\begin{itemize}
\item Kinematic boost and fission anisotropy; this will affect the relative fraction of fragments leaving the target at any given angle,
\item Fragment stopping power; the variation of stopping power with fragment energy and fragment mass will affect the slope of the distribution as a function of emission angle, since fragments at different angles travel through different amounts of target material on average,
\item Electronics saturation; at very forward angles ($\cos(\theta)>0.90$) saturation of pad-plane amplifiers occurs since such tracks occupy few pad-plane pixels.
\end{itemize}

At  emission angles close to parallel with the target,  the primary processes that affect the shape of these distributions include:
\begin{itemize}
\item Fragment stopping power; the variation of stopping power with fragment energy and fragment mass will strongly affect the tail-off of the distribution as a function of emission angle, since fragments in this angular range traverse a significant amount of target material,
\item Target roughness; if the target surface is rough, fragments could have large probability of re-entering the target material,
\item Fragment scattering; at emission angles close to the target surface, scattering processes (\eg{} multiple Coulomb scattering, nuclear recoil scattering) are more likely to result in the loss of a fragment back into the target than at forward angles.
\end{itemize}

For this \pu{}/\u{} dataset we employ dynamical thresholds on angle and energy of the fragments with lower bounds of $\cos(\theta) >$ 0.2 and ADC $>$ 7000.
In effect, the ``clean'' portion of the distribution at forward angles is used to determine the effect of kinematic boost and fission anisotropy and the model is used to determine the fraction of events lost at low  $\cos(\theta)$ values.  That is, the kinematic anisotropy effects are the primary determinant for how fission fragments are distributed in $\cos(\theta)$. By matching the model to data in a broad $\cos(\theta)$ region where straggling and surface effects are sub-dominant, we are able to correctly account for the effect of a selection cut excluding upper and lower bounds on $\cos(\theta)$. 

As will be discussed in Sec.~\ref{sec:results} and \ref{sec:validations} this approach is validated, and residual uncertainties estimated, by performing variational studies, observing the sensitivity of the cross-section ratio result to adjustment of the lower $\cos(\theta)$ and energy bounds up to $\cos(\theta) > 0.3$ and ADC $<$ 10000.  The cut variation ranges are labeled as ``dynamic" cuts in Fig.~\ref{fig:Angle_vs_Energy_Explain}.
To assess the efficiency we therefore proceed to use Monte Carlo simulations to recreate the measured $\cos(\theta)$ vs. energy distribution in a restricted $\cos(\theta)$ range.  The parameters are found by performing a multi-dimensional fit to minimize a $\chi^2$ comparison of data with the Monte Carlo simulation. 

\subsubsection{Efficiency simulation}\label{sec: efficiency simulation}
The simulation is constructed by considering the fragment mass and energy and its path from inside the target to the active area of the \ftpc{}.
In our previous work \cite{Casperson2018} the fragment mass was determined empirically using the energy of the fission fragments traveling at forward angles, which have minimal energy loss (for 0.775 $< \cos(\theta) <$ 0.95).  This energy distribution was narrowed with a deconvolution to offset the effects of target energy loss, and scaled to provide a fragment kinetic energy distribution. Fragment mass was estimated to kinematically correspond to a particular kinetic energy. 

The updated model reported here uses the FREYA \cite{FREAY2015,FREYA2018} event generator code, which supplies full fragment distribution information for incident-neutron energies up to 20 MeV. The FREYA generator produces a very good match to the data up to the turn-on of second-chance fission at approximately 6 MeV, above which there is increasing discrepancy between the model and data. Above an incident energy of 20 MeV, the FREYA model for 20 MeV is used.  The results derived using FREYA are consistent with our previous, empirical method.

The simulation must account for energy loss of the fission fragments resulting from transport through the target material.  The current model considers two methods, which provide internal cross validation.  With the first method, energy loss is described with a continuous-slowing-down-approximation (CSDA) model, using an empirical parameterized fit for ionization $dE/dx$ in UF$_4$ (derived from SRIM \cite{ZIEGLER20101818} stopping powers) as a function of fragment mass, charge, and energy. Random-walk small-angle scattering is applied at each tracking step proportional to $\sqrt{dE}$. Discrete large-angle scatters and straggling are neglected.

The second method utilizes Geant4~\cite{geant4} (version 4.10.05) which provides a physically detailed model of fragment transport.  While the default Geant4 electromagnetic physics code performs poorly when considering heavy ion scattering, we use a specialized model for heavy ion interactions, provided in the advanced examples supplied with Geant4, ``electromagnetic/TestEm7/”, which correctly handles the screened Coulomb scattering nuclear recoil contributions to ion stopping. The target material on both sides is modeled as UF$_4$ without other components.  We rely on variations in the parameterized areal density to absorb small differences in stopping power between the uranium and plutonium targets.  The results from the two methods are consistent within the uncertainties.

The target surface roughness must be considered to account for the difference in the average surface normal relative to the target plane.  The contribution to any roughness is largely a result of the target backing having roughness on a scale greater than the actual actinide deposit thickness. The surface roughness is described using a fractal noise model generated with Perlin noise fields~\cite{Perlin}. 
In contrast to our $^{238}$U/\u{} cross-section ratio determination~\cite{Casperson2018}, the impact was found to be minor. The difference between these data-sets is due to the targets used. The thin-backed $^{238}$U/\u{} target exhibited much greater roughness as a result of the carbon backing not being smooth, effectively obscuring the myriad of other scattering effects at lower values of $\cos(\theta)$.  By restricting to angles with $\cos(\theta) > 0.2$ in this analysis the effect is minimized.

The initial angular distributions of the fission fragments must be considered to determine their path length through the target.  Fission fragments are emitted isotropically in the center-of-mass frame.  Anisotropy effects are considered as a perturbation to the final angular distribution. At high incident-neutron energies not all of the incident kinetic energy is transferred to the fragment pair; some is transferred to undetected nucleons knocked out at the point of impact.  This effect was measured with an analysis of thin-backed $^{238}$U/\u{} \ftpc{} data fragment opening angles and reported in Ref.~\cite{hensle2020}.

\begin{figure}[ht]
\centering
\includegraphics[width= 1.\linewidth]{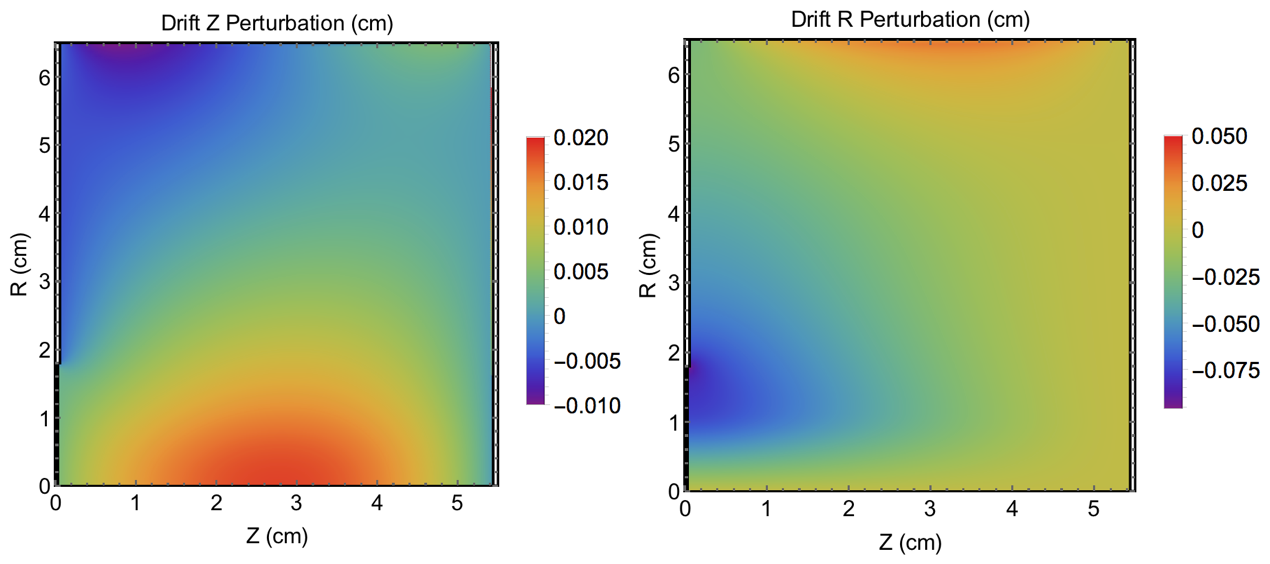}
\caption{Calculated electric drift field perturbations in the $z$- and $r$-directions resulting from ion back-flow space charge effects.  The displacement depends on the test-charge start position.}
\label{fig:DriftPerturbation}
\end{figure}

An additional effect that is present with high-activity targets like \pu{} is the build-up of space charge due to the substantial ion current drifting through the gas volume.  A finite element Poisson equation solver was used to calculate the ion density in the \ftpc{}, which showed that ions back-flowing from the MICROMEGAS gain region through the grid generate substantial perturbations to the electric drift field.
The reconstructed position perturbation is plotted in Fig.~\ref{fig:DriftPerturbation} for different initial charge positions.

This perturbation in the drift field results in a perturbation of the fragment angular distributions.  A separate correction for this angular perturbation has been integrated into the efficiency model, with the impact found to be negligible as it essentially appears as an additional anisotropy.  As shown in Fig.~\ref{fig:DriftPerturbation}, the magnitude of the radial perturbation is quite large near the target ($z$ = 0), and affects the track start vertex locations used in this measurement.  The impact of electric field distortions on the beam-target overlap calculation and its contribution to the cross-section ratio uncertainty is discussed in Sec.~\ref{sec:overlap}.

Parameters that are varied in the simulation include the target thickness, the roughness, total fission energy, and an additional scattering angle due to fragment interactions with argon in the \ftpc{} drift gas that can cause the detection angle to be different from the emission angle.  The perturbation on the emission angles for the fragments are distributed according to a second-order Legendre polynomial to account for fission anisotropy effects.  

\subsubsection{Model-Data Comparison}\label{sec:model-data comparison}

\begin{figure}[ht]
\centering
\includegraphics[clip=true, trim=10mm 10mm 7mm 0mm,width=1.\linewidth]{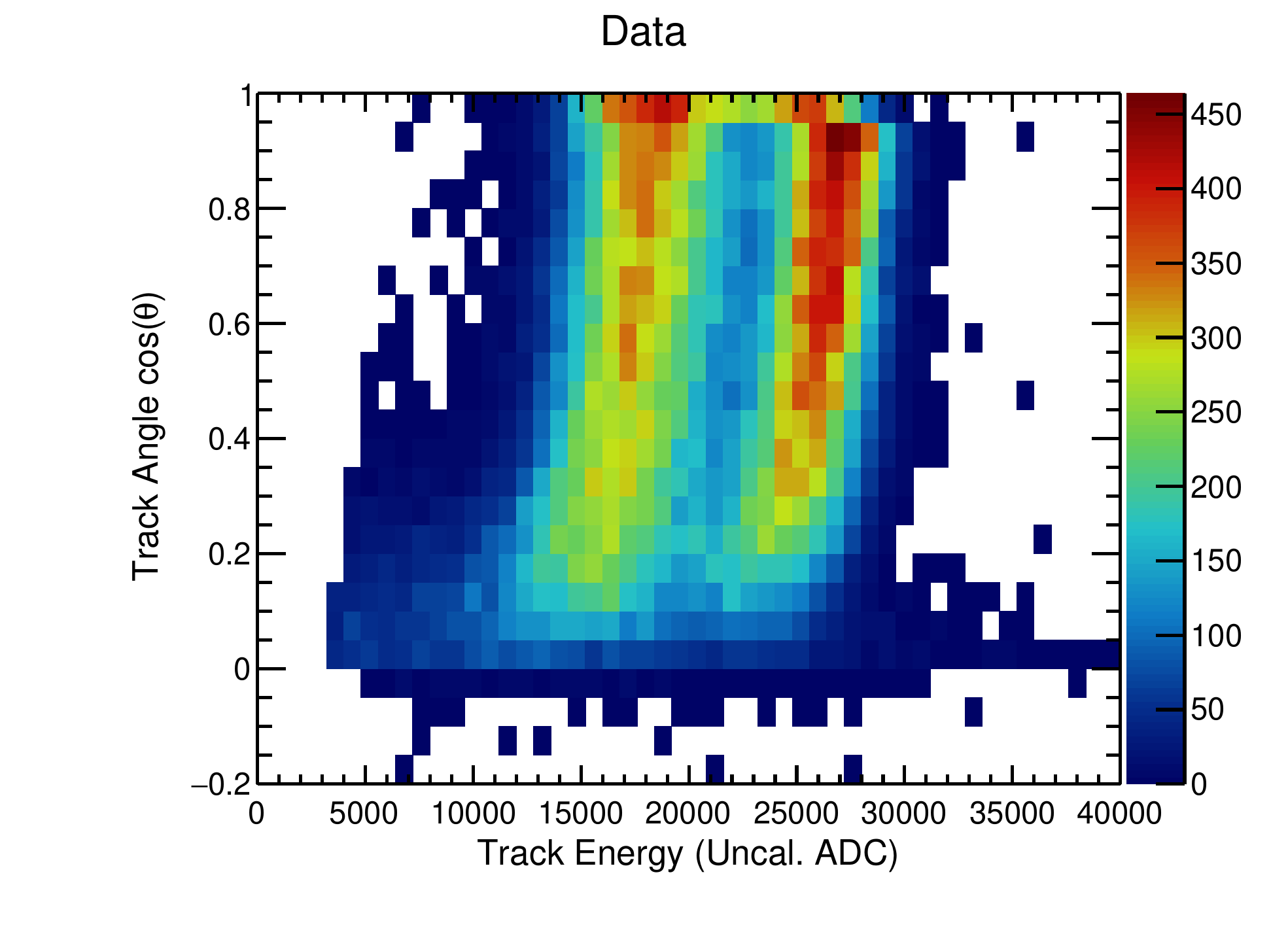}
\includegraphics[clip=true, trim=10mm 10mm 7mm 0mm,width=1.\linewidth]{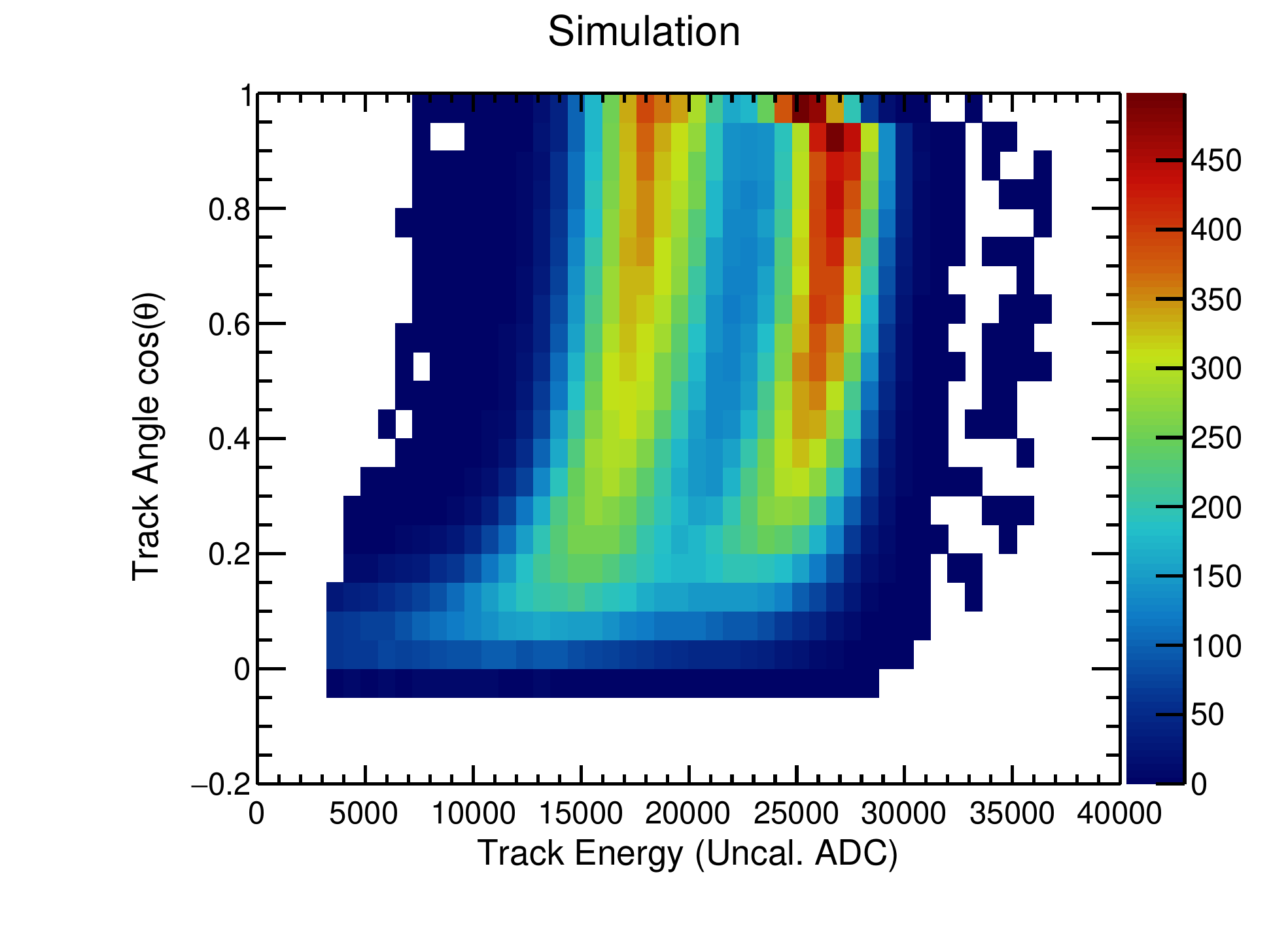}
\caption{\label{fig:EfficiencySimCompare} Fission fragment angle of emission vs.~energy for data and simulation from the \u{} side of the target and neutron energies between 10.0 and 10.6 MeV.  The two bands represent the heavy and light fission fragment distributions.}
\end{figure}

\begin{figure}[ht]
\centering
\includegraphics[clip=true, trim=20mm 12mm 13mm 0mm,width=1.\linewidth]{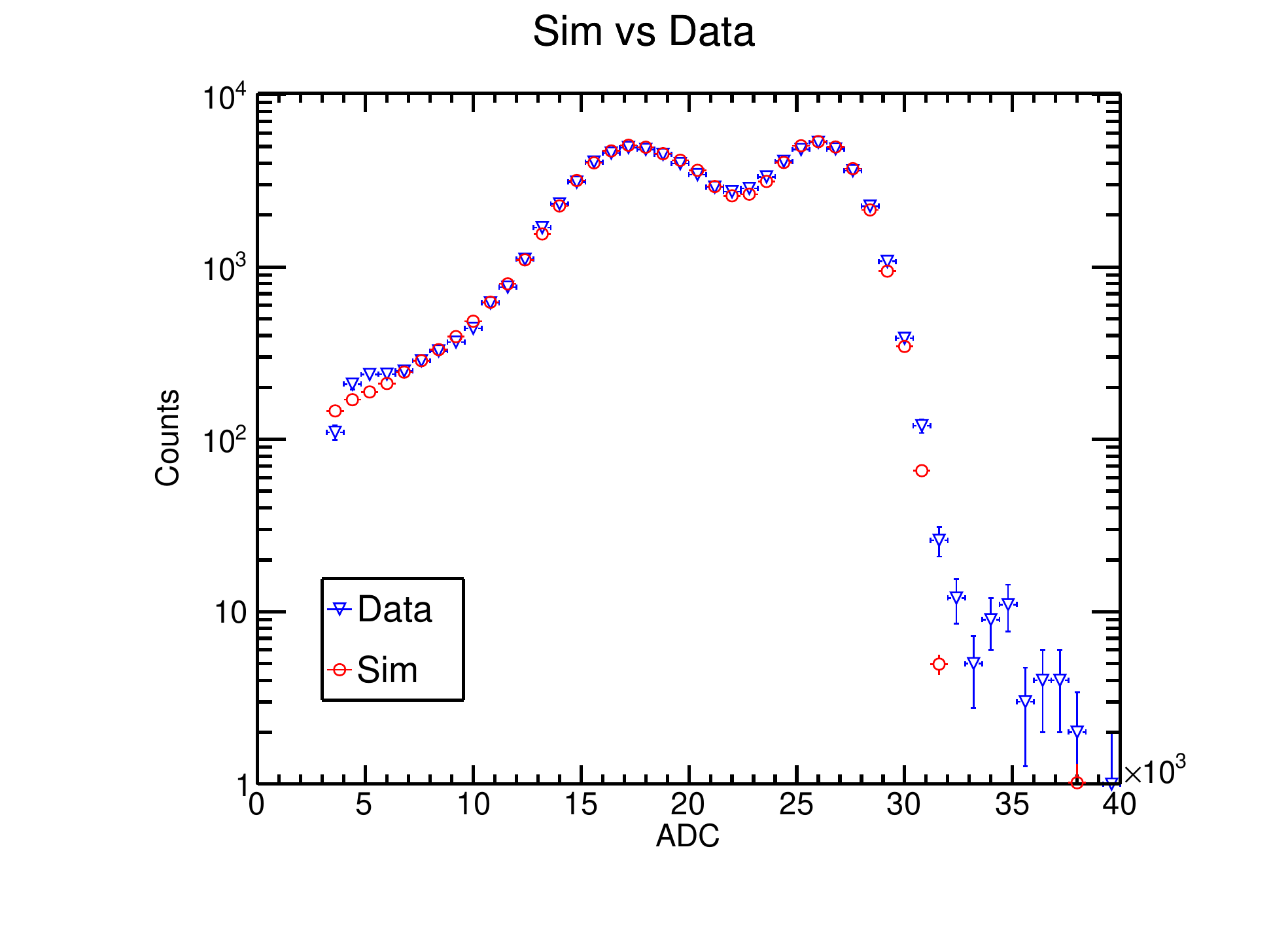}
\includegraphics[clip=true, trim=20mm 12mm 13mm 0mm,width=1.\linewidth]{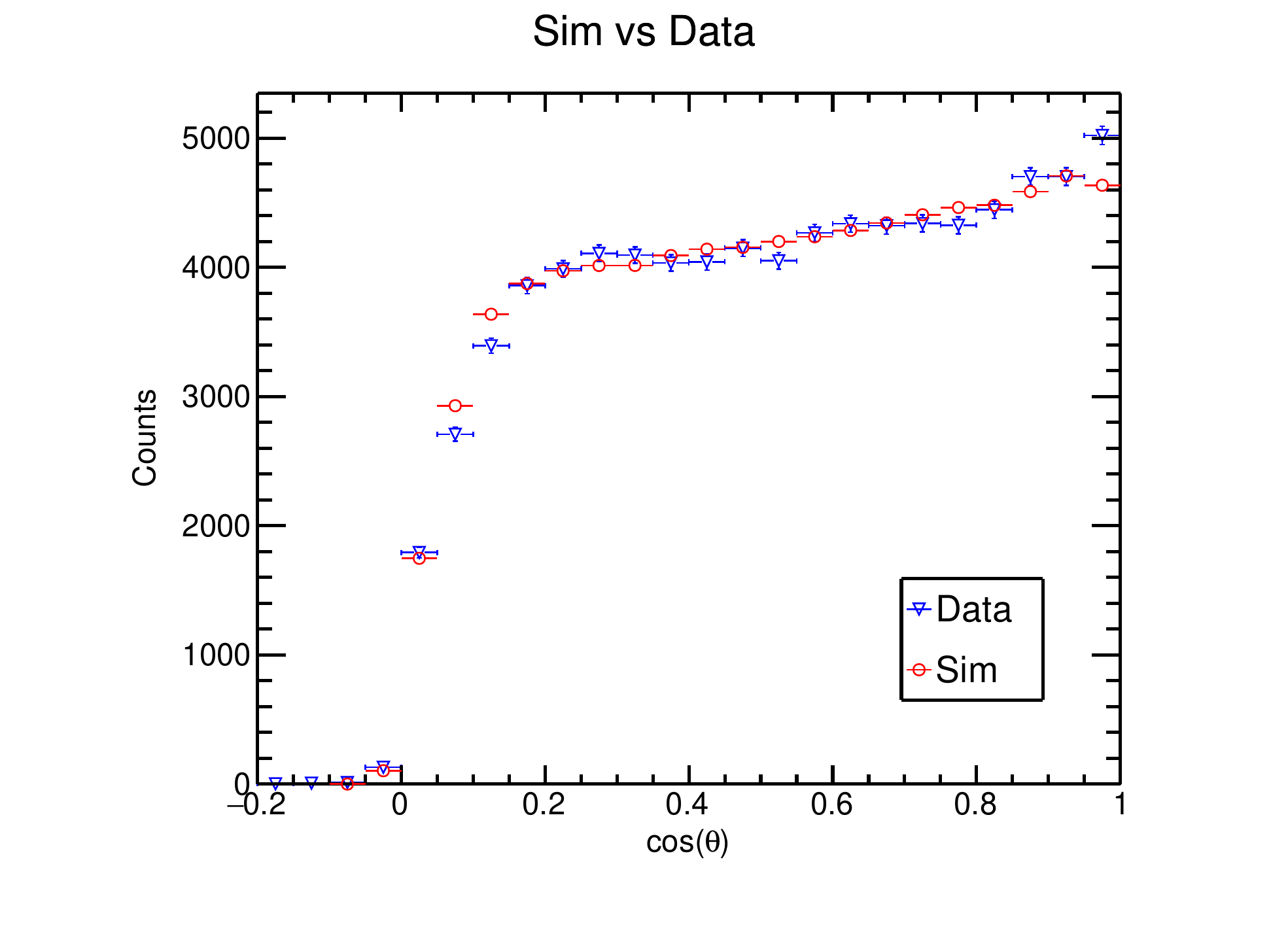}
\caption{\label{fig:EfficiencySimCompare2} Fission fragment energy and angle distributions comparing data and simulation.}
\end{figure}

The results of a representative simulation and comparison to data are shown in Fig.~\ref{fig:EfficiencySimCompare} and \ref{fig:EfficiencySimCompare2}. Fig.~\ref{fig:EfficiencySimCompare} shows the full 2-D parameter space of energy and detection angle for the data and simulation, while Fig.~\ref{fig:EfficiencySimCompare2} projects the 2-D distributions along each axis to show the level of agreement between simulation and data.  These distributions are for tracks originating from the \u{} side of the target for neutron energies between 10.0 and 10.6 MeV.  The two bands represent the heavy and light fission fragment distributions.  At angles near $\cos(\theta)=1$ the effects of electronics saturation can be observed, particularly for the higher energy fragment band.  At angles approaching $\cos(\theta)=0$ the bands bend towards lower energy resulting from losses in the target material.  To evaluate the efficiency, histograms are made for each isotope and each neutron energy bin.  The best-fit parameters that result from the $\chi^2$ minimization for the fragment transport (energy loss) and angular distributions are extracted and the expected fission fragment distributions as a function of incident-neutron energy are calculated.  The transport and anisotropy model provides the fraction of fission fragments entering the \ftpc{} that would pass analysis selection cuts.  

\begin{figure}[ht]
\centering
\includegraphics[width= 1.\linewidth]{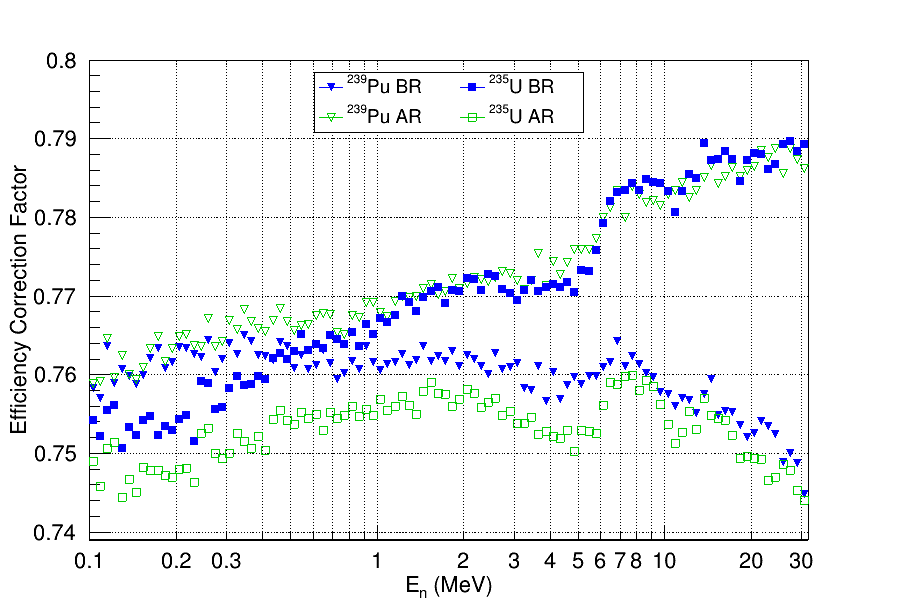}
\caption{\label{fig:Efficiency} Calculated fission fragment detection efficiency for the \pu{} and \u{} targets before (BR) and after (AR) 180 degree rotation of the \ftpc{}. The difference in the efficiency curves for the two orientations primarily reflects kinematic and anisotropy effects that are described in the text.}
\end{figure}

The efficiency as a function of neutron energy is shown in Fig.~\ref{fig:Efficiency}.  The results are shown for before and after the \ftpc{} was rotated 180 degrees with respect to the beam direction.  The orientation of the \ftpc{} in the beam has a significant impact on the angular distribution of the fission fragments and therefore on the efficiency. Further details on the analysis of the rotated data and its impact on the cross-section ratio are discussed in Sec.~\ref{sec:validations}.  A Monte Carlo procedure is used to estimate the uncertainty in the cross section resulting from the efficiency correction, which is detailed in Sec.~\ref{sec:results}.

The efficiency values are lower than those for the $^{238}$U/\u{} target reported in \cite{Casperson2018} since a stricter cut on $\cos(\theta)$ has been applied here that reduces the overall acceptance. 
The general trend in the efficiency as a function of energy is a result of the kinematic boost from neutron momentum transfer.  
For the fission fragment detection in the volume that is downstream of the neutron beam, the momentum transfer pushes the fragments in the forward direction, away from the target plane, and therefore increases the number of fragments entering that volume, and \textit{vice versa} for the upstream volume. 
The energy-dependent structure in the efficiencies is a consequence of the fission anisotropy, which changes rapidly with the onset of second chance fission.

The fragment energy cut is varied over a range of values, where the minimum value (ADC $>$ 7000) is above the $\alpha$-particle and recoil contaminants, and the maximum value (ADC $<$ 10000) removes a small fraction of fission events in the fission distribution. 
Similarly, the lower bound of the $\cos(\theta)$ selection region is varied over a range of values where the minimum value ($\cos(\theta) >$ 0.2) is above the onset of severe straggling degradation, and the maximum value ($\cos(\theta) <$ 0.3) does not significantly reduce the bulk of the fission fragment statistics.  As these cuts are varied the calculated efficiency varies accordingly, and the resulting cross-section ratio should be stable if the efficiency correction is accurate. The variation in the cross-section ratio resulting from the variation of these two selection cuts has been evaluated and is found to be small.  The effect of the variational analysis has been incorporated into the uncertainty assigned to the cross-section ratio that is presented in Sec.~\ref{sec:results}.



%% file: Shape/Overlap.tex
\label{sec:overlap}

The number of fission fragments recorded depends on the spatial distribution of the target atoms, the neutron beam, and how these overlap.  This dependence must be understood in order to use the fission fragment events to infer the cross-section ratio. In a typical fission chamber measurement an attempt is made to place two actinide targets in the same neutron beam so that when constructing the cross-section ratio the neutron flux term divides out, eliminating the need to measure it.  Attempts are also made to maximize the uniformity of both the target and beam.
When \emph{both} the beam and target have non-uniformities, a  correction to the overall normalization must be applied. If the beam shape varies with neutron energy, so too will this correction. It is the case for this measurement that both the \pu[239] target and the beam were nonuniform.

\begin{figure}[ht]
\centering
\includegraphics[width=0.9\linewidth]{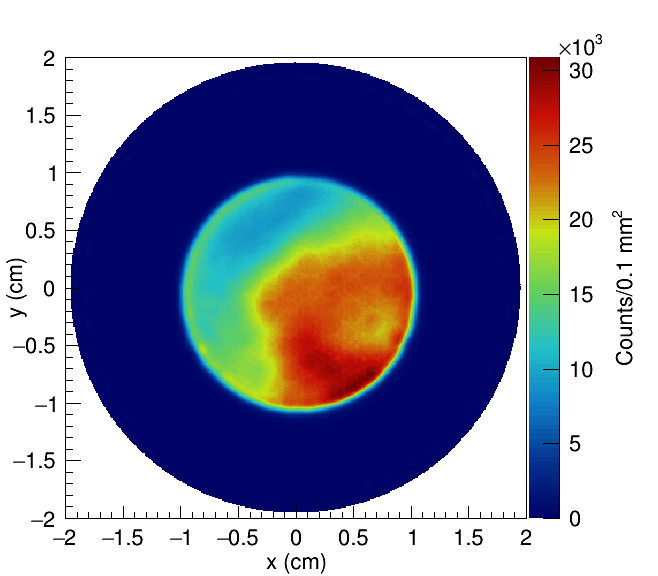}
\includegraphics[width=0.9\linewidth]{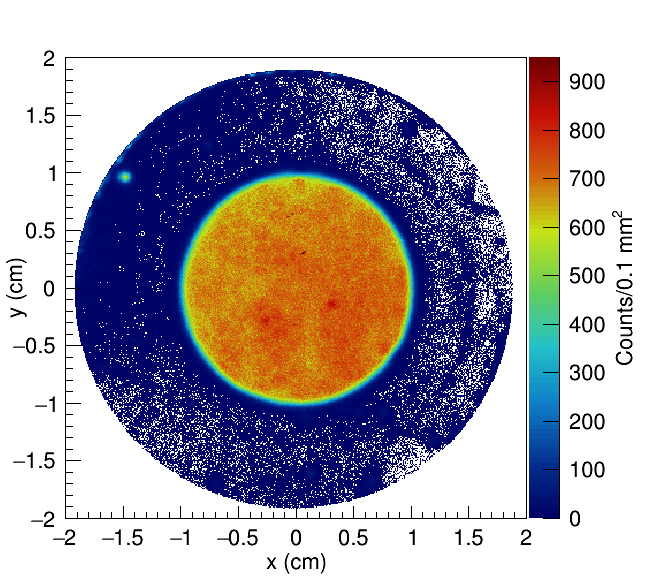}
\caption{Spontaneous \talpha-decay track start vertices for the plutonium and uranium targets. The rate of \talpha-decay is proportional to the target thickness.  The plutonium target has a significant non-uniformity, while the uranium target is uniform at the 5\% level}
\label{fig:p9Target}
\end{figure}

Evidence of the target spatial non-uniformity is depicted in Fig.~\ref{fig:p9Target}, which shows track start vertex locations for \talpha-particles arising from spontaneous decays of the target nuclei.  The rate of spontaneous \talpha-decay is proportional to the target thickness. The \pu{} target shows significant non-uniformity while the \u{} target is uniform at the 5\% level.  The background outside the target radius of 1 cm is a consequence of misidentified vertices resulting from pile-up tracks.  The high decay rate of \pu{} results in an overall increased background relative to \u{}.  
An additional background can be observed on the \u{} side as a result of small amounts of \pu{} contamination on the cathode that occurred during target insertion and detector assembly.  These backgrounds are cut from the analysis and have no effect on the measured relative distribution of the target material. 

\begin{figure}[ht]
  \centering
  \includegraphics[width=0.9\linewidth]{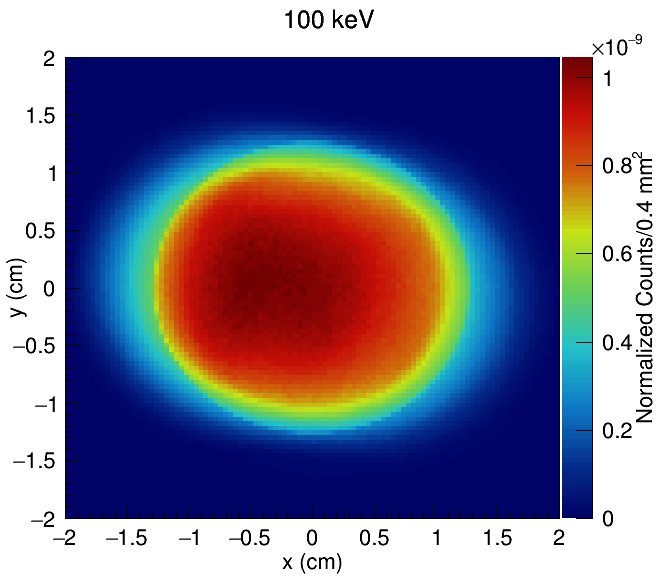}
  \includegraphics[width=0.9\linewidth]{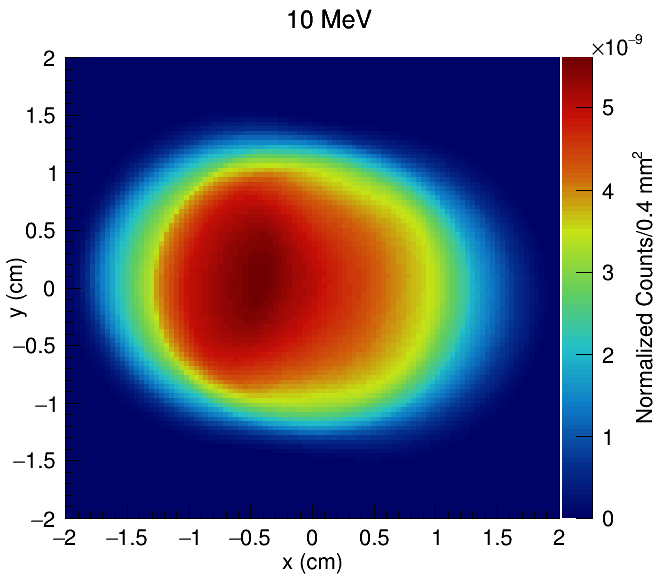}
  \caption{The spatial distribution of neutrons incident upon the upstream actinide target in the \ftpc{}, as determined by a simulation of the 90L beam line using MCNP: $E_n=$100 keV and  $E_n=$10 MeV. The non-uniformity is the result of the choice of collimator design which creates a pinhole camera effect. Note the change in the shape along the $x$-axis is parallel to the direction of the proton beam incident on the spallation target.}
  \label{fig:beamShapeMCNP}
\end{figure}

The non-uniformity in the neutron beam was modeled with a simulation of the neutron intensity in the 90L beamline at LANSCE using MCNP\cite{MCNP} and is shown for two different neutron energies ($E_n=$100 keV and  $E_n=$10 MeV) in Fig.~\ref{fig:beamShapeMCNP}. The non-uniformity is the result of the choice of collimator design which creates a pinhole camera effect.  The shape varies along the $x$-axis, which is parallel to the direction of the proton beam incident on the spallation target.  The non-uniformity in the beam has an energy dependence which is a result of the spallation-inducing protons losing energy as they traverse the $\sim$7 cm length of the tungsten target.

An energy-dependent correction term for the beam-target overlap must therefore be determined to account for the spatial variation in the rate of neutron-induced fissions.  For example, if the most intense regions of the beam and target coincide, the fission rate is enhanced relative to the case where the beam and target intensities are anti-correlated.  Not accounting for this effect could result in a systematic shift in the measured (n,f) cross section.

The \ftpc{}'s ability to measure \talpha-particle and fission fragment track vertices enables us to characterize the non-uniformity in both the beam and target with data so that an overlap correction can be applied to the measured cross-section ratio.  Although the beam shape can also be calculated with a Monte Carlo model, as described above, the simulation was not used to determine the correction to avoid any unknown systematic uncertainties in the simulation being propagated to the final result.

The overlap term in Eq.~\eqref{eqn:xsCalc} is a multiplicative correction factor that can be determined by binning the vertex data within the area of overlap and normalizing by the total number of neutrons and target atoms.  The normalized binned equation is given by

\begin{align}
\mbox{OT} &= \mbox{Overlap Term} = \sum_{XY}(\phi_{XY}\cdot n_{XY}) \nonumber \\
    \label{eqn:basic overlap term 2}
~ &= B \frac{1}{\sum_j n_j} \frac{1}{\sum_k T_k} \sum_i n_i T_i \,\text{,}
\end{align}

\noindent where $B$ is the total number of bins into which the data is divided, and the spatial bin index $XY$ is replaced by bin number index $i$, $j$, or $k$.
The number of neutrons traversing bin $i$ in any given length of time is $n_i$, and $T_i$ is the number of target nuclei in bin $i$.  The overlap term sums over the product of normalized total atoms and normalized total neutrons in each bin $(n_i/\sum_j n_j) \cdot (T_i/\sum_k T_k)$.
Since the total number of neutrons and target atoms are independent sums over the spatial bins, these terms can be brought outside the summation over $i$ as in Eq.~\eqref{eqn:basic overlap term 2}.

The summation over $n_j$ is a measure of the total number of neutrons in the beam, and can be obtained from Monte Carlo simulation, as shown in Fig.~\ref{fig:beamShapeMCNP}, or directly from the data. The direct data approach uses the shape of the target, given by the \talpha-track start vertices, and the distribution of fission fragment start vertices originating from the same target, to infer the beam shape.  Once known, the beam shape derived from this method can be applied to any {\it in situ} beam target, allowing us to determine the beam shape from one target and apply it to a second target in the same beam.  In other words this gives the option to use the relatively uniform \u{} deposit to determine the beam shape and apply it to the \pu{} overlap correction.  It also provides a cross validation whereby the \u{} and \pu{} should reproduce the same beam shape in the data.

One can verify that \ftpc{} experimental observables can be used to determine the overlap correction by beginning with the well-known equation describing the neutron-induced fission rate

\begin{equation}
    \label{eqn:basic induced fission}
    f = \sigma n L \frac{T}{V} \,\text{,}
\end{equation}

\noindent
where $\sigma$ is the interaction cross section, $n$ is the neutron flux, $L$ is the thickness of the target, $T$ is the number of target nuclei, and $V$ is the volume.  
The ratio $T/V$ is simply the number density of the target actinide (a physical constant), while the combination $L T / V$ is the areal density of the target.  

Eq.~\eqref{eqn:basic overlap term 2} can be recast in terms of the experimental observables $f_i$, the number of fission events detected in bin $i$ in a given length of time, and $\alpha_i$, the number of alpha particles detected over that same length of time in the same bin by re-writing Eq.~\eqref{eqn:basic induced fission} and solving for $n_i$,

\begin{align}
\label{eqn:fission observable}
n_i = \frac{1}{\sigma} \frac{f_i}{L_i} \frac{1}{\rho} \,\text{,}
\end{align}

\noindent
where the ratio $T/V$ has been replaced by the constant $\rho$. The number of target nuclei in any given bin can be related to the thickness of the target in that location:

\begin{align}
\label{eqn:target observable}
T_i = \rho L_i A_{bin} \,\text{,}
\end{align}

\noindent
where $A_{bin}$ is the area of bin $i$. Substituting Eqs.~\eqref{eqn:fission observable} and \eqref{eqn:target observable} into Eq.~\eqref{eqn:basic overlap term 2} gives

\begin{align}
\label{eqn:OT term almost there}
\mbox{OT}&= \frac{1}{\sum_j \frac{f_j}{L_j}} \frac{1}{\sum_k L_k} ~ B ~ \sum_i \frac{f_i}{L_i}L_i \,\text{,}
\end{align}

\noindent
All of the common constants $\sigma$, $\rho$, and $A_{bin}$ cancel, leaving a term purely governed by the shapes of the target and beam.

The spatially-dependent target depth, $L_i$, can be defined in terms of the spontaneous \talpha-decay rate from any given region of the target via

\begin{align}
\label{eqn:rate sub}
L_i &= \frac{\tau_{1/2}}{ln 2} \frac{1}{\rho A_{bin}} R_i \,\text{,}
\end{align}

\noindent
where $\tau_{1/2}$ is the half-life and $R_i$ is the spontaneous \talpha-decay rate in bin $i$.  Substituting Eq.~\eqref{eqn:rate sub} into Eq.~\eqref{eqn:OT term almost there} results in an overlap term that depends on the observed spontaneous alpha and fission fragment counts measured in time $t$ in a given spatial bin as

\begin{align}
\label{eqn:OT term}
\mbox{OT}&= B \left [ \sum_j \frac{f_j}{\alpha_j} \sum_k \alpha_k \right ]^{-1}  \sum_i \frac{f_i}{\alpha_i}\alpha_i \,\text{.}
\end{align}

\noindent where again the common constants of the \talpha-decay cancel. Note that the $\alpha_i$ term in Eq.~\eqref{eqn:OT term} cancels for the particular case of using the data from a single target, however it is included to illustrate that the beam shape and target shape can be derived from different sources.  This can be interpreted as
\begin{align}
\mbox{OT}& = \mbox{Normalization} \times \sum_i \mbox{Beam}_i \cdot \mbox{Target}_i \,\text{.}   
\end{align}

For example, if the target-shape terms are taken from a plutonium target, but the beam-shape terms are taken from a correlated uranium target, the overlap term would be given by 

\begin{align}
\label{eqn:OT example 2}
\mbox{OT} &= B \left [ \sum_k \alpha_k^{Pu}
\sum_j \frac{f_j^U}{\alpha_j^U} \right ]^{-1}
\sum_i \frac{f_i^U}{\alpha_i^U} \alpha_i^{Pu} \,\text{.}
\end{align}

\noindent
As was discussed previously in this section, the reason for such an approach is that
the relatively uniform uranium deposit requires little correction to show the shape of the beam (the $f^U_i/\alpha^U_i$ term is largely constant across bins) and it was assumed any potential systematic uncertainty from tracking bias would be minimized.

Taking the ratio of OT$^{Pu}$ and OT$^{U}$ results in a cancellation of the sum over number of spatial bins, $B$, and in the OT$^U$ sum the $\alpha$ term will also cancel, leaving the final ratio as

\begin{equation}
\label{eqn:OT worked example}
\frac{\mbox{OT}^{Pu}}{\mbox{OT}^U} = \frac{\sum_{XY}(\phi_{Pu,XY}\cdot n_{Pu,XY})}{\sum_{XY}(\phi_{U,XY}\cdot n_{U,XY})} = \frac{\frac{1}{\sum_k \alpha_k^{Pu}} \sum_j \frac{f_j^u}{\alpha_j^U} \alpha_j^{Pu}}{\frac{1}{\sum_i \alpha_i^U} \sum_l f_l^U} \,\text{,}
\end{equation}

\noindent
with all the physical constants and non-experimental terms cancelling, leaving a purely shape-dependent expression for the overlap term in Eq.~\eqref{eqn:xsCalc}.

\begin{figure}[ht]
\centering
\includegraphics[width= 1.\linewidth]{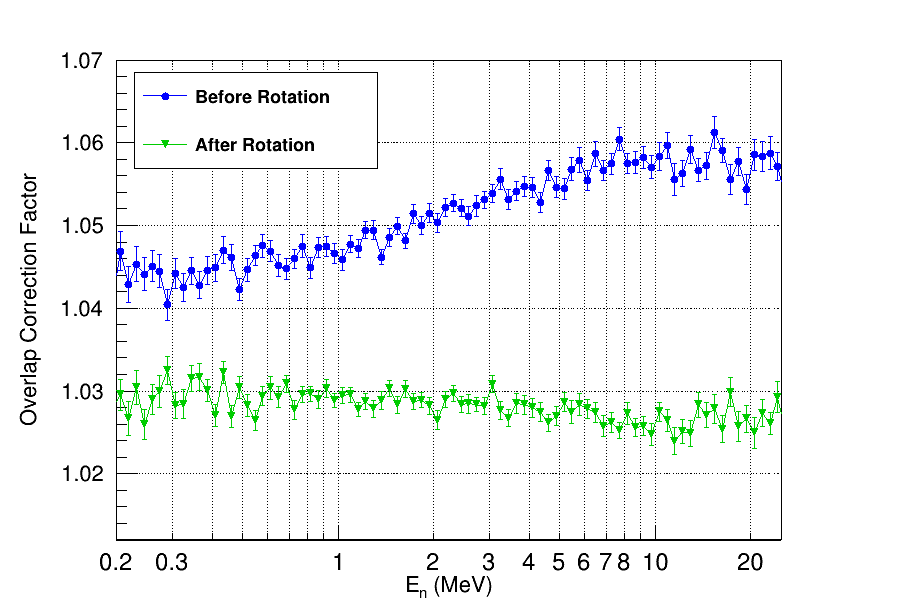}
\caption{The overlap correction for ``Before" and ``After" rotation of the \ftpc{}.  The two orientations were rotated 180$^\circ$ about the vertical axis with respect to the beam direction, which had the effect of reversing the beam and target maxima with respect to each other.  The overlap ratio was calculated using Eq.~\eqref{eqn:OT worked example}, where the uranium side is used to measure the beam for both sides of the target.}
\label{fig:overlap}
\end{figure}

Although the beam shape is assumed to be stable throughout the run, due to the target non-uniformity, the overlap term can be made to vary by rotating the target within the beam, where regions of greater or lesser target density overlap with regions of greater or lesser neutron beam flux.  The result of such a rotation is shown in Fig.~\ref{fig:overlap}.  The method used here relied purely on experimental data and did not incorporate any MCNP calculations for the beam flux.  In this case the fragments from the uranium side were used to infer the beam shape for both sides as previously discussed.  A fully self-corrected overlap term, whereby the plutonium data is used to infer its own beam shape, was also calculated and agreed within uncertainties, which acts in part as a validation of the method.  The error bars were determined with a Monte Carlo calculation assuming Poisson statistical variations of the counts in the track start vertex spatial bins. 

Similar to the efficiency calculation (Sec.~\ref{sec:efficiency}), agreement in the final cross-section ratio for the two orientations of the \ftpc{} with respect to the beam is one of the primary validations of these results and is further discussed in Sec.~\ref{sec:validations}.

\subsubsection{Impact of space charge distortion on the overlap term}
\label{sec:SpaceChargeCorrection}

As discussed in Sec.~\ref{sec:efficiency}, the build-up of space charge due to drifting ions distorts the electric drift field, leading to a perturbation in the observed fragment angular distribution.  This effect is accounted for in the fission fragment efficiency model.  
Electric field distortions also impact the track vertex distributions for both \talpha-tracks and fragments, which are drawn towards the center of the target, producing a target distribution that appears smaller in the data than the target is in reality.  As this effect primarily impacts the \pu{} target, the calculated overlap of the neutron beam is affected.

\begin{figure}
\centering
\includegraphics[width= 1.\linewidth]{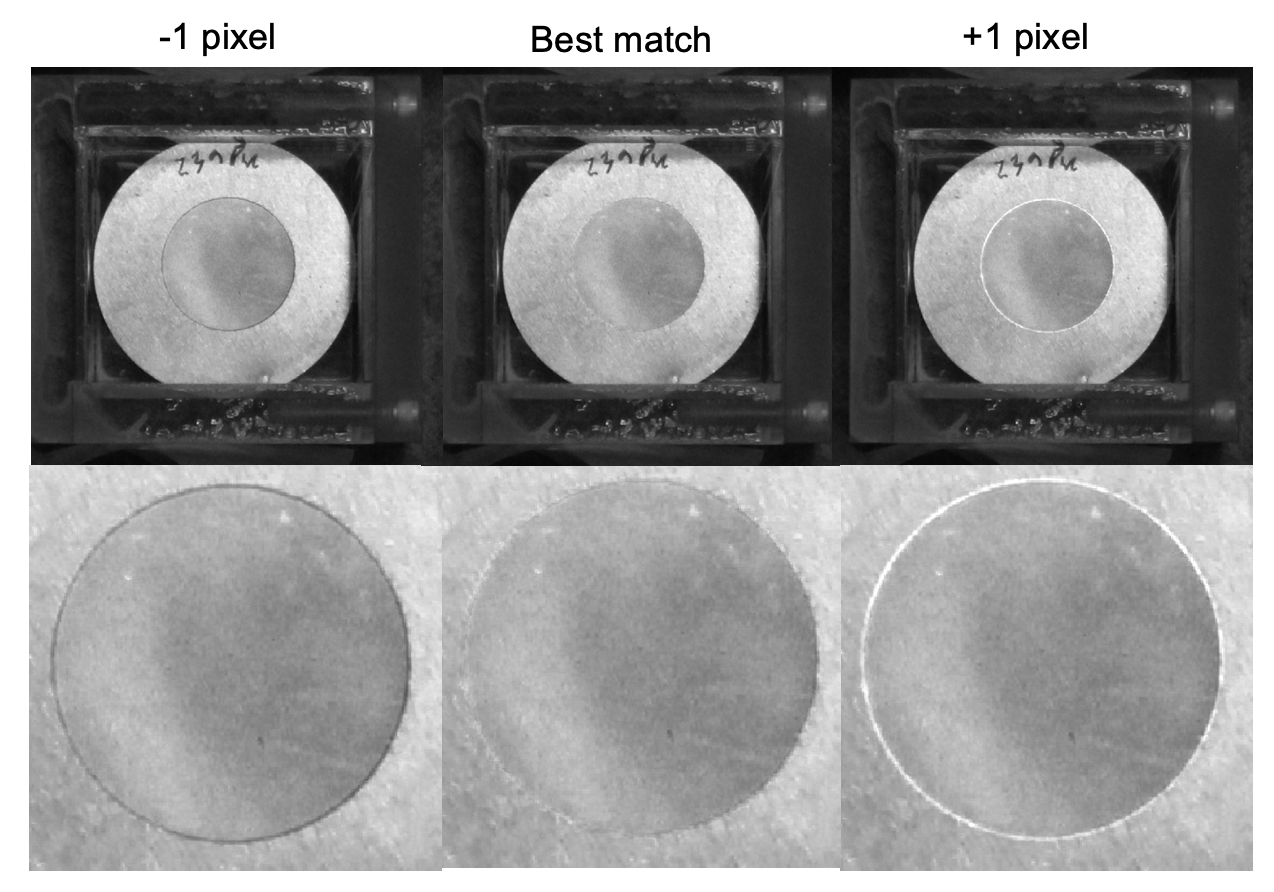}
\caption{
Photographs of the vapor-deposited \pu{} target and the backing disk (top row), and a more detailed examination of the \pu{} target (bottom row). In each the \pu{} target deposit is compared to a circle with radius given in pixels. The absence of edge artifacts (middle) indicates a matching radius.  The backing disk has a well known diameter, thereby providing an absolute length calibration for precisely determining the size of the \pu{} target.
}
\label{fig:targetRadius}
\end{figure}

Photographs of the target deposits were analyzed and compared to the inferred radii from the data (Fig.~\ref{fig:targetRadius}).  
Agreement was found between the \u{} photograph and data, while the radius inferred from the \pu{} data was found to be 2.6\% smaller than the photograph.  
This apparent shrinking in the \pu{} data is consistent with the model of electric field distortions due to space charge.  
A correction was applied directly to the track vertices that were used in the beam-target overlap calculation to account for the space charge effect in the overlap term.

\subsubsection{Alignment of Actinide Target Deposits}
\label{sec:target_alignment}

A slight relative misalignment of the pad-planes of the two volumes of the \ftpc{} of $<$ 200 $\mu$m is possible due to the detector's physical construction.  Such a misalignment could introduce a systematic shift in the beam and target overlap term since track start vertex finding is done independently in each drift volume and is referenced to the physical pad-plane geometry.  Beam-induced particles with sufficient energy to ``punch-through" the cathode result in tracks in both volumes that point to a common coordinate on the cathode.  These were used to determine the degree of misalignment so that a correction for the start vertex distributions could be applied.
The length vs.~energy distribution of such punch-through tracks indicates that these are primarily highly-energetic tracks from (n,\talpha) reactions, originating from behind the upstream MICROMEGAS mesh, and passing all the way through both volumes.

By selecting coincidence tracks matching the characteristic length vs.~energy distributions of punch-through, the coordinate offsets between the two volumes can be mapped out.
The alignment analysis was performed on high-gain data, in which the relatively low $dE/dx$ punch-through tracks are well reconstructed. 
Similar to what was done for the space charge effect, a correction was applied directly to the track vertices that were used in the  beam-target overlap calculation so that a separate alignment correction is not required for the overlap term.
The alignment correction applied to volume 1 was found to be -167 $\mu$m along the x-axis and -131 $\mu$m along the y-axis, within the expected 200 $\mu$m design specification.

%% file: Shape/Beam_Correlated.tex
In addition to detecting neutron-induced fission fragments from the actinides of interest, the \ftpc{} also records other beam-induced events which must be corrected for as shown in Eq.~\eqref{eqn:xsCalc}. These corrections include nuclear recoils and nuclear reactions from neutron scattering on the detector components ($C_r$) and (n,f) reactions on contaminant isotopes ($C_b$).  Wraparound neutrons (neutrons assigned an incorrect nToF) are also considered a beam-correlated background ($C_w$). 

\subsubsection{Nuclear Recoils Correction, $C_r$}
\input{Shape/Nuclear_Recoils}

\subsubsection{Wraparound Correction, $C_w$}
\input{Shape/Wraparound}

\subsubsection{Contamination Correction, $C_b$}
\input{Shape/Contamination}

%% file: Shape/Nuclear_Recoils.tex
\label{sec:nuclear_recoils}
The \ftpc{} detects beam-induced events such as nuclear recoils from scattering on the various detector components and nuclides in the gas (\eg{} \hyd, $^{40}$Ar, $^{12}$C). These recoils, as well as other beam-induced processes like (n,\talpha) reactions and spallation reactions from high energy neutrons, also produce detectable events that could be mistakenly identified as fission fragments. These contributions, which contaminate the $C_{ff}$ count, can be assessed directly by measuring a ``blank'' aluminum target with no actinide deposits. Approximately 43 hours of data were taken with such a target during the beam cycle, which is approximately 5\% of the total time with an active target in the beam.

\begin{figure}[ht]
  \centering
  \includegraphics[width=1.\linewidth]{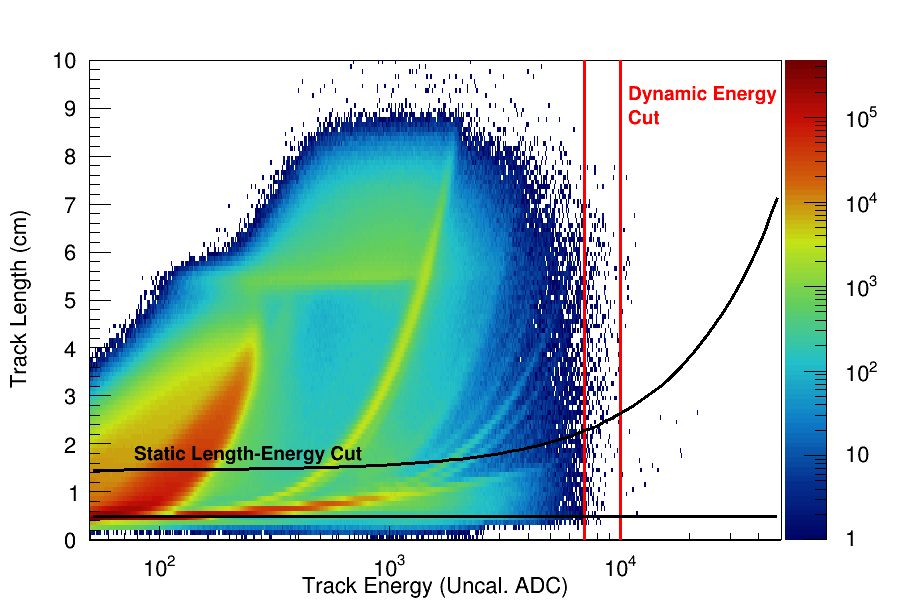}
  \caption{Track length vs.~particle energy with a blank Al target, for 43 hours of data collection. There are a negligible number of events in the fission fragment selection region.}
  \label{fig:blank_lvadc}
\end{figure}

Fig.~\ref{fig:blank_lvadc} shows the length vs.~energy distribution for
upstream and downstream \ftpc{} volumes with the blank target. The \talpha-particle band arises primarily from (n,\talpha) reactions.  The bands to the right of the \talpha-particle band are due to recoils of nuclei of increasing atomic number, as predicted by simulations. An insignificant fraction of observed recoils have sufficient energy to pass the fragment selection requiring a minimum ADC value of 7000.  Even after scaling these counts up by a factor to account for the relative time of blank target vs.~active target beam time shows their effect is insignificant.  The scaled value represents merely hundreds of background counts relative to 15+ million total fission events.  

Nuclear recoils registering with ADC $>$ 7000 are primarily induced by energetic neutrons above $\sim$50 MeV. This was estimated through simulations and was also observed in the neutron energy distribution of the limited number of tracks in the blank data set that passed the fission fragment cuts.
While only an approximate time-of-flight calibration was possible for this data as there was no photo-fission peak, it is evident that nuclear recoils are not a significant background for the current analysis. 

This conclusion is further supported by the observation that an event needs to induce a signal on the cathode, allowing event time-of-flight to be determined, to contribute to the cross-section ratio.  Only the central region of the cathode with radius $<$~3 cm is instrumented, so the cathode quickly loses sensitivity for nuclear recoils produced in the gas even a small distance away from the cathode surface. The cathode therefore intrinsically selects fission fragments (for which it is close to 100\% sensitive) over nuclear recoils in the gas.

Based on the detection rate of events that passed the fission fragment selection cuts in the blank data, nuclear recoils are estimated to be negligible below 30 MeV and a $<$0.1\% effect above 50 MeV.


%% file: Shape/Wraparound.tex
\label{sec:wraparound}

\begin{figure}[ht]
  \centering
  \includegraphics[width=1.\linewidth]{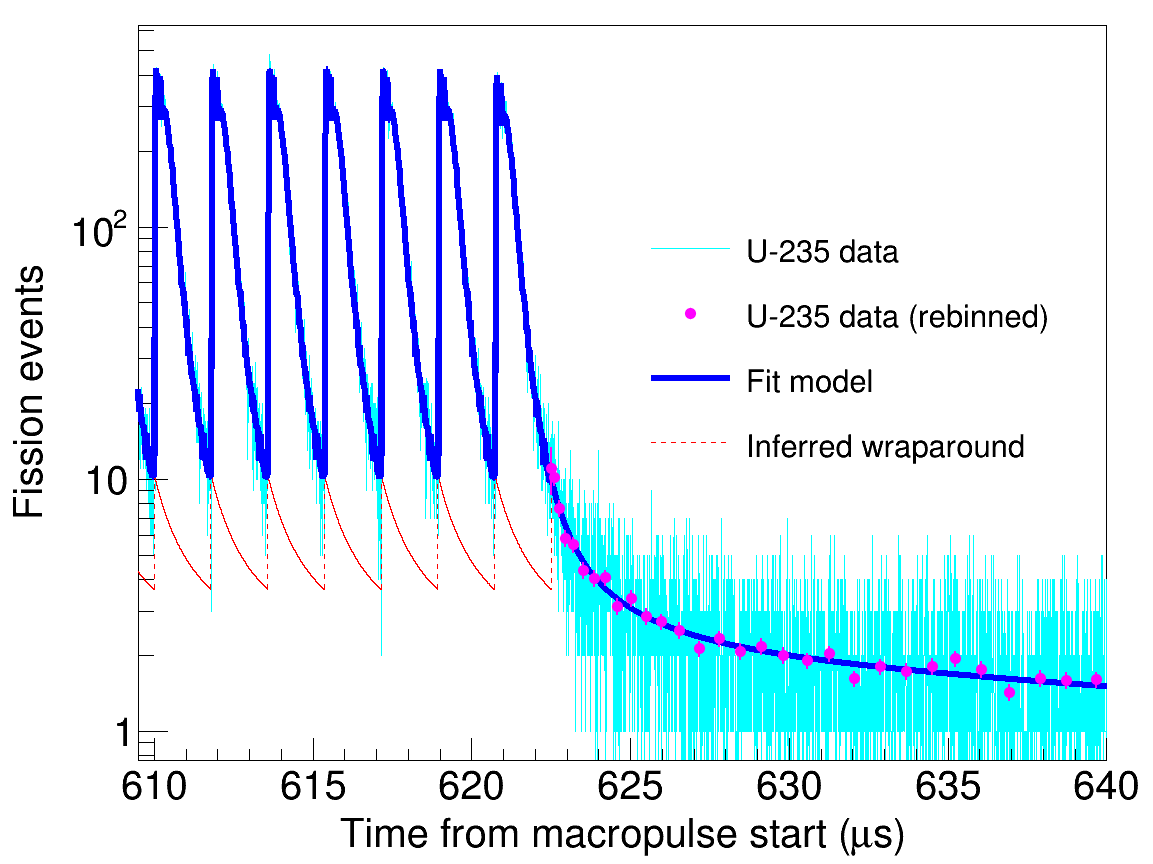}
  \includegraphics[width=1.\linewidth]{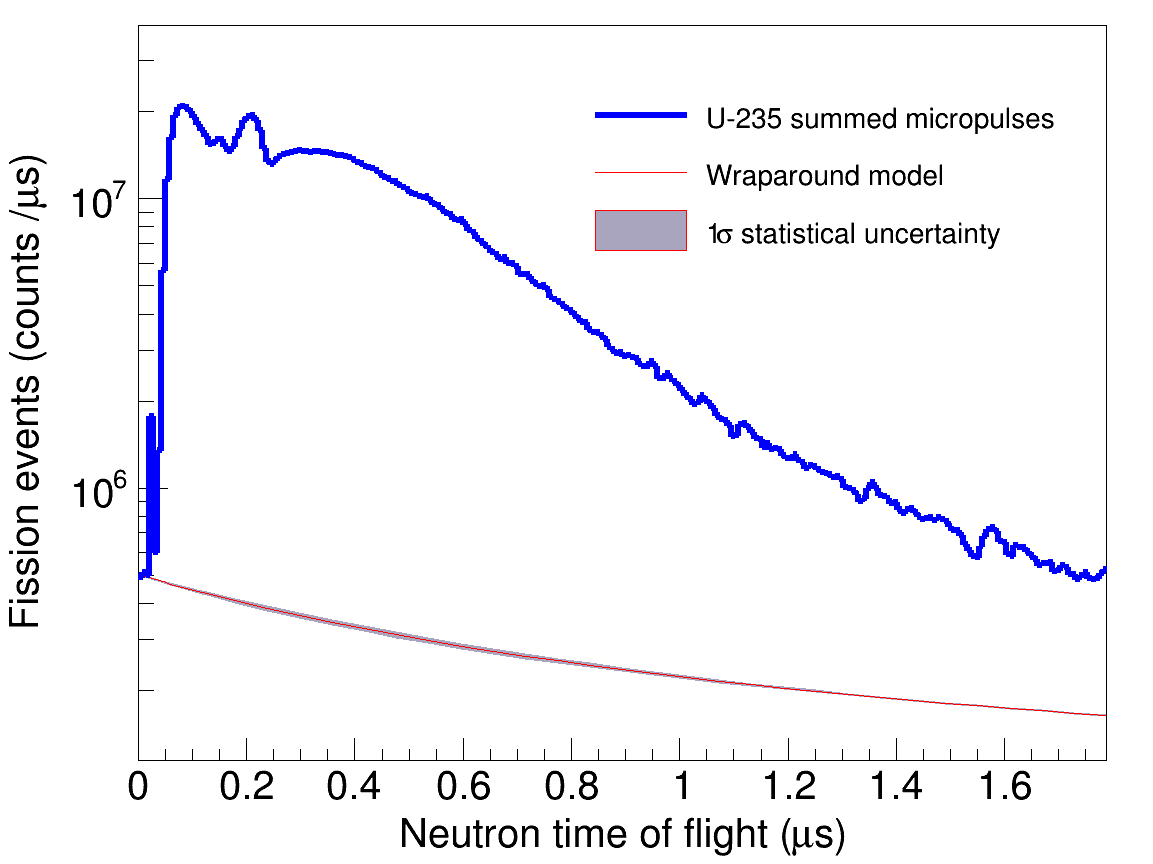}
  \caption[Determination of the wraparound correction]
  {Determination of the wraparound data for \u{}. Top: The nToF data, averaged in the low-energy tail region, are fit to determine the wraparound contribution to the fit model. Bottom: The nToF data and wraparound fit after combining all micro-pulses. The band around the fit is the fit uncertainty.}
  \label{fig:wraparound}
\end{figure}

The LANSCE proton accelerator produces a ``macro-pulse" structure that delivers beam to WNR at a frequency of typically either 40 or 100 Hz.  Within each of the macro-pulses are ``micro-pulses" or proton bunches spaced $\sim1.8~\mu s$ apart.  Low energy neutrons from preceding micro-pulses will carry over to later micro-pulses in the train.  A correction for these neutrons must be made as the fission events generated by them will be assigned the incorrect nToF.  The micro-pulse train can be ``unwrapped'' and the structure, which includes a long tail beyond the last micro-pulse, can be fit to determine the background contribution. Fig.~\ref{fig:wraparound} illustrates the micro-pulse structure within a macro-pulse, along with the fit results described below that are used to correct for wraparound. 

The macro-pulse structure is assumed to be a sum of identically-shaped micro-pulses (up to a pulse-by-pulse scale factor), shifted by constant time offsets.  Motivated by the fact that the wraparound tail is small compared to the unwrapped micro-pulse signals, an iterative procedure of successive refinements to the model-unwrapped micro-pulse shape is used in the fit.  This model shape is divided into two regions: a short timescale component in the time window between successive micro-pulses, and a long-timescale tail extending under many following micro-pulses.
    
The short-timescale region is represented with many degrees of freedom as a binned histogram of nToF since the micro-pulse start.  This histogram is produced by summing over all the micro-pulses in the macro-pulse, shifted to a common $t=0$ starting point, after subtracting off the estimated
wraparound tail contribution.
    
The wraparound tail shape is represented by fewer degrees of freedom as a smooth curve (since the data does not contain sufficient statistics to finely resolve this as in the non-wrapped region).  The curve is represented by a log-log cubic spline functional form, $y(t) = \exp[S(\ln(t))]$, where
$S(x)$ is a \texttt{ROOT TSpline3} \cite{ROOT} cubic spline curve.  The spline is controlled by degrees of freedom fixing its value at some number of ``knots'' uniformly spaced in $\ln(t)$, with intermediate values smoothly interpolated.  The values of the spline degrees of freedom are determined by a multivariate $\chi^2$-style fit to the macro-pulse shape.
    
This defines a family of smoothly varying functions to model the tail shape, without imposing strongly-model-dependent limitations on the mathematical form. The more degrees of freedom used, the more sharply the spline can bend to fit the data --- eventually over-fitting to follow nonphysical statistical fluctuations.  To counteract this tendency to over-fit, a ``smoothness hint'' is added to the $\chi^2$ minimization target term, favoring solutions with smaller second derivatives when this can be accommodated with small increases to $\chi^2$.  The iterative fitting process is started with a smaller number of degrees of freedom to allow the fitter to lock in to a reasonable initial guess, then increased in subsequent iterations using initial fit values determined from the previous round. 

%% file: Shape/Contamination.tex
\label{sec:contamination}
An energy-dependent contaminant correction factor $C_b$ is applied to the corrected fission counts in Eq.~\eqref{eqn:xsCalc} given by
\begin{equation}
	C_b = \frac{N_{x}\sigma_{\left(n,f\right)}^{x}}
    	{\sum_{i}N_{i}\sigma_{\left(n,f\right)}^{i}} \mathrm{,}
\end{equation}

where $N_{i}=N\times f_{i}$ is the number of atoms of species $i$,
$\sigma_{\left(n,f\right)}^{i}$ is the (n,f) cross section on species $i$ taken
from the ENDF/B-VII.1 database~\cite{Chadwick2011}, and $x$ designates the main isotope in each target (i.e, \u{} for the uranium target and \pu{} for the plutonium target).
The total number of atoms $N$ cancels in the ratio, and the isotopic fractions
can be taken from the (uncorrelated) mass spectroscopy measurements.  

Two representative samples of both the \u{} and \pu{} target materials were analyzed by mass spectrometry to obtain relative abundances for the target constituents.   
The target used for the cross-section measurement was not analyzed directly because of the destructive nature of analysis by mass spectrometry.  A second target produced from the same stock material at the same facility was analyzed instead.  

The relative abundances for the uranium sample are presented in Table \ref{table:uIsotopics}. The uncertainties on the mass spectrometry data were assigned using an established, well-documented method \cite{RossUncert}.  A discrepancy exists in the content of $^{234}$U and $^{238}$U between the two samples.  The deviation is 25 standard deviations for $^{238}$U and three standard deviations for $^{234}$U.   

\begin{table*}[htb]
\begin{center}
\begin{tabular}{c|c|c|c|c}
  \hline\hline
  Isotope & Sample 1 (\%) & Uncertainty & Sample 2 (\%) & Uncertainty  \\
  \hline
  $^{233}$U & 0.01886 & 0.00004 & 0.01893 & 0.00008 \\
  $^{234}$U & 0.03448 & 0.00032 & 0.03536 & 0.00004\\
  $^{235}$U & 99.677 & 0.002 & 99.634 & 0.014   \\ 
  $^{236}$U & 0.1701 & 0.00177 & 0.1763 & 0.0005  \\
  $^{238}$U & 0.0998 & 0.0005 & 0.1355 & 0.0014  \\
  \hline\hline
\end{tabular}
\end{center}
\caption{Percent abundances for the isotopes of uranium in the \u{} sample, as measured with mass spectrometry.  Sample 1 was taken from the raw stock material, while Sample 2 was taken from a target prepared in the same manner as the one used for this cross-section ratio measurement.}
\label{table:uIsotopics}
\end{table*}

These deviations were explained by the origin of the two samples.  Sample 1 came from raw stock of the material used for target preparation.  Sample 2 came from a spare target prepared from the raw stock.  We hypothesize that a minute amount of natural uranium was introduced to the material during target preparation, which is done via vacuum vapor deposition.  This hypothesis is supported by the introduction of both $^{234}$U and $^{238}$U in a ratio consistent with the composition of natural uranium.

The discrepancy in the content of $^{238}$U has little impact.  The content of $^{238}$U is less than 0.1\%, so it contributes only a small portion of the fission yield at high energies and an even smaller portion below the fission threshold at 1 MeV.  
The relative abundance of $^{234}$U in the sample is also less than 0.1\%, so it does not contribute significantly to the fission yield.  
The half life of $^{234}$U is sufficiently short, however, that it contributes roughly one third of the total \talpha-activity of the uranium sample.  This will have an impact on a measurement of the absolute normalization which is discussed in Ref.~\cite{Monterial2021}.

The relative abundances for the plutonium sample are shown in Table \ref{table:puIsotopics}.  Similar to the uranium sample there are discrepancies that are larger than the quoted uncertainties. Sample 2 is assumed to be more accurate as it was prepared in the same manner as the target used for the cross-section ratio measurement.  
It is not possible to extract the $^{238}$Pu from the mass spectrometry as a result of background $^{238}$U contamination in the system.  The $^{238}$Pu content was measured using \talpha-spectroscopy also described in Ref.~\cite{Monterial2021}.  

\begin{table*}[ht]
\begin{center}
\begin{tabular}{c|c|c|c|c}
  \hline
  Isotope & Sample 1 (\%) & Uncertainty & Sample 2 (\%) & Uncertainty \\
  \hline\hline
  $^{238}$Pu & \multicolumn{4}{c}{not included in analysis} \\
  $^{239}$Pu & 99.1323 & 0.0024 & 99.1213 & 0.0008 \\
  $^{240}$Pu & 0.8675 & 0.0023 & 0.8770 & 0.0008  \\
  $^{241}$Pu & \multicolumn{2}{c}{not detected} & 0.001427 & 0.000010 \\ 
  $^{242}$Pu & 0.000242 & 0.000042 & 0.000250 & 0.000006 \\
  \hline\hline
\end{tabular}
\end{center}
\caption{Percent abundances for the isotopes of plutonium in the \pu{} sample, as measured with mass spectrometry.  Sample 1 was taken from the raw stock material, while Sample 2 was taken from a target prepared in the same manner as the one used for this cross-section ratio measurement.}
\label{table:puIsotopics}
\end{table*}

\begin{figure}
	\centering
    \includegraphics[width= 1.\linewidth]{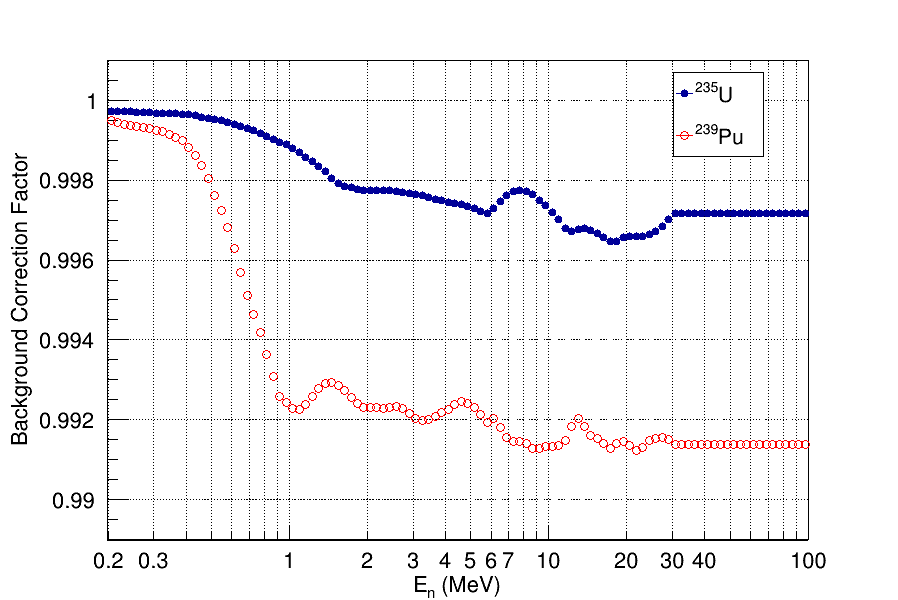}
    \caption[Contamination correction]
    {The size of the contamination correction, $C_b$, as a function of energy for both actinides. The uncertainties are driven primarily by uncertainties in the ENDF/B-VII.1 cross section.}
    \label{fig:contaminants}
\end{figure}

Fig.~\ref{fig:contaminants} shows the size of the contaminant correction, $C_b$, as a function of neutron energy for each target actinide.
Covariance matrices from the ENDF/B-VII.1 \cite{Smith2011} database for the $\sigma_{\left(n,f\right)}^{i}$ cross sections and
(uncorrelated) uncertainties for the $f_{i}$ isotopic fractions have been used to generate Monte Carlo realizations of the contaminant correction factor for each energy bin to determine the uncertainty in the correction.

%% file: Shape/Beam_Uncorrelated.tex
\label{sec:beam_uncorrelated}
Aside from beam-induced non-fission interactions, certain beam-uncorrelated events could also introduce a source of background in the $C_{ff}$ count. In particular, the relatively high \pu{} spontaneous \talpha-decay rate (roughly 1 MBq) is a cause for concern. While the \pu{} \talpha-particles have an energy of only $\sim$5 MeV, pile-up of multiple \talpha-particles needs to be carefully evaluated.  This potential source of background is accounted for in Eq.~\eqref{eqn:xsCalc} by the $C_\alpha$ term. 
The \ftpc{} tracking and particle identification performance are key to ensuring that this contribution is negligible.

\begin{figure}[ht]
  \centering
  \includegraphics[width=1.\linewidth]{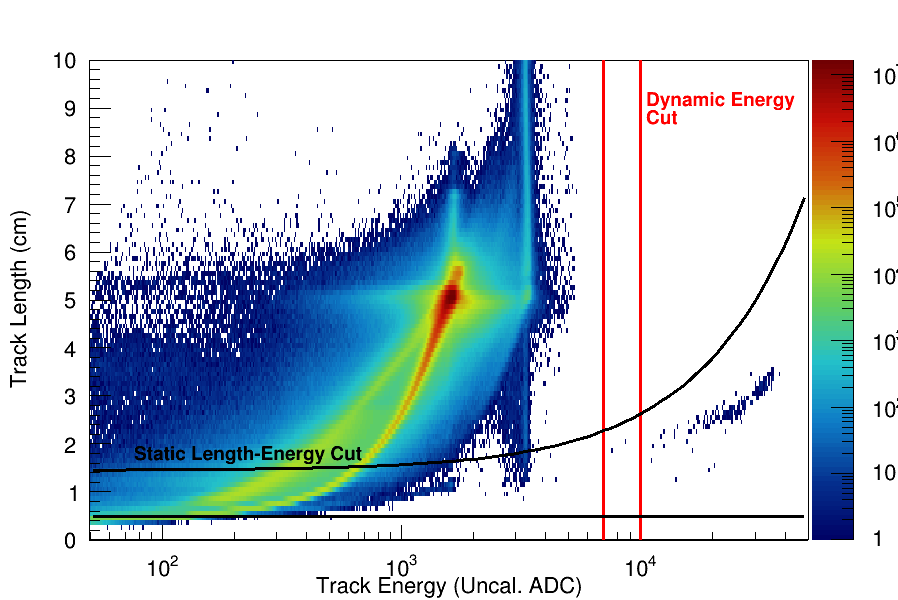}
  \includegraphics[width=1.\linewidth]{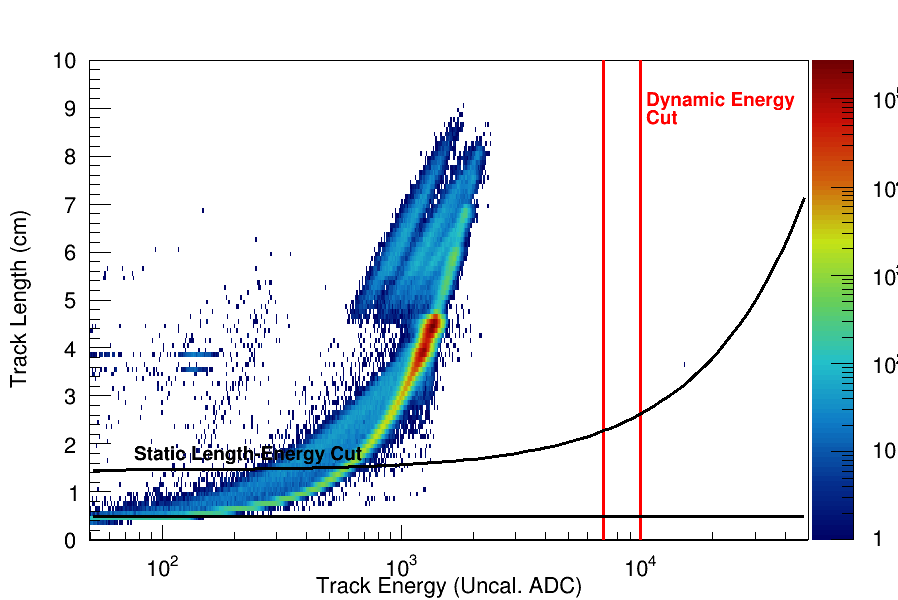}
  \caption{Track length vs.~particle energy during beam-off data collection for \pu{} and \u{}.  A negligible fraction of counts appear in the fission fragment cut selection region.}
  \label{fig:lvadc_autorad}
\end{figure}

The effect of \talpha{}-particle contamination was studied during beam-off periods dedicated to target radiograph measurements.  Approximately 400 hours of beam-off data for the uranium target and 45 hours for the plutonium target were collected.   Fig.~\ref{fig:lvadc_autorad} shows the length vs.~energy distributions for the radiograph data sets for \pu{} and \u{}. The \pu{} data set produced only 206 spontaneous fission events that are above the fission cut and no \talpha{}-decay pile-up events in 45 hours of operation, while the \u{} data set produced no events within the cut selection.  The data acquisition system is designed to have triggers inhibited or masked  
during the periods between beam macro-pulses, meaning the detector is live for fission triggers for only 7\% of the total detector up-time.  A false beam inhibit was also applied during the beam off runs to mimic this behavior and keep the data rate from the spontaneous \talpha-decay of \pu{} at a manageable level.
Within the unmasked data region it is estimated that the spontaneous fission detection rate is $\sim$4.6 fragments per hour 
which would result in the detection of approximately 2300 total spontaneous fission events in the approximately 500 hour run period (for each rotation), or about 0.1\% of the data set of $\sim$15 M beam-induced fission events.  These events would appear randomly in time and would be removed as part of the wraparound correction (see Sec.~\ref{sec:wraparound}).

Based on these observations we conclude that mis-identification of spontaneous \talpha-decays for fragments or detection of spontaneous fission are negligible contributions to the analysis presented here and no correction is applied.  This negligible contribution results from the \ftpc{}'s ability to track individual \talpha-particles, greatly reducing pile-up even under high rate conditions.

%% file: Shape/Livetime_L2.tex
\label{sec:livetime}
The \ftpc{} was designed to have 100\% live time and uses a triggering scheme that avoids traditional sources of dead time. The data pipeline remains open for new data as new triggers are received.  Data is stored based on the trigger condition and a second trigger that occurs during data storage simply extends the storage window.  This is in contrast to more conventional data acquisition systems that have dead time while an event is processed.  All measurements that have been implemented to date indicate that 100\% live time has been achieved, so no corrections have been made.  This section presents evidence that supports this assertion.

In the case of a ratio measurement, it is the ratio of the live times that matters, but it is easier to investigate the absolute live time and use the fact that the ratio of live times is closer to unity than the individual live times.  
The definition of live time is the percentage of time the \ftpc{} is actually live when it is expected to be live.  The \ftpc{} is expected to be live when the inhibit signal to the \ftpc{} is logical low.  The definition of live is that if a fission occurs, enough information is collected to identify the event as a fission.  When making cross-check measurements of live time, one has to be careful because efficiency and live time can be confused; using large events like fission that are far above the trigger threshold helps separate these effects and isolate the live time component.

There are a number of possible live time effects that could exist even in a digital system designed for no dead time.  The effects that were considered are:

\begin{itemize}
\item Analog pile-up,
\item Fission fragment and \talpha-particle pile-up,
\item Saturation / Spark,
\item Filled buffers on the digital pipeline,
\item Loss of an ethernet packet,
\item Events occurring at the enable/disable of the trigger,
\item Analysis/data processing errors,
\item Hardware construction or configuration issues.
\end{itemize}

Analog pile-up occurs when two fission events happen within the rise time of the amplifier and are not seen as distinct in time.  Analog pile-up can occur both on the anode and cathode, but the anode has thousands of channels while the cathode is instrumented with only one.  The probability for overlapping tracks on the anode is at least an order of magnitude lower than for the cathode because the tracks would have to line up perfectly in space as well as time, so only the cathode case is considered.  The rise time of the amplifier is about 250 ns and the average fission event rate is less than 100 Hz (10 ms).  Using Poisson statistics one can calculate the probability for two events to both happen within the amplifier rise time.  

\begin{equation}
P = \frac{\lambda^k e^{-\lambda}}{k!} \approx 3 \times 10^{-10} \,\text{,}
\end{equation}

\noindent with $k$ = 2 and $\lambda$ = 250 ns/10 ms = 25$\times$10$^{-6}$.   The rise time would have to be an exceedingly slow 450 $\mu$s for the effect to even rise to the 0.1\% level.  A careful analysis of the cathode waveform amplitudes could be used to distinguish pile-up events from single fission events but since the probability is so low, this has not been pursued and this effect is dismissed.  

Fission fragments and \talpha-particles do pile up on the cathode, especially for high specific activity targets like \pu{}.  The \talpha-particle energies are typically of order 5 MeV compared to the $\sim$90 MeV of a fission fragment.  Overlapping fragments and \talpha-particles would only increase the energy of the detected track so it is impossible to conceive of a mechanism that would cause a change in the live time.

A spark can cause an amplifier to latch into a saturated state for some time and while in that state it would not respond to input charge.  Sparks produce several distinct observations in the \ftpc{} when they occur, including very distinctive waveforms.  Most observations of sparking occurred early in the development of the \ftpc{} when the hardware was still being refined.  These events are now rare, much less than one per 20-minute data collection run.  Assuming one spark occurs at a time that is much less than 10 ms in length, at most one fission event would be lost, corresponding to a 20 ppm effect.  Saturation from a fission fragment could in principle cause a similar effect, but the fragments do not saturate the electronics nearly as much as the sparks, and there is no latch-up or other bad behavior from saturation caused by fission fragments.

The electronics pipeline includes buffers that could fill up if there were extraordinary bursts of particles.  The electronics are designed to handle such a situation gracefully.  When a buffer is full, the trigger is inhibited and a ``busy'' packet is 
issued that indicates precisely how much time the system was not live.  Inspection of the data shows that busy packets are rare and therefore not a significant source of dead time.

Data is sent over ethernet from the detector to the DAQ computers.  While a mature protocol, ethernet is a lossy transport that could lose packets. The ethernet packets are numbered sequentially so that lost packets are readily apparent.  Online controls include counters for these ``missingEDFs''.  These have been investigated and are a negligible source of dead time.

As was discussed briefly in Sec.~\ref{sec:beam_uncorrelated},
the data acquisition system is designed to have triggers inhibited or masked during the periods between beam macro-pulses.  The mask prevents the needless recording of \pu{} spontaneous \talpha-decays for the approximately 93\% of the time that no beam is present. The data rate is further reduced by the implementation of a ``level-2" triggering system.  The cathode is live for the entire beam pulse extending out to 700 $\mu$s, while the anode (pad-plane) remains inhibited.  The drift time of a fragment track in the \ftpc{} gas volume is slow enough such that the anode inhibit can be released for 10 $\mu$s upon the detection of a fission fragment on the cathode, in time to record the fission fragment signal on the anode.  The level-2 triggering system is disabled (no anode trigger inhibit) every 6$^{th}$ macro-pulse to provide a subset of data with which to validate the performance of the level-2 triggering scheme.  The nToF spectra of data with and without the level-2 trigger implemented were compared.  The ratio of fission fragment counts for the two data subsets was found to be 5.003 $\pm$ 0.003 as expected, and no difference in the shape of the nToF spectra were observed.
It is possible that the time immediately after the \ftpc{} is enabled or immediately before it is disabled could contribute to dead time. The clock period is 20 ns and the live window is either 700 $\mu$s or 10 $\mu$s depending on the level-2 trigger state.  The system should only have an ambiguity set by the rise time of the trigger inhibit signal or one clock cycle.  The events should record normally inside of this time.  The percent of live time in question is $\frac{20 ns \times 2}{700 \mu s} \approx 0.006\%$.  The cathode is live for all of the beam macro-pulse windows, but the anode cards are live for 10 $\mu$s in some cases (when the level-2 is triggered).  Nevertheless, this anode live-time is synchronous with the cathode so the potential 0.006\% dead time applies the same for the anode.  Furthermore any potential cathode dead time would be shared for each volume of the \ftpc{}, \ie{} $\omega_s/\omega_x = 1$ in the cross-section ratio.

Finally, the possibility of software and hardware issues that could cause the system to behave differently than expected was considered.  The anode is instrumented with thousands of channels so it is unlikely that all of them are dead at the same time.  A comparison of the signals on the anode and cathode from the same tracks showed agreement to better than 99\%. However, because the cathode is only instrumented over a 3-cm radius around the center of the target, its efficiency depends on track position.  Any live time effect due to differences between the cathode and anode signals would therefore be much smaller than the effect due to cathode efficiency.

%% file: Shape/Attenuation.tex
\label{sec:attenuation}
There are two related corrections, scattering and attenuation of beam neutrons, that have to be taken into account for measuring the rates of interaction on the front and back of the target. Scattering can occur anywhere between the tungsten spallation target and the actinide target, primarily in beam collimation and the detector housing.  Attenuation of the neutron beam in the 0.25 mm thick aluminum target backing causes the flux on the front and the back of the target to be different. Scattering affects the estimate of the neutron energy, which is measured via neutron time-of-flight (nToF). Near-target neutron scatter will cause an underestimate of the neutron energy, and is expected to dominate over scattering farther from the target.

\begin{figure}[ht]
\begin{center}
\includegraphics[width= 1.\linewidth]{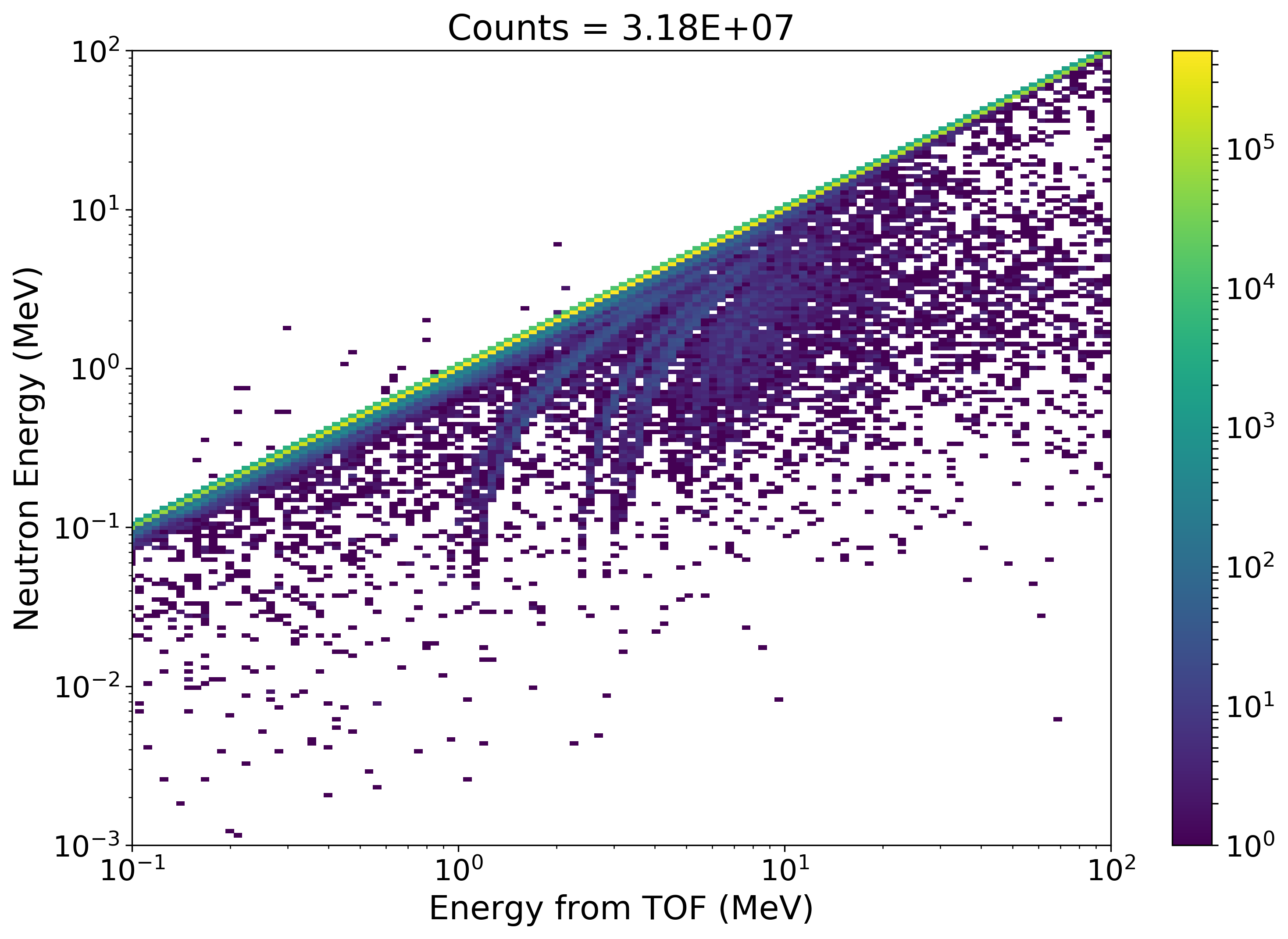}
\end{center}
\caption{Neutron flux matrix $\phi(E_T,E_T)$ from an MCNP simulation of the 90L beamline.}
\label{fig:katt_diagonal}
\end{figure}

In order to capture both of these effects an MCNP simulation of neutrons originating at the tungsten target and transported through the collimator toward the \ftpc{} was conducted. Neutron flux tallies were placed at the front and back of the target, and the tallies were subdivided by energy and time. The results enabled the construction of a matrix of flux given by the relationship between the MCNP generated neutron energy ($E_N$) and the nToF-calculated neutron energy ($E_T$), as shown in Fig.~\ref{fig:katt_diagonal}.  The diagonal represents perfect agreement between the true neutron energy and the nToF-calculated energy. The finger-like projections below the diagonal are caused by scattering and attenuation due to specific excited states of the (n,n$^\prime$) interaction in the aluminum backing.  This matrix is constructed for both front and back of the target, where ``front" is defined as the side of the target backing facing the beam and ``back" is facing away.  Therefore it captures both the scattering and attenuation effects. 


The flux matrix can be used to estimate the observed fission reaction rate, $R'(E)$, as
\begin{equation}
    R'(E) = N \sum_{E_N}\int_{E1}^{E2}\phi(E_T,E_N)\sigma(E_N)dE \,\text{,}
\end{equation}

\noindent where $\sigma(E_N)$ is the input fission cross section and $N$ is the total number of atoms in the target. In a similar way, the expected fission reaction rate, $R(E)$, can be computed when the true neutron energy is equal to the nToF-calculated energy as 
\begin{equation}
    R(E) = N \int_{E1}^{E2}\phi(E_T = E_N)\sigma(E_N)dE \,\text{.}
\end{equation}

\noindent The ratio of the expected fission rate to the observed rate from the MCNP simulation is the attenuation and scattering correction factor
\begin{equation}
    \kappa = \frac{R}{R'}\,\text{.}
\end{equation}

\begin{figure}[ht]
\begin{center}
\includegraphics[width= 1.\linewidth]{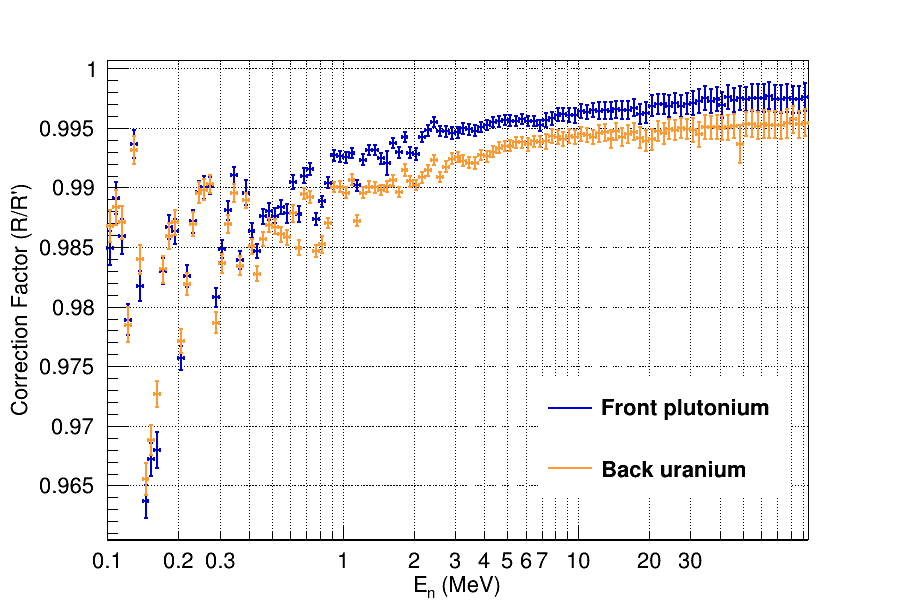}
\end{center}
\caption{Scattering and attenuation correction for both the front and back of the target, where ``front" is defined as the side of the target facing the beam and ``back" is facing away}
\label{fig:katt_correction}
\end{figure}

\noindent This correction can be estimated for both the front (facing the beam) and back (away from the beam) of the target, as shown in Fig.~\ref{fig:katt_correction}.

\begin{figure}[ht]
\begin{center}
\includegraphics[width= 1.\linewidth]{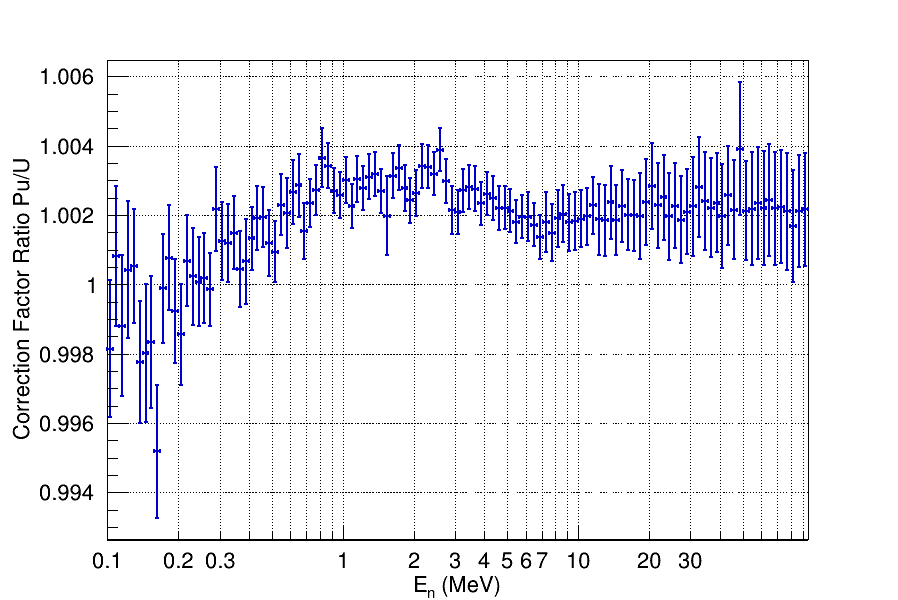}
\end{center}
\caption{Ratio of the corrections from Fig.~\ref{fig:katt_correction} that incorporate both the effects of scattering and attenuation.}
\label{fig:katt_ratio}
\end{figure}

Although the magnitude of these corrections are large on their own, the ratio of corrections for the \pu{} and \u{} cross section is generally much less than 1\%, as shown in Fig.~\ref{fig:katt_ratio}.


%% file: Results.tex
\label{sec:results}
The primary goal of the NIFFTE collaboration was to create an appropriate framework for quantifying all of the experimentally accessible uncertainty contributions to the cross-section ratio as a function of neutron energy, including the full energy correlation matrix, rather than just the uncertainties for each energy bin. The full correlation matrix allows for better analysis of the final evaluated cross-section uncertainties when combining with other data. The measured fission cross-section ratio data of \pu{} relative to \u{} as a function of neutron energy from 0.2--100 MeV are shown in Fig.~\ref{fig:XS_GLS} and are tabulated in \ref{app:Results}.  The quoted uncertainty is the total absolute uncertainty from appropriate combination of the contributions from all sources.  The contributions to the uncertainty are shown in Fig.~\ref{fig:TotalUncertainty} and the bin-to-bin energy correlation matrix is shown in Fig.~\ref{fig:TotalCorrelation}.  The partial uncertainties are tabulated in \ref{app:uncertResults}.  The values of the correlation matrix are provided as supplemental material with the electronic version of this article. The total uncertainties are below 1\% over the entire energy range for the cross-section \emph{shape}.  Discussion of the absolute normalization, and comparison of the cross-section ratio measurement with ENDF/B-VIII.0 and other experiments are presented in Sec.~\ref{sec:compare}.

\subsection{Uncertainty Calculation Methodology}
\label{sec:uncert_method}
There are two primary ways to generate the full energy covariance matrix of Eq.~\eqref{eqn:xsCalc}: first, one can determine the covariance of each term separately, and then combine them analytically. This approach has a shortfall, in that it relies on a Taylor approximation of the dependence of the ratio on each term.  This approximation neglects higher-order correlations, which may be non-negligible.

The second approach, and the one adopted for the this measurement, is via Monte Carlo.  Data-driven covariance matrices for each component of the cross-section ratio have been assigned. From these covariance matrices, random sampling of each component (\eg{} counting statistics, target isotopics, beam-target overlap, \etc) can be generated, and a corresponding realization of the cross-section ratio calculated. This Monte Carlo approach produces a full energy covariance matrix.

The general approach is to assign every term in the cross-section ratio calculation Eq.~\eqref{eqn:xsCalc} an uncertainty, based on the method used to calculate each term, (\eg{} fit parameter uncertainties for wraparound, Poisson statistics for overlap, \etc) as a seed and generate 100 Monte Carlo (MC) samples of those terms assuming Gaussian distributed inputs.  Every term is varied simultaneously,each with an independent random seed, for each of the 100 MC samples of the cross-section ratio that are generated.  The value of each of the terms in Eq.~\eqref{eqn:xsCalc} varies with the choice of cuts that are applied in the length--energy and cos($\theta$) space. For each set of cuts, 100 variations are generated within the dynamic cut range and the cross section is calculated for each of the variations in an attempt to estimate any residual uncertainties.  These variations are referred to as \emph{dynamic cuts} and are described in Sec.~\ref{sec:frag_selection} shown in Fig.~\ref{fig:lvadc_labelled} and \ref{fig:Angle_vs_Energy_Explain}.
The MC sampling is performed for each of the cut variations, resulting in a total of 10k values of the cross-section ratio in each neutron energy bin.  The mean value and standard deviation of the ratio distribution is calculated for each bin and the covariance is calculated for each pair of energy bins.

A separate covariance matrix for the efficiency $\epsilon_{ff}$ (see Sec.~\ref{sec:efficiency}) is needed to account for the two-dimensional fit of the efficiency model to \ftpc{} counts as a function of fragment emission angle and fragment energy.  Realizations of the angle-energy data are generated by Monte Carlo sampling of the efficiency-model parameters, and compared to the actual \ftpc{} data in order to find the parameter set that minimizes the $\chi^{2}$. This procedure generates a single energy-independent covariance matrix for the $\epsilon_{ff}$ model parameters.  An anisotropy correction is then fit to each incident energy bin in the data to provide the energy-dependent contribution to the uncertainty of the efficiency model.
Similarly, for the wraparound correction $C_w$ (see Sec.~\ref{sec:wraparound}) the Monte Carlo realizations are generated from a full covariance matrix error propagation of the fit parameters. 

As discussed in Sec.~\ref{sec:contamination}, for the isotopic contamination corrections $C_b$ the covariance matrices from the ENDF/B-VII.1 \cite{Chadwick2011} database for the $\sigma_{\left(n,f\right)}^{i}$ cross sections and (uncorrelated) uncertainties for the $f_{i}$ isotopic fractions are used as inputs to generate the Monte Carlo realizations.  $C_b$ is assumed to scale directly with the fission counts $C_{ff}$ and is not recalculated for each cut variation.  Small relative differences could exist in the detection efficiency for fission fragments from contaminant isotopes as a result of differences in the fragment energy and angle distributions, but these are assumed to be negligible.

The uncertainty in the overlap correction $\sum_{XY}(\phi_{XY}\cdot n_{XY})$ discussed in Sec.~\ref{sec:overlap}, is dominated by the counting statistics in the track vertex binning structure.  Monte Carlo realizations are generated based on propagating counting statistics through the overlap calculation.  The fragment spatial distribution is assumed to be uncorrelated with fragment energy and angle distributions and therefore it is not recalculated for the cut variations.  This assumption is not entirely true as the energy loss of the fission fragments will vary with the thickness of the target material.  The effects of energy loss in the backing however are minimized with the forward angle cut and the angular distributions are dominated by kinematics.  

Several of the corrections in Eq.~\eqref{eqn:xsCalc} were shown to be negligible and their uncertainties are also considered to be negligible.  Each such term was explored in its own section of this article.  These include the live time $\omega$, discussed in Sec.~\ref{sec:livetime}, which was shown to be 100\%; the background component from beam-induced recoils or other beam-induced charged particle reactions $C_r$ discussed in Sec.~\ref{sec:nuclear_recoils};  the beam-uncorrelated background correction resulting from the spontaneous \talpha-decay of the actinide targets $C_{\alpha}$ discussed in Sec.~\ref{sec:beam_uncorrelated}; and the beam attenuation correction $\kappa$ resulting from scattering in the aluminum target backing presented in Sec.~\ref{sec:attenuation}.

The uncertainty on the target atom number ratio ${N_s}/{N_x}$, derived from the  silicon detector measurements, is as described in Ref.~\cite{Monterial2021}.  Random, uncorrelated realizations of the target atom number ratio were obtained using the central value and uncertainty quoted. Since the covariance matrices for the total number of target atoms and contaminants is independent of energy (that is, the target is constant), these realizations are generated for the fragment cut variations and are used for each energy bin of the calculation.

\subsection{Combining Data Sets}
\label{sec:combine}
As will be discussed in detail in Sec.~\ref{sec:validations}, the \ftpc{} was rotated with respect to the neutron beam direction as a method of validating the efficiency and overlap corrections.  These two subsets of data were about evenly split to each contain half of the total statistics and each subset was analyzed as an independent measurement.  It is desirable then to combine the results to take advantage of an overall reduction in statistical uncertainty.  It is not appropriate to simply average the results however, as they are obviously highly correlated.  The approach chosen was to combine the data using the formalism of Generalized Least Squares (GLS) \cite{GLS}.  The linear regression model is written as
\begin{equation}
    \vec{y} = X\vec{\beta} + \vec{\epsilon}\,\text{,}
\end{equation}
where $\vec{y}$ is the $2n\times1$ vector of the observed values, where $n$ is the length of a single observed vector, or in this case the number of energy bins. $X$ is the $2n\times n$ design matrix, $\vec{\beta}$ is the $n\times 1$ vector of unknown constants that is to be estimated from the data, and the error term $\vec{\epsilon}$ is a random vector with mean 0 and covariance matrix $\Omega$.  The solution is then
\begin{equation}
\label{eq:betaHat}
    \hat{\beta} = \left(X^T\Omega^{-1}X\right)^{-1}X^T\Omega^{-1}\vec{y}
\end{equation}
and
\begin{equation}
\label{eq:Cov_betaHat}
    \text{Cov}\left[\hat{\beta}|X\right] = \left(X^T \Omega^{-1}X\right)^{-1} \,\text{.}
\end{equation}

It is instructive to consider the example of combining two vectors $\vec{y_a}$ and $\vec{y_b}$ of length 2.  In this case the vector of observed values and the design matrix are constructed as
\begin{equation}
    \vec{y} = \begin{pmatrix}
y_{1a} \\ y_{2a} \\ y_{1b} \\ y_{2b} 
\end{pmatrix}\,\text{,}
\hspace{0.5cm}
X = 
\begin{pmatrix}
1 & 0 \\ 0 & 1 \\ 1 & 0 \\ 0 & 1
\end{pmatrix} \,\text{.}
\end{equation}
The covariance matrix can then be written as
\begin{equation}
\Omega = 
\begin{pmatrix}
\sigma^2_{1a} \hspace{0.4cm}& \sigma_{1a,2a} \hspace{0.4cm}& \sigma_{1a,1b} \hspace{0.4cm}& \sigma_{1a,2b} \\
\sigma_{2a,1a} \hspace{0.4cm}& \sigma^2_{2a} \hspace{0.4cm}& \sigma_{2a,1b} \hspace{0.4cm}& \sigma_{2a,2b} \\
\sigma_{1a,1b} \hspace{0.4cm}& \sigma_{2a,1b} \hspace{0.4cm}& \sigma^2_{1b} \hspace{0.4cm}& \sigma_{1b,2b} \\
\sigma_{1a,2b} \hspace{0.4cm}& \sigma_{2a,2b} \hspace{0.4cm}& \sigma_{2b,1b} \hspace{0.4cm}& \sigma^2_{2b} \\
\end{pmatrix}\,\text{.}
\end{equation}
In this example the upper left $2\times 2$ sub-matrix is the covariance matrix of the data set $\vec{y_a}$ while the lower right sub-matrix is the covariance for data set $\vec{y_b}$.  The off diagonal sub-matrices are then the covariance matrices between the data sets and are the transpose of one another. For the case of the actual data sets one can simply expand these matrices $n$ to the number of energy bins.  The covariances are calculated for each pair of energy bins using the 10k realizations of the Monte Carlo sampling and the cut variations as described earlier in this section.

\begin{figure}[ht]
\centering
\includegraphics[width= 1.\linewidth]{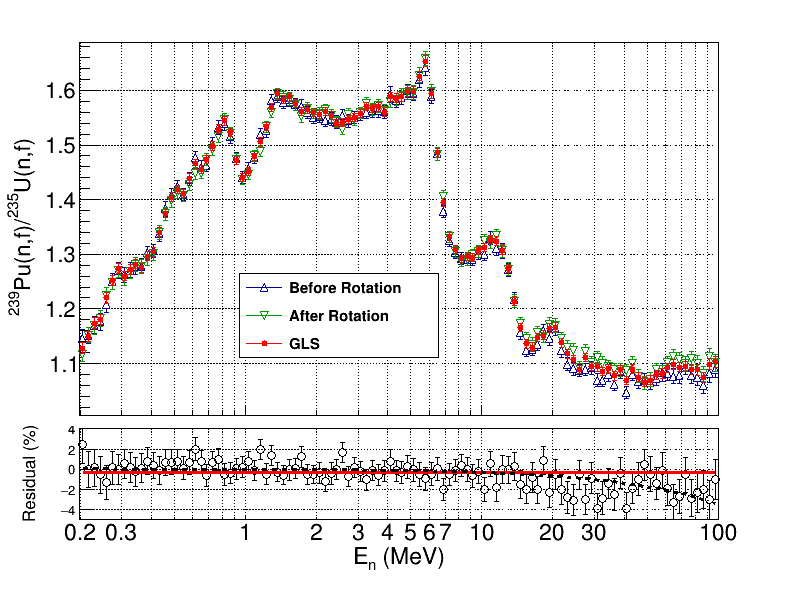}
\caption{The \ftpc{} measured \pu(n,f)/\u(n,f) cross-section ratio as a function of neutron energy.  Two subsets of data, referred to as ``before" and ``after" rotation, are shown along with the Generalized Least Squares (GLS) combination of the data.  The lower panel shows the residual between the two rotation data sets ((before-after)/before).  The error bars shown on the residual are simply the propagation of the uncertainties from the data sets treated as if they were independent measurements with no correlations. The solid red and dashed black lines are fits of a zeroth order and first order polynomial respectively.}
\label{fig:XS_GLS}
\end{figure}

Fig.~\ref{fig:XS_GLS} shows the two data sets from before and after the rotation and the resulting GLS combination of the two. Also shown is the residual between the two rotation data sets. The GLS central values are the results of the vector in Eq.~\eqref{eq:betaHat} while the error bars shown are the square root of the diagonals of the covariance matrix of Eq.~\eqref{eq:Cov_betaHat}.

The $\chi^2$ of the combined data can be written in matrix notation as
\begin{equation}
\label{eq:chi}
    \chi^2 = r^T W r \,\text{,}
\end{equation}
where $r$ is the vector of residuals and the weight matrix $W$ is the inverse of the covariance matrix $\Omega$.  The reduced $\chi^2$ agreement over the full energy range of 0.2--100 MeV is equal to 1.33 indicating some systematic disagreement between the before and after rotation data sets.  The reduced $\chi^2$ over a narrower energy range of 0.2--20 MeV is 1.04.  This indicates that the disagreement is at higher energies, which can be observed in Fig.~\ref{fig:XS_GLS}.  This could be the result of some small systematic uncertainty in the efficiency model at higher energies.  The averaging of the data should accurately account for the systematic discrepancy which will be discussed further in Sec.~\ref{sec:validations}.

\subsection{Partial Uncertainties}
To determine the partial uncertainties of the before and after rotation data sets the Monte Carlo and cut variations are performed for the single term of interest while the other terms are left static and no cut variations are performed.  The cut variation is treated as its own partial uncertainty.  While this method works well for the individual rotation data sets it is a challenge for the GLS method as the covariance matrix $\Omega$ is singular for most of the partial uncertainty calculations and cannot be properly inverted as is required in Eq.~\eqref{eq:betaHat}.  To overcome this challenge and avoid the use of ``pseudo-inversions" a second method for combining the data sets as a weighted average was used to provide an estimate of the partial uncertainties.
The combined cross-section ratio is defined as 
\begin{equation}
\begin{aligned}
C_{i}\left(E_{j}\right) & \equiv\omega_{A}\left(E_{j}\right)A_{i}\left(E_{j}\right)+\omega_{B}\left(E_{j}\right)B_{i}\left(E_{j}\right)
\,\text{,}
\end{aligned}
\end{equation}
where $A$ and $B$ are the two rotation data sets, the index $i$ labels the individual Monte Carlo and cut variation realizations and $j$ is the energy bin index.  The weights are subject to the condition 
\begin{equation}
    \omega_A + \omega_B = 1 \,\text{.}
\end{equation}
Note that the weights can be different for different energy bins.  The weights are determined by minimizing the variance of $C$ using the method of Lagrange multipliers.  While this method does not use the full covariance matrix $\Omega$ of the GLS method, it does nevertheless consider correlations between the two data sets.  The resulting weights are determined to be
\begin{equation}
\begin{aligned}\omega_{A} & =\frac{\sigma_{B}^{2}-\textrm{cov}\left(A,B\right)}{\sigma_{A}^{2}+\sigma_{B}^{2}-2\textrm{cov}\left(A,B\right)} \,\text{,}\\
\omega_{B} & =\frac{\sigma_{A}^{2}-\textrm{cov}\left(A,B\right)}{\sigma_{A}^{2}+\sigma_{B}^{2}-2\textrm{cov}\left(A,B\right)} \,\text{.}
\end{aligned}
\label{eq:weights}
\end{equation}

\begin{figure}[ht]
\centering
\includegraphics[width= 1.\linewidth]{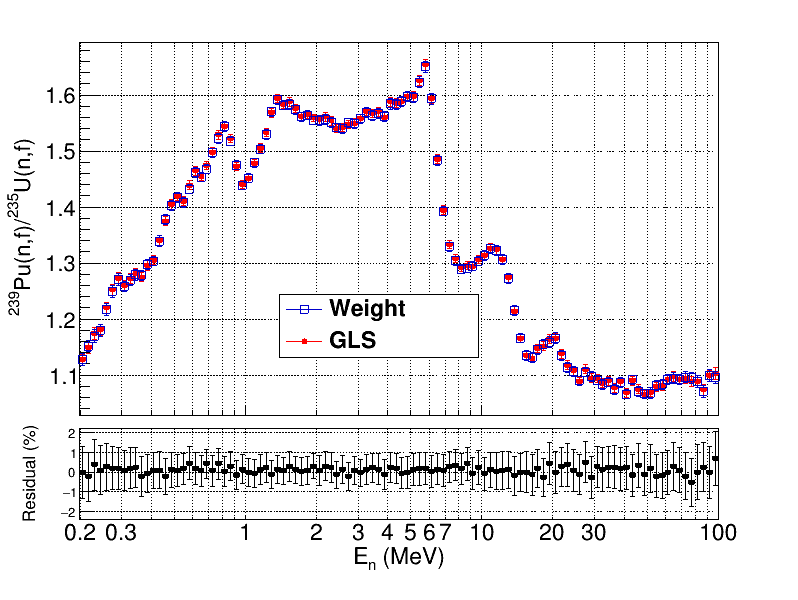}
\caption{A comparison of the \pu(n,f)/\u(n,f) cross-section ratio results of the two methods for combining the before and after rotation data.  
The lower panel shows the residual between the two methods  ((GLS-Wgt)/GLS).}
\label{fig:XS_Wgt_GLS}
\end{figure}

\begin{figure}[ht]
\centering
\includegraphics[width= 1.\linewidth]{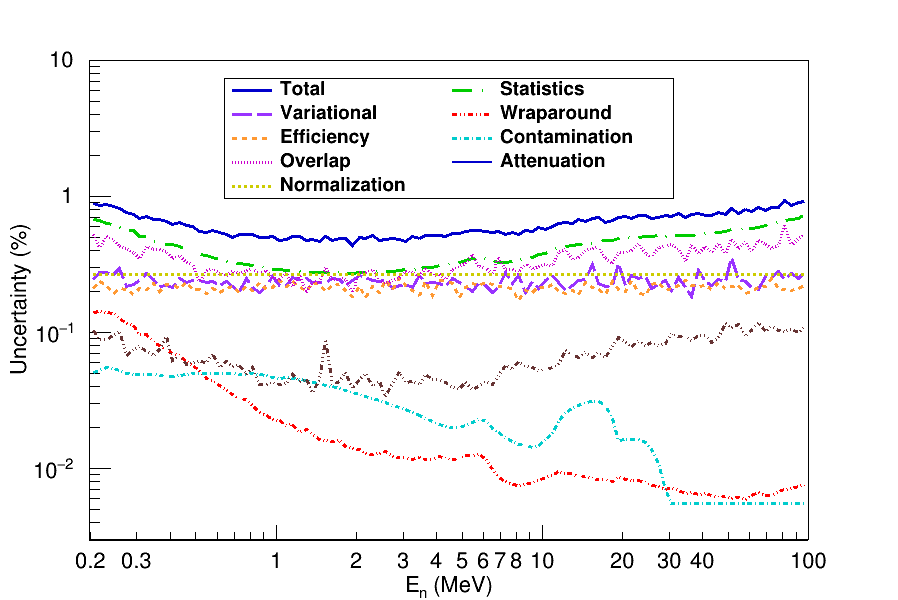}
\caption{Uncertainty contributions for each term in the cross-section ratio as a function of incident-neutron energy.  The total uncertainty is determined with the GLS method of Eq.~\eqref{eq:Cov_betaHat} while the partial uncertainties were determined using the weighted method of Eq.~\eqref{eq:weights}. }
\label{fig:TotalUncertainty}
\end{figure}

\begin{figure}[ht]
\centering
\includegraphics[width= 1.\linewidth]{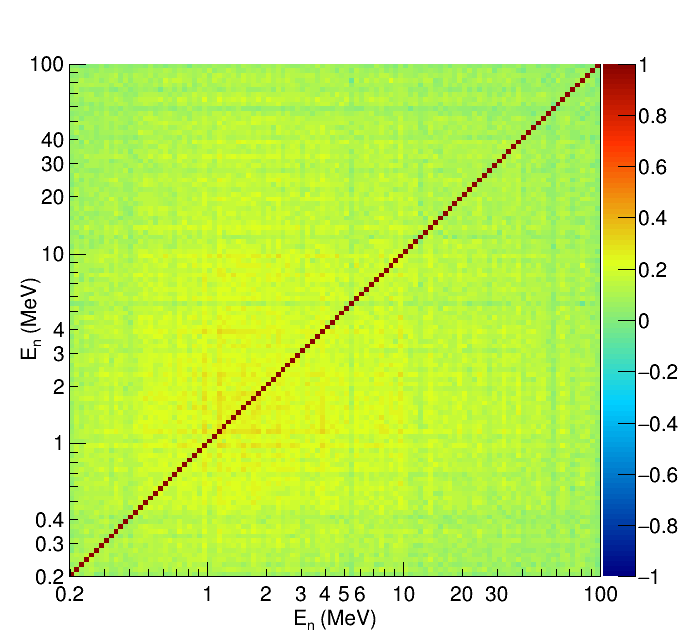}
\caption{Neutron energy correlation matrix of the cross-section ratio measurement calculated with the GLS method.}
\label{fig:TotalCorrelation}
\end{figure}

Fig.~\ref{fig:XS_Wgt_GLS} shows a comparison between the results of the GLS method and the weighted method.  The two methods produce very similar results with any differences generally being much less than 1\%.  The similarity between the two methods gives confidence that the weighted method produces sufficiently accurate estimates of the partial uncertainties.  Furthermore the GLS method is able to produce a partial uncertainty for the statistical component which was found to agree with that produced by the weighted method to better than 1\% relative. The partial uncertainties are shown in Fig.~\ref{fig:TotalUncertainty}.  
The uncertainties for a single rotation data set are generally higher than for the combined data set, except for normalization and the contamination correction, which are fully correlated between before and after rotation.  The correlation matrix for the GLS combined data is shown in Fig.~\ref{fig:TotalCorrelation}.

%% file: Validations.tex
\label{sec:validations}
While many of the correction terms in Eq.~\eqref{eqn:xsCalc} are negligible, the efficiency and overlap corrections are not, and the approaches used to determine each was novel when compared to previous fission cross-section measurements. For this reason we developed procedures to validate our methods and the accuracy of the derived corrections.  
Each of the effects studied was shown to have a well-controlled correction procedure that keeps the cross-section ratio stable and introduces minimal additional uncertainty. 

The validations include assessing the impact of rotating the \ftpc{} in the beamline, binning the data according to location on the target, applying radial cuts and assessing the impact of track vertex resolution and bias, confirming the alignment of the targets, and assessing the sensitivity to space-charge scaling.  Each of these validation methods will be discussed in the next several sections.

While various tracking biases can cause systematic shifts, the overall shape of the overlap correction (and therefore the cross-section ratio) is the result of relatively smoothly varying target and beam profiles that are well captured by the data and is largely insensitive to tracking, binning and alignment biases.  All of the sensitivity studies presented in the next several sections indicate that any biases from tracking would be at worst up to 1\% effects if no corrections were applied.  These effects would appear as an overall normalization shift in the overlap correction, having negligible impact on the energy-dependence of the cross section.  The radial cut and spatial binning validations (see Sections \ref{sec:radcut} and \ref{sec:bin2bin_Validation}) require all of the tracking elements to work in concert.  The fact that these validations show agreement within uncertainties is a strong indication that tracking biases are a negligible source of systematic uncertainty in the \ftpc{} measurements.

\subsection{Rotation Validation}\label{sec:rotval}
The back-to-back configuration of the actinide deposits means that an assumption of the efficiency canceling in ratio does not hold.  The target facing the beam will tend to have the fragments boosted into the target and backing with a decrease in efficiency with increasing beam energy, while the opposite is true for the target facing away from from the beam.  This effect was illustrated in Fig.~\ref{fig:Efficiency} in Sec.~\ref{sec:efficiency}.  Rotating the \ftpc{} also means that regions of greater or lesser target density are inserted into regions of greater or lesser neutron beam flux, necessitating the need for the overlap correction described in Sec.~\ref{sec:overlap}.
Efficiency and overlap corrections also impact other methods for measuring fission cross sections to some degree, so those methods are reviewed briefly to provide context for the rotation validation.

The most common methods for measuring actinide cross-section ratios are either the use of a Parallel Plate Ionization Chamber (PPIC) \cite{Wender1993} or a Gridded Ionization Chamber (GIC) \cite{Budtz1987}.  These instrument designs are relatively simple, compact and robust, providing typical measurement uncertainties on the order of 3--5\% \cite{Tovesson2015}.  In the case of a PPIC, multiple foils ($>2$) with thin actinide deposits, typically 100--200 $\mu$g/cm$^2$, are arranged in the chamber in a parallel fashion and are placed in a neutron beam perpendicular to the beam direction. When utilized to make a fission cross-section ratio measurement the different foils will have different actinide deposits.  The foils on which the actinides are deposited are thick enough such that the fission fragments cannot penetrate them and only the fragments which are emitted in a direction away from the backing foil are observed in the detector.
In the case of a GIC the thin actinide deposits are arranged on a single foil in a back-to-back fashion and are placed in a neutron beam perpendicular to the beam direction. A GIC can be used with a single actinide deposit on a thin backing foil to observe the fragment pairs, but this arrangement is not used for cross-section ratio measurements.    The GIC design is most similar to the \ftpc{} in that there is a central cathode and two volumes, one for each actinide.

When performing ratio measurements with these instruments the efficiency for detecting the fission fragments from the different deposits must be evaluated.  While the efficiencies are typically very high, they are less than 100\% as a result of fission fragments being absorbed in the actinide deposit.  The rate of absorption is dependent on the angular distribution of the fragments relative to the surface of the deposit and the material composition and thickness of the deposit.  The angular distributions for two different actinides are not necessarily the same.  Fission fragment angular distributions have an energy dependent anisotropy (\ie{} not isotropic in the lab frame) that is not the same for different isotopes, and are boosted by the linear momentum transfer of the incident neutron \cite{Geppert-Kleinrath2019, hensle2020}.  Furthermore the composition of the deposits and the method of deposition are often different, particularly when comparing uranium and plutonium deposits.  Uranium deposits are typically vapor deposited in the form of UF$_4$ while plutonium deposits are often electroplated.  The vapor deposits are smooth and uniform while the electroplated deposits can be subject to surface roughness, flaking and non-uniformities \cite{Neudecker2020,Poenitz1983}.  Corrections for the energy dependent detector efficiency resulting from fission fragment anisotropy were developed by G.~Carlson \cite{Carlson1974}.

The PPIC is designed such that all of the deposits face in the same direction and are oriented away from the beam direction, which maximizes detection efficiency as the kinematic boost pushes fragments away from the backing foil and deposit.   By constructing the deposits such that they have similar thicknesses it is often assumed that the efficiency correction will cancel out in the cross-section ratio \cite{Tovesson2015,Staples1998}.  This assumption is challenged by the fact that differing isotopes will still have different fission anisotropy, and different material compositions, even with the same thickness, can still have different stopping powers.  These effects, however, will generally be less than 1\% at low energy and negligible at energies greater than 20 MeV.

When measurements are made with GICs, data is collected with the detector in two different orientations and the data are averaged \cite{Meadows1983}.  This solution still requires efficiency corrections to account for variable thickness between the targets and differing anisotropy of the isotopes.  

For this work, rather than average the data from the two orientations, the efficiency model calculates the efficiency for each target in a given orientation and takes into consideration the target thickness, roughness, and anisotropy as model parameters, as described in detail in Sec.~\ref{sec:efficiency}.  The resulting cross section is then compared for the two orientations of the \ftpc{} and interpreted as a validation of the efficiency model.  The results from the two orientations are averaged together for the final result, but this is to reduce statistical uncertainty rather than an attempt to bypass the need for an efficiency correction.  The effect of the target-beam orientation shown in Fig.~\ref{fig:Efficiency} is considerable, particularly at high energies.  The general trend of decreasing or increasing efficiency is primarily a result of the momentum transfer, while the structure is a result of reaction anisotropy.

Fig.~\ref{fig:XS_GLS} shows the agreement between the cross-section ratios for the two orientations of the \ftpc{} and targets with respect to the beam direction.  The lower panel of the figure shows the residual of the two data sets.  A zeroth order and a first order polynomial fit to the residuals are shown to illustrate the trend of agreement with beam energy.  The first order fit suggests that there is some disagreement above 20 MeV. This systematic disagreement was also indicated by the reduced $\chi^2$ of Eq.~\ref{eq:chi} as discussed in Sec.~\ref{sec:combine}.  This small disagreement likely indicates a potential systematic uncertainty in the efficiency model in that region. The averaging of the before and after rotation results should accurately account for any systematic discrepancy at high energy where the efficiency model is dominated by linear momentum transfer.  This is observed when the average cross-section ratios with and without the efficiency correction applied are compared.  The results are in excellent agreement at high energy, indicating that an averaging of the before and after rotation data largely captures the efficiency effects at high energy.  This was the approach used in previous GIC measurements \cite{Meadows1983} as previously stated in this section.


 Previous cross-section ratio measurements would be susceptible to the same type of orientation effect if they were made with non-uniform beams and targets.  Experimenters were generally aware of this concern and took measures to avoid the need for an overlap correction.  In the measurement reported by Staples and Morley \cite{Staples1998} a correction for beam or target non-uniformity was not required as the neutron beam was measured to be uniform to 0.5\%, while the electroplated plutonium samples were measured to be uniform to the 5\% level.  The uniformity of the uranium samples was not reported, but as they were vapor deposited it can be assumed they were uniform to better than 5\% which is typical for that type of target construction.  In the case of measurements such as that of Carlson \& Behrens \cite{Carlson1978} the threshold technique for normalizing is insensitive to non-uniformities in the beam.  In the case of the \ftpc{} measurement reported here, the circumstances were such that a nonuniformity overlap correction could not be avoided.   

The effect of the detector orientation on the overlap correction factor was shown in Fig.~\ref{fig:overlap} of Sec.~\ref{sec:overlap}.  The difference in the correction between rotations is as much as 3\% and has a dependence on neutron energy.  The agreement of the cross-section ratio data for the two orientations, shown in Fig.~\ref{fig:XS_GLS}, is therefore a validation of the correction methods applied. 

\subsection{Spatially Binned Cross Section Validation}
\label{sec:bin2bin_Validation}
Another way to validate the overlap correction is to bypass the need for the correction by taking a back-to-back cross-section ratio in each spatial bin such that the bin size is small enough that the beam and/or target is relatively uniform over the bin size.  The target-atom number ratio then must be scaled for each bin using the \talpha-decay data (see Fig.~\ref{fig:p9Target}).  The total cross-section ratio is determined by averaging over the binned cross-section ratio values.  The averaging scheme must be carefully chosen because the mean of the ratio of two quantities is not simply the ratio of the means of the numerator and denominator.   Instead, for ratios $r_i$, the geometric mean, $\left(\Pi_{i=1}^{N}r_i\right)^{1/N}$, which respects taking the inverse, is the appropriate operation.  Arithmetic and geometric mean formulations can be related through logarithms and exponentials

\begin{equation}
    e^{\sum\ln {r_i}} = \Pi r_i.
\end{equation}

The generic weighted geometric average formulation is then
\begin{equation}
    \ln{\frac{\sigma_0}{\sigma_1}} = \frac{\int w(x) \ln\frac{F_0(x)\rho_1(x)}{F_1(x)\rho_0(x)}d\mathbf{x}}{\int w(x) d\mathbf{x}},
\end{equation}
where the subscripts 0 and 1 refer to the back-to-back targets, $w(x)$ is a weight, $F_v(x)$ are the fission fragment counts and $\rho_v(x)$ are the target atom densities in each volume $v$, all with spatial coordinate $x$.  The target atom density $\rho_v(x)$ is proportional to the radiograph counts $\alpha_v(x)$ where the proportionality constant is determined from the atom number counting measurements.  The weight was chosen such that bins with the lowest uncertainty have the highest weight.  The main uncertainty that varies between bins is the counting statistics of the fragment and alpha counts, $F_v$  and $\alpha_v$, respectively.  The statistical uncertainty on a particular bin due to counting fluctuations is, for example in $F_0$, $\partial F_0 = \sqrt{F_0}$ and
\begin{equation}
    \delta^2_{F_0} = \left(\partial F_0\frac{\partial}{\partial F_0}\ln\frac{F_0\rho_1}{F_1\rho_0}\right)^2 = \left(\frac{\partial F_0}{F_0}\right)^2 = \frac{1}{F_0}.
\end{equation}
The total statistical variance is the sum of the independent counting variances and the weight for each bin is $1/\delta^2$, thus
\begin{equation}
    \delta^2 = \frac{1}{F_0} + \frac{1}{F_1} + \frac{1}{\alpha_0} + \frac{1}{\alpha_1}
\end{equation}
and
\begin{equation}\label{eq:weight}
     w(x) = \frac{1}{\frac{1}{F_0} + \frac{1}{F_1} + \frac{1}{\alpha_0} + \frac{1}{\alpha_1}}.
\end{equation}

This treatment requires that our binning discretization allows adequate counts for $F_v$ and $\alpha_v$ in each bin, so that small-number fluctuations are not a factor. The weighting function in Eq.~\eqref{eq:weight} assigns a weight of zero to any bins with zero counts in any category.  It does, however, introduce some systematic bias due to being correlated with the counts in the ratio.

\begin{figure}[ht]
\centering
\includegraphics[width=1.\linewidth]{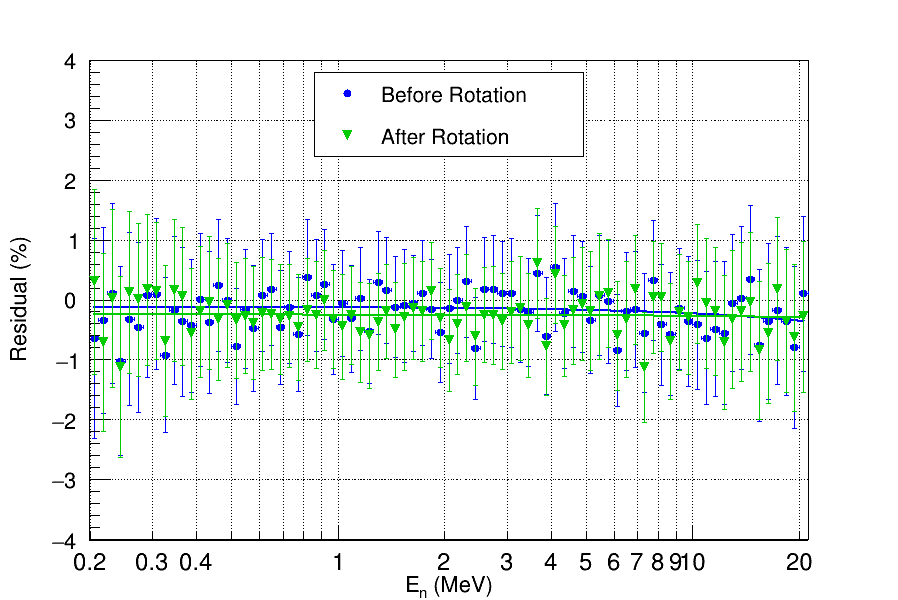}
\caption{The residual between the cross-section ratio using the overlap correction and that from the binned analysis, (overlap-corrected ratio - binned ratio)/(overlap-corrected ratio), for the two orientations of the \ftpc{}.  The results of the two methods are consistent.  The error bars are statistical only for the binned analysis.  A first order polynomial fit is included to illustrate the level of agreement.  The slope of the fit is consistent with zero while the intercept has a small systematic of $\sim$0.2 $\pm$0.15 \%}
\label{fig:BinOverResid}
\end{figure}

The residual between the cross-section ratio when applying the overlap correction (see Sec.~\ref{sec:overlap}) and when using the binned analysis described here is shown in Fig.~\ref{fig:BinOverResid} for the two orientations of the \ftpc{}.  The residual is defined as (overlap-corrected ratio - binned ratio)/(overlap-corrected ratio). A residual of zero across all energy bins indicates that the results of the two methods are consistent.  

While the two methods both rely on particle tracking, they are not mathematically equivalent.  The overlap method calculates a dot product of the beam and target shape while the binned method uses the target shape to renormalize many cross sections and average them together. 
Fig.~\ref{fig:BinOverResid} represents a cross-validation for the two methods of addressing the beam and target non-uniformity.
The final cross-section ratio reported here has been calculated using the overlap correction rather than the binned analysis because the overlap correction is not subject to increased statistical uncertainties from binning the fission fragments nor the weighting bias in the binned analysis.  

\subsection{Radial Cut Validation}
\label{sec:radcut}
\begin{figure}[ht]
\centering
\includegraphics[width=1.\linewidth]{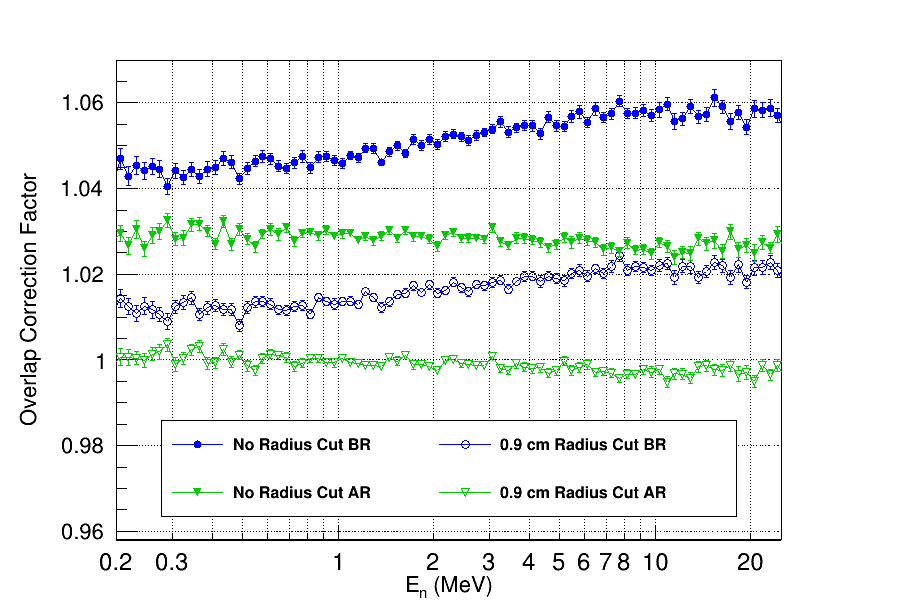}
\caption{The overlap correction for the before and after rotation data with no radial cut and a 0.9 cm radius cut.  The target radius is 1 cm.  The magnitude of the overlap correction changes significantly since the edge of the target in effect acts as a large non-uniformity.}
\label{fig:overlapRadCut}
\end{figure}

The magnitude of the overlap correction varies with the size of the target region used in the analysis because the aggregate shape of the beam and target varies with the target area.  This is particularly the case for the non-uniform plutonium target as it has no axis of symmetry (see Fig.~\ref{fig:p9Target}).  The overlap correction and the resulting cross-section ratios were calculated for several radial cuts and compared to validate the correction methodology.  Fig.~\ref{fig:overlapRadCut} shows the change in the magnitude of the overlap correction when making no radius cut compared to a 0.9 cm radius cut out of the total target radius of 1 cm.  The change in the magnitude of the overlap correction is $\sim$3\%, while the shape remains stable.  This result reflects that the edge of the target behaves like a large non-uniformity in the target shape.  The gradual change in the shape of the beam with neutron energy (see Fig.~\ref{fig:beamShapeMCNP}) is still largely captured within the central bulk of the target and so does not change with the small radial cut.

\begin{figure}[ht]
\centering
\includegraphics[width= 1.\linewidth]{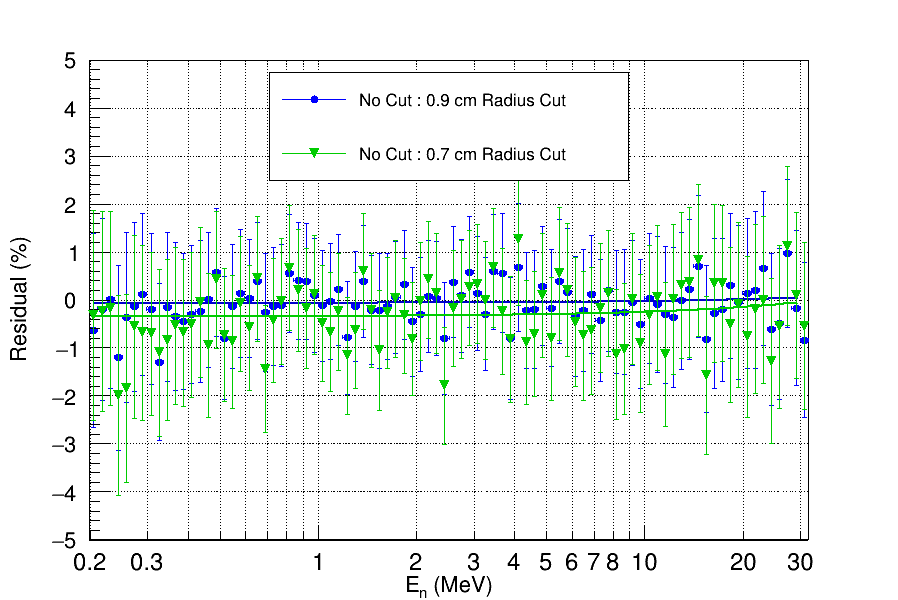}
\caption{The residual between the cross-section ratio with no radial cut and two radial cuts of 0.9 cm and 0.7 cm, (No-Cut - Radial-Cut)/(No-Cut).  The total target radius is 1 cm.  First order polynomial fits are included to illustrate the agreement.  The slope and intercept of the fit are consistent with zero.}
\label{fig:radcutResidual}
\end{figure}

Fig.~\ref{fig:radcutResidual} shows the residual between the cross-section ratio with no radial cut and radial cuts of 0.9 cm and 0.7 cm, (No-Cut -- Radial-Cut)/(No-Cut).  The results are consistent between radial cuts in both shape and overall normalization. First order polynomial fits are included to illustrate the level of agreement.   The radial cut analysis is a validation of the vertex tracking in general as two of the terms in Eq.\eqref{eqn:xsCalc} are effected by the radial cut.  Not only the overlap correction but also the target atom number ratio normalization term changes.  Because of the large non-uniformity in the plutonium target, the atom number ratio is not invariant with a change in target area.  The \talpha-decay radiograph in Fig.~\ref{fig:p9Target} is used to scale the overall normalization.  For a radius cut of 0.9 cm the atom number ratio changes by 7.7\%.  The level of agreement between the cross-section ratios for the two radial cuts validates that the radiograph accurately reflects the overall target spatial distribution.  No radial cut within the target area was performed on the final data analysis as it requires a recalculation of the normalization constant which would introduce systematic uncertainties to the final normalization which was performed with an independent measurement.

\subsection{Track Vertexing Resolution}
\label{sec:trackRes}
In the previous Sec.~\ref{sec:radcut} it was pointed out that reducing the target area with a radial cut necessitates a re-scaling of the target atom number ratio.  The scaling factor is determined using the \talpha-vertex tracking data. If the radial cut validation is to be believed, any potential additional uncertainty introduced by the track pointing (vertex) resolution on this scaling factor must be quantified.
Furthermore the choice of vertex bin size could potentially have a systematic effect on the overlap correction.  Both of these possibilities have been explored.

The targets are produced with machined metal masks to set the shape for both the vapor deposited uranium target and the electroplated plutonium target.  It is assumed that the target is circular and that the edge of the target is vertical, with any deviations from the expected shape to be much smaller than our vertex resolution.  By plotting the radial distribution of the \talpha-track vertices and fitting the edge with a Gaussian-like function the radius of the target and therefore the vertex resolution can be inferred.  The edge-fitting function was defined as

\begin{equation}
    p_0\cdot x^{p_3}\left(1 - \text{erf}\frac{x-p_1}{\sqrt{2}p_2}\right)+p_4
    \label{eqn:erf}
\end{equation}

\begin{figure}[ht]
\centering
\includegraphics[width= 1.\linewidth]{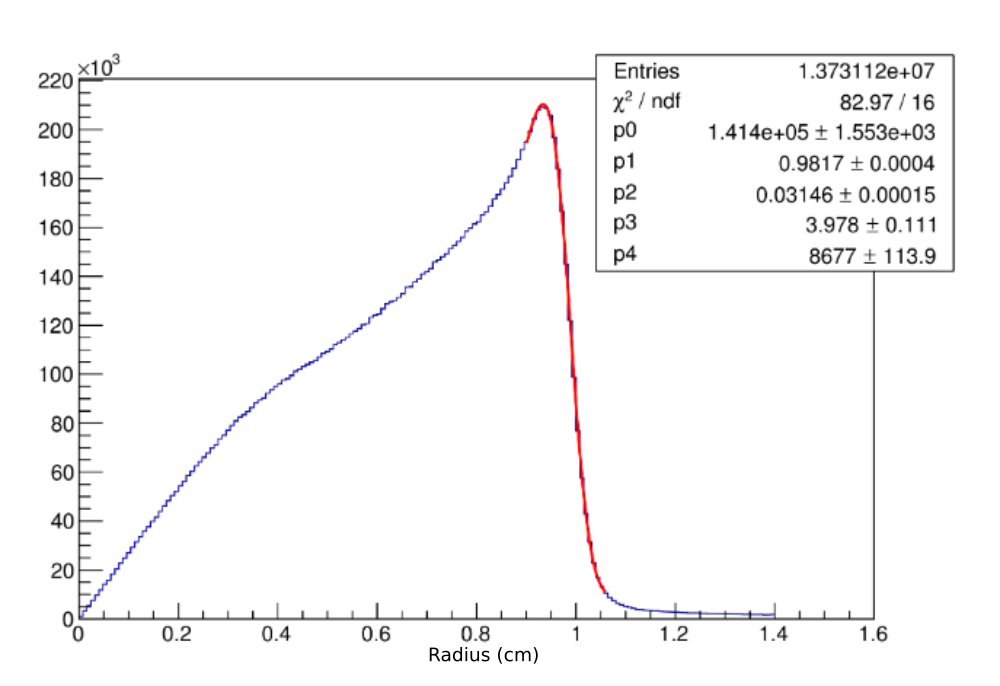}
\caption{Distribution of \talpha-track vertex radial locations for the plutonium target, fitted to determine the target radius ($p_1$) and vertex resolution ($p_2$), which is $\sim$0.31 mm.  The results for the uranium target were the same within uncertainties.}
\label{fig:edgeFit}
\end{figure}

Fig.~\ref{fig:edgeFit} shows the distribution of \talpha-track vertex radial locations and the results of the fit for the plutonium target.  The vertex resolution ($p_2$) is found to be $\sim$0.31 mm with the uranium target resolution the same within uncertainties.

\begin{figure}[ht]
\centering
\includegraphics[width=0.48\linewidth]{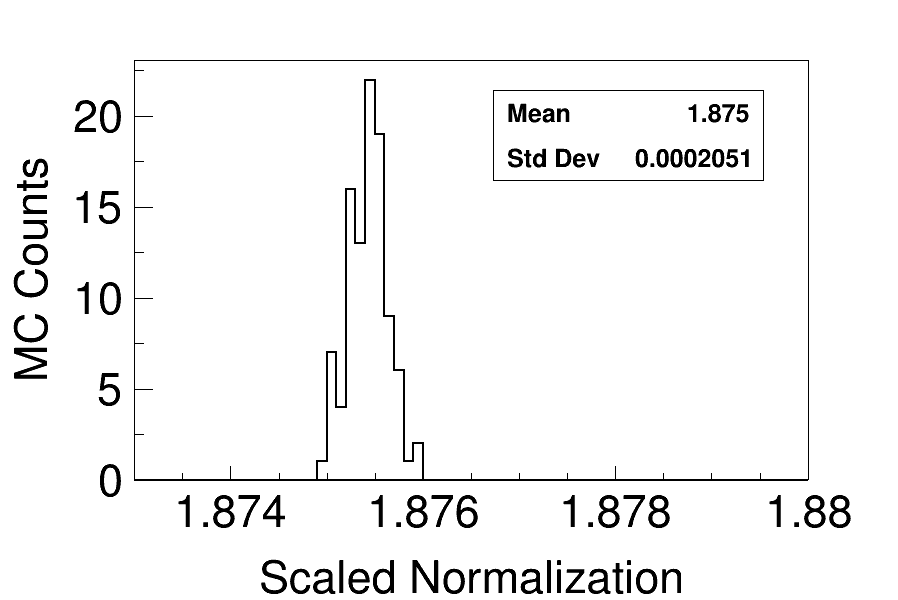}
\includegraphics[width=0.48\linewidth]{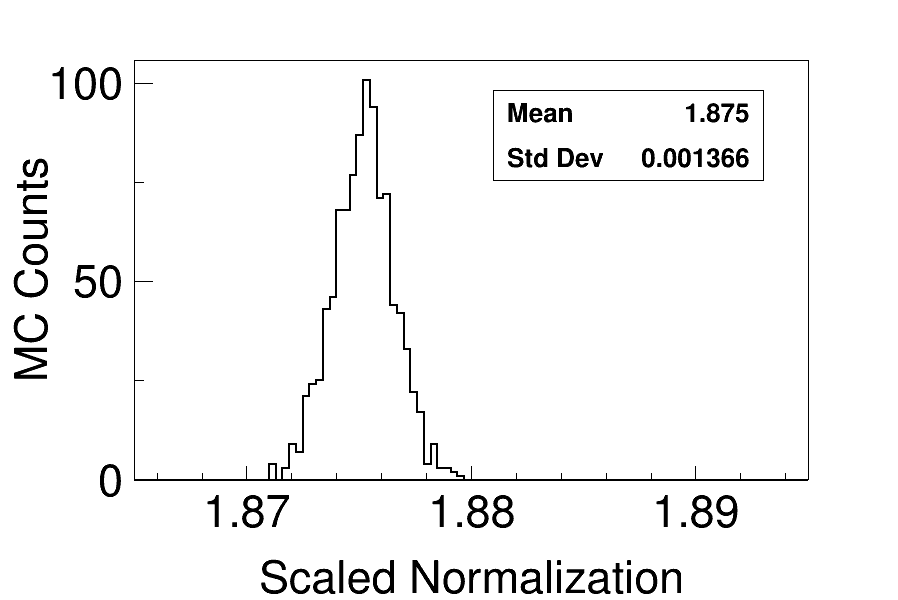}
\caption{(Left) The scaled target atom number ratio for the 0.9 cm radial cut generated from 100 Monte Carlo realizations varying the vertex bin counts with Poisson statistics. (Right) The scaled target atom number ratio for the 0.9 cm radial cut generated from 1000 Monte Carlo realizations with 1M tracks each with a Gaussian smearing of 0.31 mm applied.}
\label{fig:MC_Normalization}
\end{figure}

Two approaches were taken to evaluate the uncertainty in the target atom number ratio scaling needed for the radial cut analysis.  The first involved performing a Monte Carlo calculation of the scaling factor by varying the counts in a given \talpha-track vertex bin with Poisson statistical uncertainties.  The second approach used the \talpha-track vertex data as a probability distribution to generate vertices which were then smeared according to a Gaussian resolution.  The first panel of Fig.~\ref{fig:MC_Normalization} shows the distribution of the calculated scaled target atom number ratio for the 0.9 cm radial cut from 100 Monte Carlo realizations, varying the statistics in the vertex bins assuming Poisson statistics.  The second panel of Fig.~\ref{fig:MC_Normalization} shows the distribution for the 0.9 cm radial cut calculated from 1000 realizations of 1M vertex distribution with 1M counts each using a Gaussian smearing of $\sigma$=0.31 mm.  

The first method results in an uncertainty of 0.01\% while the second method yields an uncertainty of 0.07\%.  It is not surprising that the second method has a higher uncertainty as it relies on the data for the probability distribution to generate realizations of the target shape, which inherently includes the tracking resolution.  Then it is further degraded by applying the additional resolution smearing.  Both methods show that the uncertainty on the scaling of the target atom number ratio for radial cuts by using the \talpha-track vertex distribution is negligible if one assumes Poisson statistics and Gaussian resolution.

\begin{figure}[ht]
\centering
\includegraphics[clip=true, trim=10mm 10mm 10mm 10mm,width=1.\linewidth]{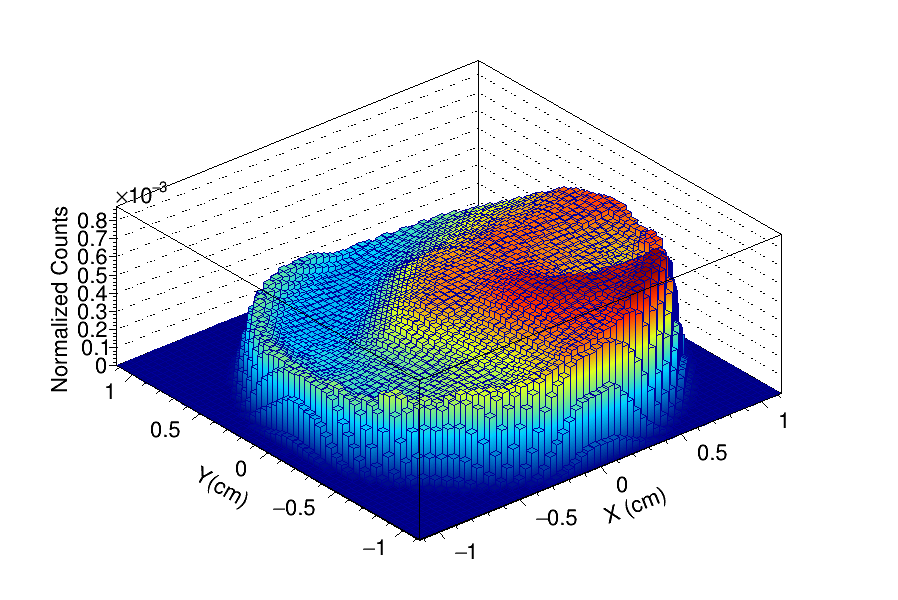}
\includegraphics[clip=true, trim=10mm 10mm 10mm 10mm,width=1.\linewidth]{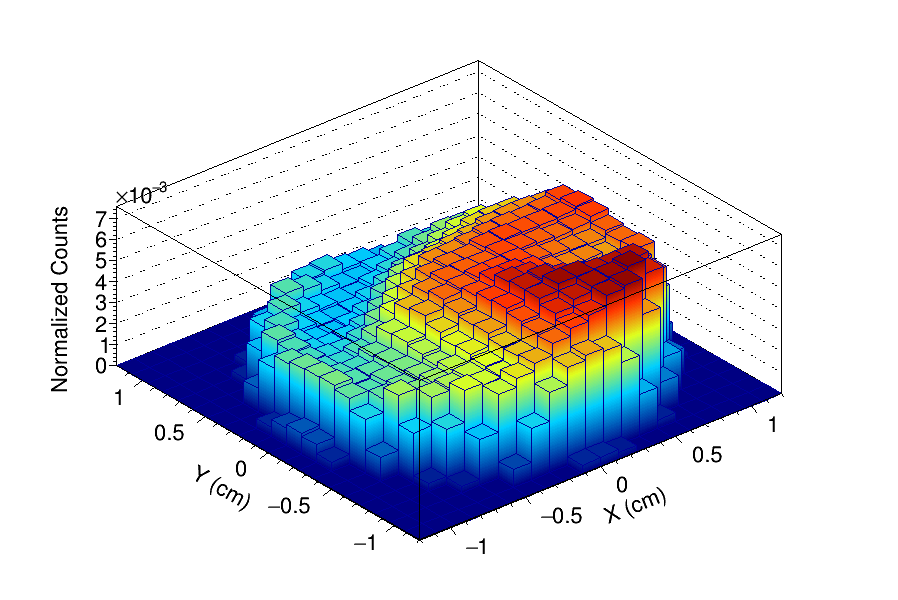}
\caption{The plutonium target shape plotted with 0.4 mm and 1.2 mm bin sizes.  As the bin size increases the target edge is less accurately represented.  The vertex resolution is 0.31 mm.}
\label{fig:vertexBin}
\end{figure}

\begin{figure}[ht]
\centering
\includegraphics[width=1.\linewidth]{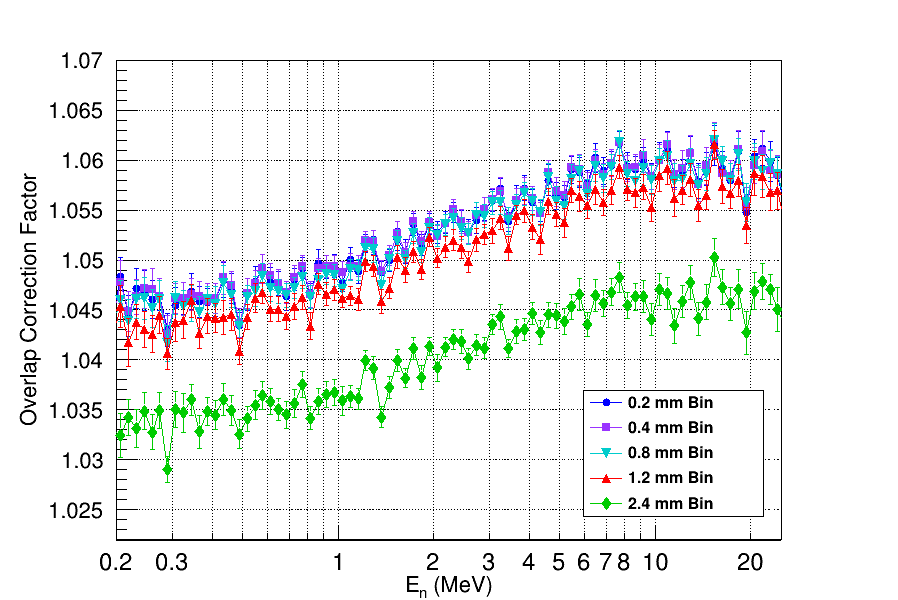}
\includegraphics[width=1.\linewidth]{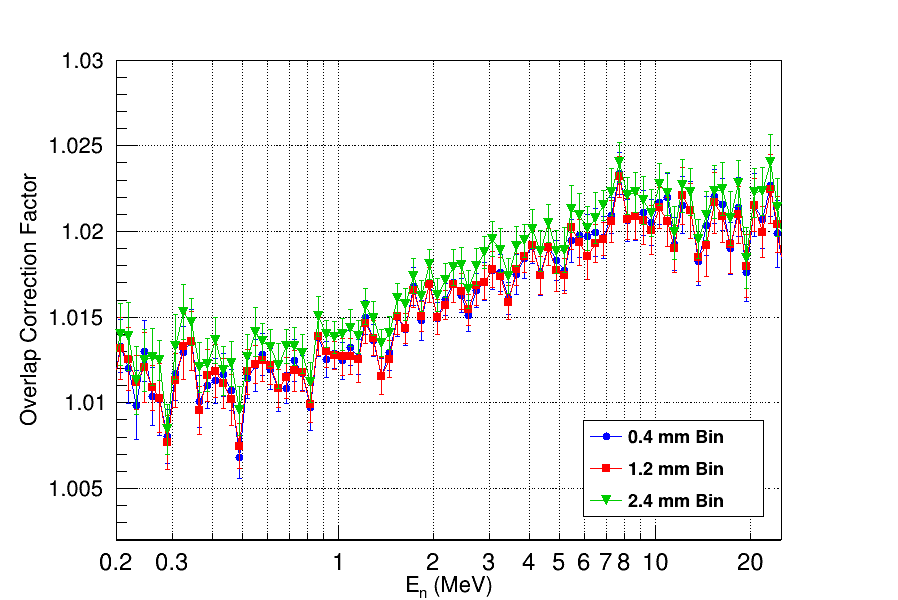}
\caption{The calculated overlap correction for several bin sizes with no radial cut and a 0.9 cm radius cut.  The overlap has a systematic sensitivity to bin size.  When the target edge is cut out the sensitivity is greatly reduced.}
\label{fig:overlapRebin}
\end{figure}

Features in the target or beam distribution obscured by the vertex resolution could result in a systematic uncertainty in the overlap correction.  This is not expected however since the processes and physics dictating the beam and target shape are expected to vary smoothly.  The vertex bin size sensitivity study suggests that any potential effect is negligible.  The overlap correction was calculated for several vertex bin sizes.  Fig.~\ref{fig:vertexBin} shows the plutonium distribution plotted in 0.4 mm and 1.2 mm square bins.  Fig.~\ref{fig:overlapRebin} shows the overlap correction plotted for several bin sizes.

As the bin size increases the overlap correction systematically shifts down while the shape is unaffected.  For a bin size of 2.4 mm the overlap correction shifts by $\sim$1\%.  When the target edge is cut from the data the systematic effect is removed.  This again illustrates how the target edge itself acts as a large non-uniformity.  The overlap correction is insensitive to $<$0.3\% for binning below a bin size of 1.2 mm or 4 times our pointing resolution of 0.31 mm.  Since the target has a radius of 1 cm, a bin size of 2.4 mm corresponds to $\sim$1.8\% of the total target area but represents 24\% of the total target radius.  In other words, it requires an unreasonable degradation of the vertex resolution by a factor of 8, to the point where only features 1/4 the size of the total target can be resolved, to see even a 1\% effect on the overlap calculation. 
Other than the target edge, sharp irregular features in the target shape are not expected to result from either the vapor deposition process or the electroplating process.  While there is the potential for a systematic uncertainty resulting from vertex resolution, this sensitivity study indicates it is negligible.

\subsection{Track Vertex Bias Sensitivity}
\label{sec:VertexBias}

\begin{figure*}[ht]
  \centering
  \includegraphics[width=0.3\linewidth]{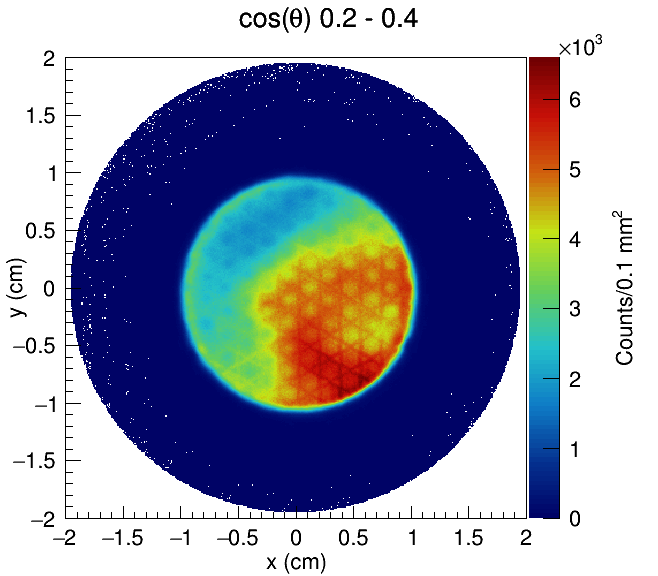}
  \includegraphics[width=0.3\linewidth]{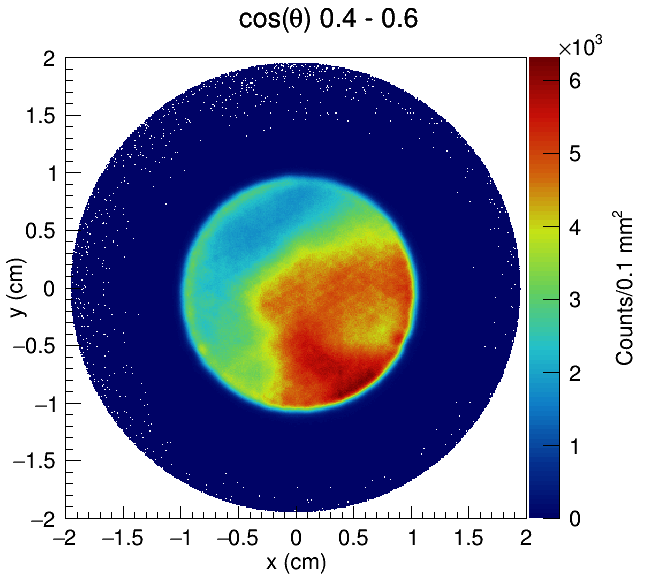}
  \includegraphics[width=0.3\linewidth]{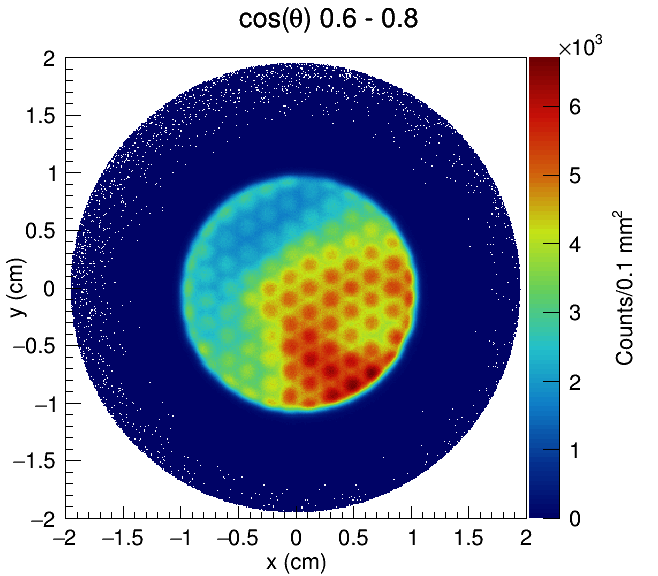}
  \caption{The tracking algorithm has the potential to bias vertices toward the edges or centers of the hexagonal anode pads of the \ftpc{}.  The algorithm is adjusted so that this effect is averaged out for all tracks but selecting a subset based on polar angle cuts can make the biasing apparent.  The three panels show track vertices for cos($\theta$) cuts in the range of 0.2--0.4 (left), 0.4--0.6 (middle), and 0.6--0.8 (right) out of the full range of 0--1.}
  \label{fig:p9_theta}
\end{figure*}
\begin{figure}[ht]
\centering
\includegraphics[clip=true, trim=12mm 0mm 25mm 0mm,width=1.\linewidth]{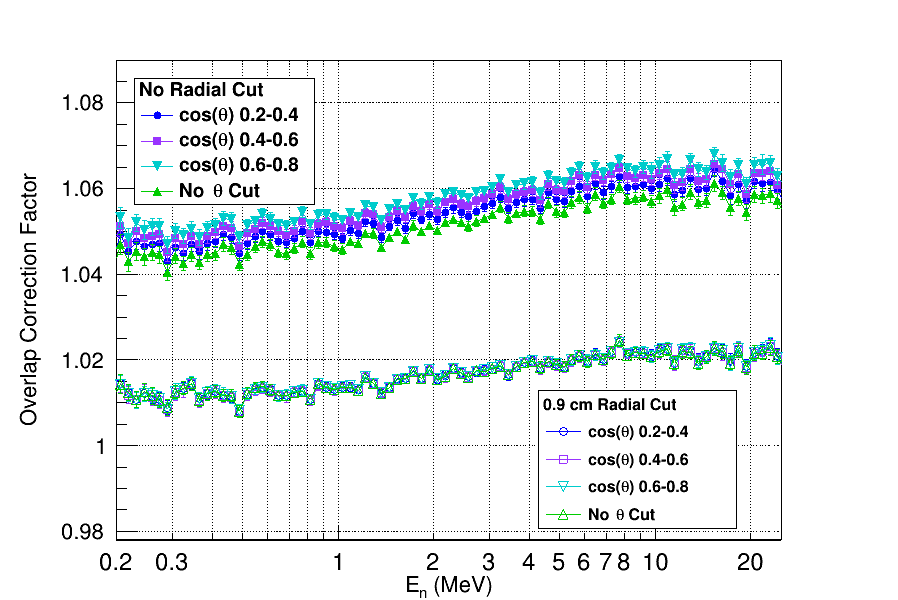}
\includegraphics[clip=true, trim=10mm 0mm 25mm 0mm,width=1.\linewidth]{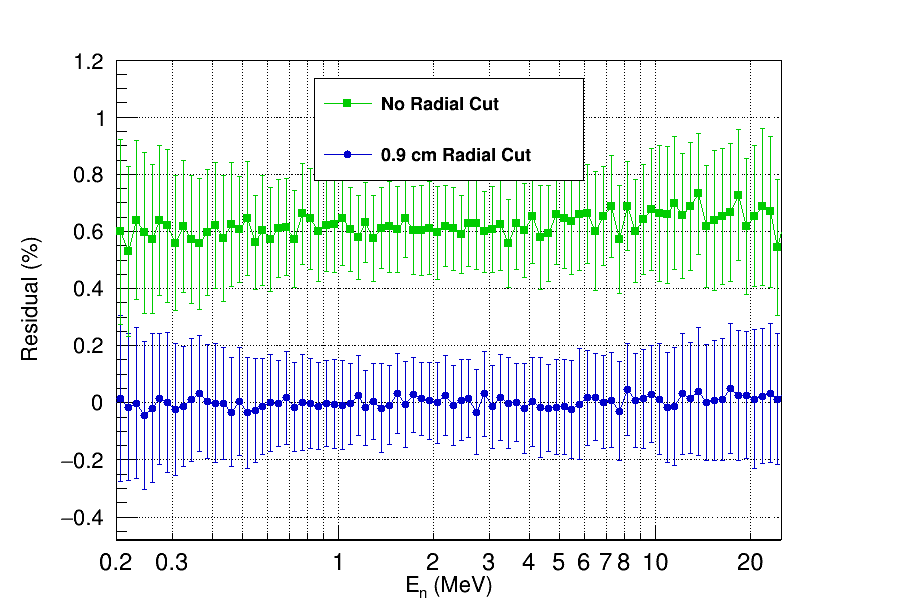}
\caption{The overlap correction calculated when applying cos($\theta$) cuts to the vertex data with and without a radial cut. Also shown is the residual in the overlap correction between no $\theta$ cut and the cos($\theta$) 0.6--0.8 range, the largest shifted example, with and without a radial cut of 0.9 cm.  With no radial cut there is a maximum of 0.6\% effect on the overlap while with a 0.9 cm radius cut the effect is negligible.}
\label{fig:resid_ThetaCut}
\end{figure}

The radial cut validation is also sensitive to any bias in the vertex distribution used to calculate the target atom number ratio scaling factor.  The track vertex reconstruction is subject to a bias resulting from the 2 mm pitch hexagonal pad structure of the anode.  The algorithm has a focusing procedure that on average eliminates the vertex bias, but for tracks in certain polar angle ranges the bias becomes apparent.  A sensitivity study was performed on three track polar angle ranges that show the potential effect in the most extreme cases of the vertex biasing.  Fig.~\ref{fig:p9_theta} illustrates the vertex biasing on the plutonium target distribution, where the hexagonal pattern is readily visible in two of the three polar angle ranges, with the worst biasing occurring for tracks with 0.6$<\cos(\theta)<$0.8.
The first panel of Fig.~\ref{fig:resid_ThetaCut} shows the overlap correction (OT) calculated for the different regions of $\cos(\theta)$ and integrated over all angles with and without a radial cut of 0.9 cm.  The second panel shows the residual on the overlap correction, calculated as (OT$_{no~cut}$ - OT$_{\cos\theta ~cut}$)/OT$_{no~cut}$, for the most extreme vertex bias window (0.6$ ~<~ \mbox{cos}(\theta)~<~$0.8)  with and without the radial cut.  The residual is larger for the case of no radial cut, on the order of 0.6\%, with no apparent energy dependence, and becomes negligible when the radial cut is applied.  Fig.~\ref{fig:resid_ThetaCut} highlights again the sensitivity of the overlap correction to target non-uniformities such as the target edge.  While the radial cut removes the residual in the overlap correction due to the vertex biasing, as discussed in Sections \ref{sec:radcut} and \ref{sec:trackRes}, the trade-off is that the target atom number ratio must be re-scaled using a prescription that is also sensitive to the vertex bias.  In fact, if the scaling for a radial cut is calculated using the most extreme vertex biased data of Fig.~\ref{fig:p9_theta} there is a sensitivity of up to 1\%.  

Despite the sensitivities of the analysis to the radial cuts and the track vertexing resolution and bias, the stability of the cross-section ratio in both its energy-dependence and overall normalization under extreme variations of these effects is a validation of both the overlap correction procedure and the tracking algorithm.  Since a correction is applied to compensate for the bias and the effect is small, its uncertainty is considered to be negligible.

\subsection{Target Alignment Sensitivity}
\label{sec:targetAlignValidate}
The overlap correction and binned analysis could also be sensitive to a misalignment between the two volumes of the \ftpc{}.  While the anode pad planes are precisely constructed printed circuit boards and are identical from one volume to the other, their alignment is not necessarily perfect as a result of the way the \ftpc{} is assembled.  The anodes are held in place by dowel rods that have approximately 200 $\mu$m tolerance.  This means that the coordinate system from one volume is not necessarily aligned with the other, which would have an obvious impact on the overlap correction and binned analysis.  The misalignment is corrected for by observing beam induced \talpha-tracks that have enough energy to pass through the cathode plane and deposit energy in both volumes.  By observing the offset of the track vertices where they should intersect at the cathode plane the alignment correction can be determined.  The alignment correction as discussed in Sec.~\ref{sec:target_alignment} was found to be -167 $\mu$m along the x-axis and -131 $\mu$m along the y-axis, within the expected 200 $\mu$m maximum.

\begin{figure}[ht]
\centering
\includegraphics[clip=true, trim=9mm 0mm 25mm 0mm,width=1.\linewidth]{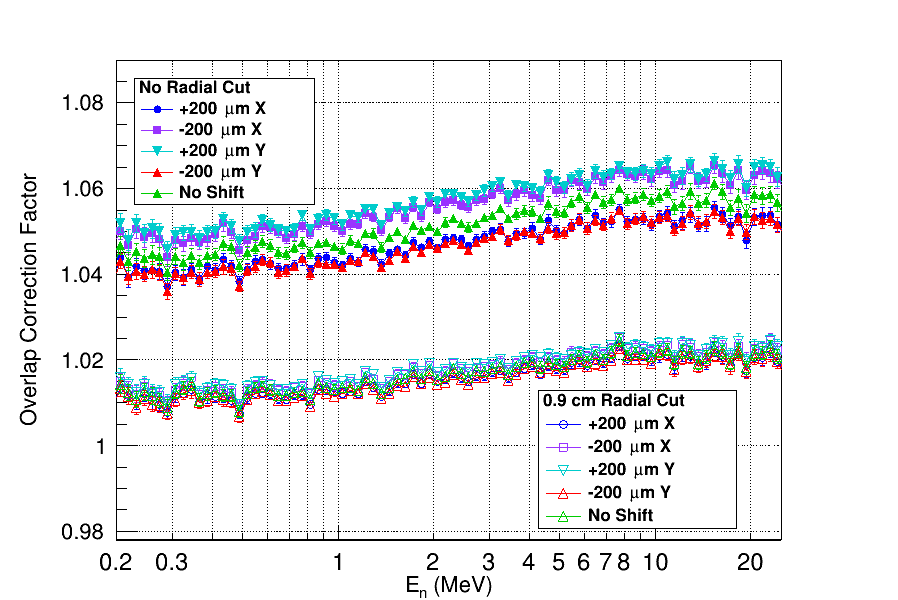}
\includegraphics[clip=true, trim=9mm 0mm 25mm 0mm,width=1.\linewidth]{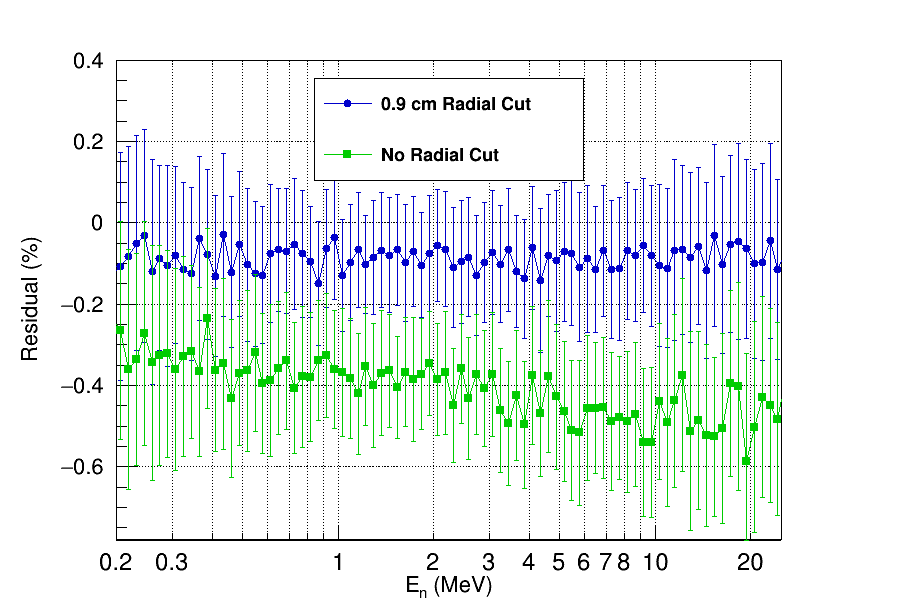}
\caption{The overlap correction calculated when applying registration shifts to the vertex data with and without a radial cut.  Also shown is the residual calculated as (unshifted - shifted)/unshifted between no shift and the +200$\mu$m X shift. This shows a maximum of 0.5\% effect on the overlap correction if no correction for target alignment is performed.}
\label{fig:resid_Shift}
\end{figure}

To understand the potential impact of an inaccurate alignment measurement, a sensitivity study was conducted by introducing a bias to the track vertex data by shifting one volume by 200$\mu$m with respect to the other.  The same shift was applied for both fragments and \talpha-tracks as the data sets for the target radiograph and cross-section ratio were taken without a disassembly of the \ftpc{} in between so no shift in the pad plane position could have occurred. Fig.~\ref{fig:resid_Shift} shows the residual between the calculated overlap for shifted and unshifted data, calculated as (unshifted - shifted)/unshifted.  The overlap correction shows up to a 0.5\% shift if no radial cut is performed and the shift reduces to a 0.1\% effect if a 0.9 cm radius cut is performed.  This reflects the fact that the target edge itself represents a large non-uniformity, but when the edge is removed with a radial cut the targets are relatively smooth over small distances.  The results are consistent for the binned analysis method.
This sensitivity study is the worst case potential misalignment for the data. Based on these results, the cross-section ratio uncertainty from target alignment bias is considered to be negligible.

\subsection{Space Charge Scaling Sensitivity}
\label{sec:SpaceChargeValidate}
Finally, we evaluate the sensitivity of the overlap correction to the electric drift field distortions caused by the build-up of space charge in the \ftpc{}.  The high \talpha-decay activity of the plutonium target is the primary source of the sustained space-charge of positive ions drifting towards the cathode in the \ftpc{} drift volume.
The space-charge distortion causes the start vertices of the fission fragment and \talpha-particle distributions to be drawn towards the center of the \ftpc{} so that the target distribution appears smaller in the data than it is in reality.  Using an analysis of a photograph of the target, the plutonium data is corrected by scaling the track vertices out radially.  The correction, described in Sec.~\ref{sec:SpaceChargeCorrection}, is 0.26 mm or 2.64\% of the total radius.  

\begin{figure}[ht]
\centering
\includegraphics[width= 1.\linewidth]{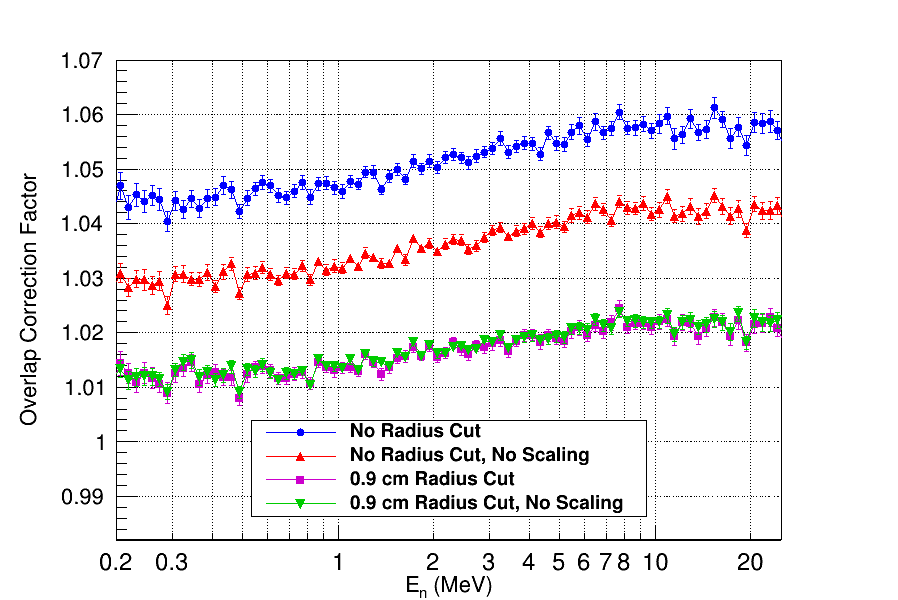}
\caption{The overlap correction calculated with and without the space-charge scaling factor applied and with and without radial cuts.  The space-charge scaling has no effect when a radius cut is applied and $\sim$1.5\% effect with no radius cut.  Only the overlap correction for before rotation is shown.  The effect for after rotation is the same.}
\label{fig:overlap_NoScale}
\end{figure}

For this sensitivity study the space-charge correction was simply turned off.  Fig.~\ref{fig:overlap_NoScale} shows the overlap correction for data with and without the space-charge scaling applied and with and without radial cut.  The space-charge correction has an energy-independent $\sim$1.5\% systematic effect on the overlap correction.  However, when the 0.9 cm radial cut is applied, eliminating the target edge, the effect disappears.  

The overlap correction normalizes the target distribution within the cut radius. Therefore when a radial cut is applied inside the scaling correction the general shape of the target is largely unaffected and the target density is unchanged.  In other words, turning off the space-charge scaling correction breaks the radial cut stability described in Sec.~\ref{sec:radcut} and illustrated in Fig.~\ref{fig:radcutResidual}.  This is interpreted as a cross-validation that the space-charge scaling correction is needed and is being applied correctly.  

The radius of the plutonium target was determined to be 10.11 $\pm$ 0.05 mm based on an analysis of a digital photograph, while the \ftpc{} vertex resolution was determined to be 0.3 mm.  The vertex resolution, the effect of which was found to be negligible on the overall uncertainty, is larger than the target radius uncertainty. Consequently, any systematic uncertainty in the overlap correction resulting from the space-charge correction or radius uncertainty is considered to be negligible.

%% file: Comparisons.tex
\begin{figure}[ht]
\centering
\includegraphics[width=1.\linewidth]{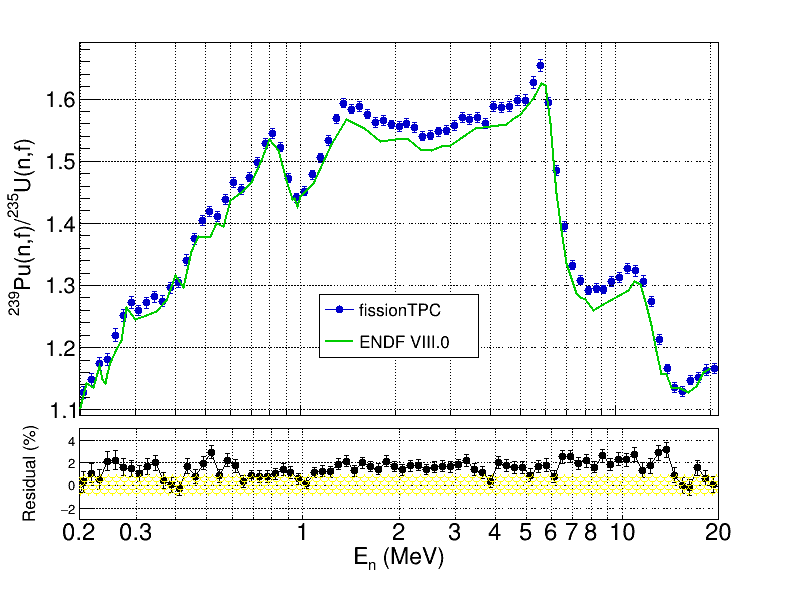}
\caption{The \ftpc{} measured \pu(n,f)/\u(n,f) cross-section ratio as a function of neutron energy compared to the ENDF/B-VIII.0 evaluation.  The lower panel shows the residual ((data-ENDF)/data) between the data and evaluation.  The yellow band is $\pm$ 1\% from a zero residual.  The error bars on the residual are from the \ftpc{} data only.}
\label{fig:fTPC_ENDF8}
\end{figure}

\label{sec:compare}
Fig.~\ref{fig:fTPC_ENDF8} shows the \ftpc{} \pu(n,f)/\u(n,f) cross-section ratio as a function of neutron energy from 0.2 to 20 MeV along with the ENDF/B-VIII.0 evaluation \cite{ENDF8}.  The lower panel of the figure shows the residual between the data and the evaluation, calculated as (data - ENDF)/data.  There is a strong indication that the \ftpc{} normalization is in disagreement with that of ENDF at the 2\% level.  The measurement is largely in agreement with the ENDF energy-dependent shape.

\begin{figure}[ht]
\centering
\includegraphics[width=1.\linewidth]{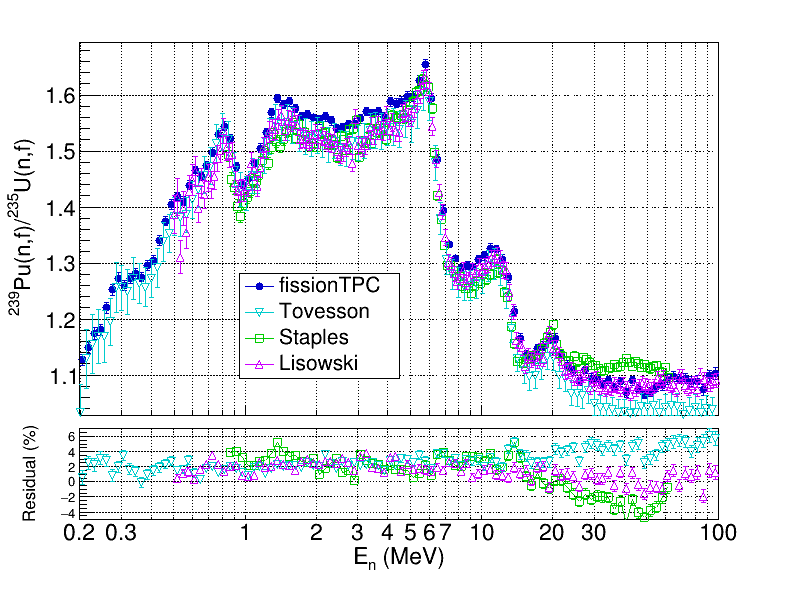}
\caption{The \ftpc{} measured \pu(n,f)/\u(n,f) cross-section ratio as a function of neutron energy compared to data from references \cite{Staples1998,Lisowski,Tovesson}. The lower panel shows the residual (calculated as (\ftpc{} - reference)/\ftpc) between the \ftpc{} data and the various references. The error bars on the residual are from \ftpc{} data only.}
\label{fig:fTPC_Staples}
\end{figure}

Fig.~\ref{fig:fTPC_Staples} compares the \ftpc{} cross-section ratio as a function of neutron energy from 0.2 to 100 MeV to published data that were also measured at LANSCE-WNR \cite{Staples1998,Lisowski,Tovesson}.  The lower panel of the figure shows the residual between \ftpc{} measurements and the three published results, calculated as (\ftpc{} - reference)/\ftpc.  The level of disagreement in the normalization is similar to that of the disagreement with ENDF/B-VIII.0, with the \ftpc{} data $\sim$2\% systematically high relative to the other results.  This is not surprising as the data shown, amongst others, contribute to the ENDF evaluation.  Above 20 MeV there are significant shape disagreements between all of the data sets.  

\subsection{Discussion}
There are two obvious concerns as to why the \ftpc{} results are systematically different from the ENDF evaluation.  The first concern is the large non-uniformity of the \pu{} target and the corresponding non-uniformity of the beam shape.  While this did present a challenge in the analysis, the \ftpc{} is uniquely capable of directly measuring and correcting for the effect, which is detailed in Sec.~\ref{sec:overlap}. A variety of analysis validations were explored in Sec.~\ref{sec:validations} to test the correction.  In particular, the rotation of the \ftpc{} and the so-called binned analysis, provide strong evidence that the non-uniform beam--target overlap was accurately corrected for. Furthermore, the effect of the overlap has a shape, \ie{} it varies as a function of neutron energy as the shape of the beam changes.  If the overlap term were the cause of the discrepancy with ENDF, it would not likely appear to be so systematic across energy bins.  While one could argue that the \ftpc{} measurements show shape disagreements with other published results, particularly above $\sim$30 MeV, the three examples to which the \ftpc{} data are compared (see Fig.~\ref{fig:fTPC_Staples}) all disagree with each other at higher energies and there is no definitive case to be made for the cause of these discrepancies or which result should be trusted more.  In fact we would argue that the ability of the \ftpc{} to directly address the effects of angular distributions and beam-induced backgrounds, all of which would be compounded at higher energies, could lend greater credence to our shape results above 30 MeV.  
There is some consideration that proton contamination in the LANSCE beam for energies $>$50 MeV could exist and that proton induced fission could result in a systematic uncertainty.  These protons may come from (n,p) scatters in the beamline or from high-energy protons from the spallation target down-scattering along the beamline. This could be investigated with detailed simulations and potentially a \ftpc{} analysis, however it is beyond the scope of this article.  The 90L flight path at LANSCE WNR does contain a permanent magnet to suppress protons in the beam.

The second concern, and the one most likely to be the ultimate source of the discrepancy with ENDF, is that the measurement of the target normalization is possibly biased. The target atom number ratio for this work was measured with a dedicated instrument. 
Details of the silicon detector setup and measurements used to assess the target atom number ratio for this experiment are documented in Ref.~\cite{Monterial2021}.
Based on the evaluation of the silicon detector measurements and the mass spectrometry data, the target atom number ratio correction, or $N_s/N_x$ is 1.7343 $\pm$ 0.0050.

The measurement was performed with a silicon detector, with the design of the apparatus guided by the prescriptions of Pomme \cite{Pomme2015} and Denecke \etal~\cite{Denecke}.  Those studies make detailed recommendations and describe independently confirmed measurements with accuracies better than 0.1\%.  References in those articles highlight the precision achievable by these means and the maturity of the field with respect to establishing and validating uncertainties.

The target was measured on two separate occasions and the data was analyzed by three separate teams in an attempt to validate the results.  The results from both data sets and from each of the analysts all agree within uncertainties. Our concern centers on the fact that the original target characterization data was collected $\sim$6 months after the in-beam cross-section data were collected.  It is possible that the target either underwent some form of degradation, such as oxidization of the plutonium whereby target material fell off the surface of the target, or was damaged when being removed from the \ftpc{}.  

The second set of target characterization data was collected $\sim$3 years after the first set.  Since the results of the two data sets are in good agreement it is unlikely that target degradation is the cause, as the process would likely have continued and the two measurements, separated by such a lengthy time, would have observed a further change.  This leaves the most likely possibility that the target could have been damaged upon removal from the \ftpc{}, such as coming in contact with a gloved hand or tool whereby material was removed.

Radiograph data was taken while the target was in the \ftpc{} prior to collecting beam data, which could potentially provide evidence of target damage during target removal.  The analysis of the radiograph data was originally abandoned as the primary normalization measurement because it was not believed that an uncertainty $<$1\% could be achieved, which is the primary objective of the \ftpc{} measurement.  The primary challenge in analyzing the radiograph data is the high rate of activity of the \pu{} which results in a significant amount of pile-up, the effect of which can be seen in Fig.~\ref{fig:lvadc_autorad}.  The various counts around the primary peak in the \pu{} are a result of pile-up whereby overlapping tracks can confuse the tracking algorithm and portions of the charge clouds are not correctly assigned.  The tracks at greater length and energy than the primary peak in the \u{} data are the result of high energy \talpha-particles from the uranium decay chain daughters which will not always be fully contained within the active area of the \ftpc{}. The pile-up had a negligible effect on the vertex histograms used for the overlap analysis as data selection cuts could be made to clean the spectrum.  These pile-up events however present a much greater challenge when one is attempting to count the absolute number of \talpha-tracks to better than 1\% accuracy.  It is therefore impossible to make any usefully accurate conclusions based on the radiograph data at this time.

Among the cross-section ratio normalization terms, the target atom number ratio has the largest deviation from unity and has been a leading source of uncertainty in previous cross-section measurements.  In several studies, measurement practices and resulting uncertainties for target characterization were not documented sufficiently and make it difficult to compare with the present work.  In the measurement reported by Lisowski \etal~\cite{Lisowski} it is simply stated that the normalization factor was obtained by  ``using the results of Meadows'' though it is unclear which of the specific targets measured by Meadows \cite{Meadows1983} were used.  In the measurement reported by Staples \& Morley \cite{Staples1998} it is stated that the same fission foil samples as Lisowski were used.  Staples \& Morley explain that \talpha-spectroscopy was used to count the foil mass and ``verify the isotopic contamination'' but no details on the detector design were given and the only uncertainty referred to is that of the calibration source.  The work by Meadows \cite{Meadows1983} and Poenitz \cite{Poenitz1983} are very well documented however, and our approach to counting the target is generally the same as theirs: determine the isotopic composition of the actinide and count the \talpha-decays.  It should be noted that the isotopic composition of the target reported by Staples \& Morley does not exactly match any of the target samples reported by Meadows or Poenitz.  As was discussed in Ref.~\cite{Monterial2021}, lack of accurate knowledge of the isotopic composition can lead to systematic uncertainty in the number of target atoms calculated from \talpha-counting, but the \talpha-spectroscopy results of Staples \& Morley cannot be further assessed without more details.

Other relatively recent measurements such as those of Shcherbakov \etal~\cite{Shcherbakov} and Tovesson \& Hill \cite{Tovesson} were normalized with either the cross-section threshold method \cite{Behrens1973,Behrens1981,Carlson1978} or normalized to existing evaluations.

\subsection{Conclusions}
While we are confident in the quality of our data and analysis, the $\sim$2\% absolute normalization disagreement with ENDF and previous measurements is simply too large to ignore.  The systematic nature of the discrepancy, and some circumstantial evidence give a strong indication that the absolute normalization measurement is the culprit.  The unique nature of the \ftpc{} data has allowed us to explore a variety of potential sources of systematic uncertainty not accessible to previous measurements.  The careful analysis and several validations detailed throughout this article give us a high level of confidence in the cross-section shape analysis and the associated uncertainties.  We therefore conclude that the data reported here can be utilized as precision \emph{shape} data in any future evaluation. 

At the time of this writing another \pu/\u{} target has been prepared by NIFFTE collaborators at Oregon State University.  This target utilized vacuum volatilization for both the \u{} and \pu{}, which results in a very uniform deposit.  Currently there are plans to measure said target in beam at LANSCE-WNR during the 2021 run cycle.  Steps will be taken to characterize the target before and after cross-section data collection, and adjustments to the beam collimation will be made to improve uniformity.  This measurement should provide an absolute normalization and address any remaining concerns regarding the beam and target non-uniformity.

In the mean-time, the shape data discussed here have been reported to the IAEA co-ordinated Neutron Data Standards group~\cite{Carlson2018}.
This group provides standard  nuclear data, such as for $^{235}$U(n,f) or $^6$Li(n,t) cross sections, or reference data, such as for $^{238}$U(n,f) and $^{239}$Pu(n,f) cross sections, for many reactions other experiments are frequently measured in ratio to.
The standard nuclear data were also partially included in ENDF/B-VIII.0~\cite{ENDF8}.
The measured data presented here were already incorporated in a test run in the current database underlying the Neutron Data Standards evaluation in Ref.~\cite{Neudecker2021}, 
following the standard practice for new data, the Neutron Data Standards group is considering adopting this result as part of the database for the next standards release.

%% file: Results_Table.tex
\label{app:Results}
The fission cross section of \pu{} relative to \u{} over a neutron energy range of 0.2 -- 100~MeV is given here.  The bin structure is logarithmic.  The reported energy bin value is the lower edge of the bin.  The absolute, total uncertainty is 1$\sigma$.

\begin{longtable}{c|c|c|c}
\caption{$^{239}$Pu(n,f)/$^{235}$U(n,f) cross section ratio.}
\label{table:results}\\
\hline\hline
Energy Bin (MeV) & Bin Width & $\sigma_{(n,f)}$ (ratio) & Uncertainty (abs.) \\  
\hline
\endfirsthead
\hline\hline
Energy Bin (MeV) & Bin Width & $\sigma_{(n,f)}$ (ratio) & Uncertainty (abs.)\\
\hline
\endhead
\hline\hline
\endfoot
0.1059 & 0.00628 & 0.9979 & 0.015 \\
0.1122 & 0.00665 & 0.9529 & 0.014 \\
0.1189 & 0.00704 & 0.9743 & 0.014 \\
0.1259 & 0.00746 & 0.99 & 0.012 \\
0.1334 & 0.0079 & 1.008 & 0.012 \\
0.1413 & 0.00837 & 1.042 & 0.013 \\
0.1496 & 0.00887 & 1.061 & 0.011 \\
0.1585 & 0.00939 & 1.058 & 0.011 \\
0.1679 & 0.00995 & 1.105 & 0.011 \\
0.1778 & 0.0105 & 1.091 & 0.011 \\
0.1884 & 0.0112 & 1.057 & 0.01 \\
0.1995 & 0.0118 & 1.128 & 0.01 \\
0.2113 & 0.0125 & 1.149 & 0.0098 \\
0.2239 & 0.0133 & 1.174 & 0.01 \\
0.2371 & 0.0141 & 1.181 & 0.0099 \\
0.2512 & 0.0149 & 1.22 & 0.0099 \\
0.2661 & 0.0158 & 1.252 & 0.0095 \\
0.2818 & 0.0167 & 1.273 & 0.0093 \\
0.2985 & 0.0177 & 1.26 & 0.0087 \\
0.3162 & 0.0187 & 1.272 & 0.009 \\
0.335 & 0.0198 & 1.282 & 0.0087 \\
0.3548 & 0.021 & 1.275 & 0.0086 \\
0.3758 & 0.0223 & 1.296 & 0.0084 \\
0.3981 & 0.0236 & 1.305 & 0.0081 \\
0.4217 & 0.025 & 1.341 & 0.0086 \\
0.4467 & 0.0265 & 1.375 & 0.0084 \\
0.4732 & 0.028 & 1.405 & 0.0084 \\
0.5012 & 0.0297 & 1.419 & 0.0078 \\
0.5309 & 0.0315 & 1.411 & 0.0077 \\
0.5623 & 0.0333 & 1.439 & 0.008 \\
0.5957 & 0.0353 & 1.465 & 0.008 \\
0.631 & 0.0374 & 1.455 & 0.0076 \\
0.6683 & 0.0396 & 1.474 & 0.0074 \\
0.7079 & 0.0419 & 1.498 & 0.0078 \\
0.7499 & 0.0444 & 1.529 & 0.008 \\
0.7943 & 0.0471 & 1.545 & 0.008 \\
0.8414 & 0.0499 & 1.523 & 0.0076 \\
0.8913 & 0.0528 & 1.472 & 0.0074 \\
0.9441 & 0.0559 & 1.441 & 0.0073 \\
1 & 0.0593 & 1.452 & 0.0069 \\
1.059 & 0.0628 & 1.478 & 0.0071 \\
1.122 & 0.0665 & 1.506 & 0.0076 \\
1.189 & 0.0704 & 1.534 & 0.0077 \\
1.259 & 0.0746 & 1.568 & 0.0074 \\
1.334 & 0.079 & 1.594 & 0.0076 \\
1.413 & 0.0837 & 1.584 & 0.0074 \\
1.496 & 0.0887 & 1.588 & 0.0081 \\
1.585 & 0.0939 & 1.576 & 0.0074 \\
1.679 & 0.0995 & 1.562 & 0.0076 \\
1.778 & 0.105 & 1.566 & 0.0076 \\
1.884 & 0.112 & 1.559 & 0.0068 \\
1.995 & 0.118 & 1.556 & 0.0078 \\
2.113 & 0.125 & 1.561 & 0.0076 \\
2.239 & 0.133 & 1.554 & 0.008 \\
2.371 & 0.141 & 1.54 & 0.0072 \\
2.512 & 0.149 & 1.542 & 0.0074 \\
2.661 & 0.158 & 1.547 & 0.0075 \\
2.818 & 0.167 & 1.55 & 0.0075 \\
2.985 & 0.177 & 1.557 & 0.0073 \\
3.162 & 0.187 & 1.571 & 0.0079 \\
3.35 & 0.198 & 1.567 & 0.0077 \\
3.548 & 0.21 & 1.57 & 0.0081 \\
3.758 & 0.223 & 1.56 & 0.008 \\
3.981 & 0.236 & 1.588 & 0.0082 \\
4.217 & 0.25 & 1.586 & 0.008 \\
4.467 & 0.265 & 1.588 & 0.0085 \\
4.732 & 0.28 & 1.598 & 0.0085 \\
5.012 & 0.297 & 1.598 & 0.0088 \\
5.309 & 0.315 & 1.626 & 0.0091 \\
5.623 & 0.333 & 1.654 & 0.0093 \\
5.957 & 0.353 & 1.594 & 0.0087 \\
6.31 & 0.374 & 1.486 & 0.008 \\
6.683 & 0.396 & 1.395 & 0.0077 \\
7.079 & 0.419 & 1.333 & 0.007 \\
7.499 & 0.444 & 1.309 & 0.0071 \\
7.943 & 0.471 & 1.292 & 0.0067 \\
8.414 & 0.499 & 1.296 & 0.0072 \\
8.913 & 0.528 & 1.294 & 0.0071 \\
9.441 & 0.559 & 1.307 & 0.0077 \\
10 & 0.593 & 1.313 & 0.0075 \\
10.59 & 0.628 & 1.327 & 0.008 \\
11.22 & 0.665 & 1.324 & 0.0084 \\
11.89 & 0.704 & 1.307 & 0.0083 \\
12.59 & 0.746 & 1.274 & 0.0081 \\
13.34 & 0.79 & 1.214 & 0.0081 \\
14.13 & 0.837 & 1.166 & 0.0077 \\
14.96 & 0.887 & 1.136 & 0.0078 \\
15.85 & 0.939 & 1.13 & 0.0079 \\
16.79 & 0.995 & 1.148 & 0.0074 \\
17.78 & 1.05 & 1.152 & 0.0077 \\
18.84 & 1.12 & 1.164 & 0.0082 \\
19.95 & 1.18 & 1.166 & 0.0083 \\
21.13 & 1.25 & 1.138 & 0.0078 \\
22.39 & 1.33 & 1.118 & 0.0081 \\
23.71 & 1.41 & 1.107 & 0.0079 \\
25.12 & 1.49 & 1.089 & 0.0075 \\
26.61 & 1.58 & 1.111 & 0.0078 \\
28.18 & 1.67 & 1.095 & 0.008 \\
29.85 & 1.77 & 1.094 & 0.0078 \\
31.62 & 1.87 & 1.085 & 0.0081 \\
33.5 & 1.98 & 1.09 & 0.0076 \\
35.48 & 2.1 & 1.077 & 0.0079 \\
37.58 & 2.23 & 1.089 & 0.0081 \\
39.81 & 2.36 & 1.069 & 0.0077 \\
42.17 & 2.5 & 1.09 & 0.0079 \\
44.67 & 2.65 & 1.075 & 0.0082 \\
47.32 & 2.8 & 1.066 & 0.0079 \\
50.12 & 2.97 & 1.07 & 0.0086 \\
53.09 & 3.15 & 1.079 & 0.0081 \\
56.23 & 3.33 & 1.082 & 0.0086 \\
59.57 & 3.53 & 1.092 & 0.0085 \\
63.1 & 3.74 & 1.098 & 0.009 \\
66.83 & 3.96 & 1.093 & 0.0086 \\
70.79 & 4.19 & 1.094 & 0.009 \\
74.99 & 4.44 & 1.089 & 0.0089 \\
79.43 & 4.71 & 1.088 & 0.01 \\
84.14 & 4.99 & 1.074 & 0.0091 \\
89.13 & 5.28 & 1.099 & 0.0098 \\
94.41 & 5.59 & 1.103 & 0.01 \\
\end{longtable}

%% file: Uncertainty_Table.tex
\label{app:uncertResults}
The fission cross section ratio partial uncertainties of \pu{} relative to \u{} over a neutron energy range of 0.2 -- 100~MeV are given here.  The bin structure is logarithmic.  The reported energy bin value is the lower edge of the bin.  The bin widths are the same as reported in table~\ref{table:results}. The partial uncertainties are absolute. The partial uncertainties listed are $\sigma_{C_{ff}}$ statistical , $\sigma_v$ from the variational analysis, $\sigma_{C_w}$ is the wraparound correction, $\sigma_{\epsilon_{ff}}$ is the efficiency correction, $\sigma_{C_b}$ is the contamination correction, $\sigma_{\phi_{XY}}$ is the overlap correction and $\sigma_{\kappa}$ is the beam attenuation and scattering correction.

\begin{longtable}{c|c|c|c|c|c|c|c}
\caption{$^{239}$Pu(n,f)/$^{235}$U(n,f) cross section ratio partial uncertainties.}
\label{table:uncertResults}\\
\hline\hline
Energy Bin (MeV) & $\sigma_{C_{ff}}$ & $\sigma_v$ & $\sigma_{C_w}$ & $\sigma_{\epsilon_{ff}}$ & $\sigma_{C_{b}}$ & $\sigma_{\phi_{XY}}$ & $\sigma_{\kappa}$\\
\hline
\endfirsthead
\hline\hline
Energy Bin (MeV) & $\sigma_{C_{ff}}$ & $\sigma_v$ & $\sigma_{C_w}$ & $\sigma_{\epsilon_{ff}}$ & $\sigma_{C_{b}}$ & $\sigma_{\phi_{XY}}$ & $\sigma_{\kappa}$\\
\hline
\endhead
\hline\hline
\endfoot
0.1059 & 1.2 & 0.49 & 0.38 & 0.21 & 0.099 & 0.72 & 0.14 \\
0.1122 & 1.1 & 0.53 & 0.28 & 0.21 & 0.12 & 0.59 & 0.13 \\
0.1189 & 1.1 & 0.39 & 0.24 & 0.21 & 0.13 & 0.66 & 0.14 \\
0.1259 & 0.98 & 0.38 & 0.2 & 0.2 & 0.11 & 0.56 & 0.13 \\
0.1334 & 0.92 & 0.34 & 0.17 & 0.24 & 0.082 & 0.54 & 0.13 \\
0.1413 & 0.97 & 0.37 & 0.19 & 0.22 & 0.057 & 0.53 & 0.15 \\
0.1496 & 0.84 & 0.29 & 0.16 & 0.19 & 0.057 & 0.52 & 0.13 \\
0.1585 & 0.82 & 0.3 & 0.17 & 0.18 & 0.064 & 0.55 & 0.12 \\
0.1679 & 0.78 & 0.25 & 0.17 & 0.19 & 0.066 & 0.49 & 0.11 \\
0.1778 & 0.74 & 0.26 & 0.15 & 0.23 & 0.06 & 0.53 & 0.1 \\
0.1884 & 0.75 & 0.28 & 0.16 & 0.25 & 0.053 & 0.52 & 0.11 \\
0.1995 & 0.68 & 0.24 & 0.14 & 0.21 & 0.051 & 0.53 & 0.1 \\
0.2113 & 0.67 & 0.28 & 0.14 & 0.24 & 0.053 & 0.41 & 0.088 \\
0.2239 & 0.64 & 0.27 & 0.14 & 0.22 & 0.055 & 0.51 & 0.089 \\
0.2371 & 0.62 & 0.27 & 0.14 & 0.19 & 0.055 & 0.48 & 0.096 \\
0.2512 & 0.59 & 0.29 & 0.13 & 0.21 & 0.053 & 0.43 & 0.1 \\
0.2661 & 0.56 & 0.22 & 0.12 & 0.21 & 0.05 & 0.42 & 0.068 \\
0.2818 & 0.56 & 0.23 & 0.11 & 0.19 & 0.049 & 0.39 & 0.074 \\
0.2985 & 0.51 & 0.24 & 0.096 & 0.21 & 0.049 & 0.35 & 0.079 \\
0.3162 & 0.51 & 0.27 & 0.097 & 0.21 & 0.05 & 0.43 & 0.073 \\
0.335 & 0.47 & 0.24 & 0.084 & 0.23 & 0.049 & 0.41 & 0.068 \\
0.3548 & 0.47 & 0.22 & 0.08 & 0.2 & 0.048 & 0.41 & 0.075 \\
0.3758 & 0.44 & 0.22 & 0.075 & 0.22 & 0.047 & 0.39 & 0.094 \\
0.3981 & 0.44 & 0.21 & 0.07 & 0.23 & 0.048 & 0.35 & 0.062 \\
0.4217 & 0.43 & 0.25 & 0.068 & 0.21 & 0.048 & 0.37 & 0.067 \\
0.4467 & 0.41 & 0.23 & 0.061 & 0.23 & 0.049 & 0.33 & 0.058 \\
0.4732 & 0.4 & 0.24 & 0.058 & 0.21 & 0.05 & 0.32 & 0.06 \\
0.5012 & 0.37 & 0.24 & 0.053 & 0.21 & 0.05 & 0.23 & 0.061 \\
0.5309 & 0.37 & 0.23 & 0.046 & 0.19 & 0.05 & 0.28 & 0.06 \\
0.5623 & 0.36 & 0.24 & 0.044 & 0.21 & 0.05 & 0.3 & 0.07 \\
0.5957 & 0.35 & 0.22 & 0.041 & 0.2 & 0.05 & 0.27 & 0.055 \\
0.631 & 0.33 & 0.24 & 0.038 & 0.21 & 0.05 & 0.26 & 0.062 \\
0.6683 & 0.33 & 0.22 & 0.035 & 0.22 & 0.05 & 0.28 & 0.056 \\
0.7079 & 0.32 & 0.2 & 0.033 & 0.2 & 0.049 & 0.27 & 0.057 \\
0.7499 & 0.31 & 0.25 & 0.032 & 0.22 & 0.049 & 0.29 & 0.049 \\
0.7943 & 0.31 & 0.22 & 0.029 & 0.2 & 0.05 & 0.29 & 0.056 \\
0.8414 & 0.3 & 0.2 & 0.026 & 0.23 & 0.049 & 0.27 & 0.041 \\
0.8913 & 0.3 & 0.22 & 0.025 & 0.21 & 0.048 & 0.29 & 0.041 \\
0.9441 & 0.29 & 0.26 & 0.023 & 0.21 & 0.046 & 0.25 & 0.043 \\
1 & 0.29 & 0.22 & 0.023 & 0.24 & 0.046 & 0.24 & 0.041 \\
1.059 & 0.29 & 0.22 & 0.021 & 0.22 & 0.046 & 0.28 & 0.042 \\
1.122 & 0.28 & 0.25 & 0.021 & 0.24 & 0.046 & 0.22 & 0.049 \\
1.189 & 0.28 & 0.24 & 0.018 & 0.22 & 0.045 & 0.27 & 0.044 \\
1.259 & 0.28 & 0.25 & 0.019 & 0.2 & 0.044 & 0.2 & 0.045 \\
1.334 & 0.27 & 0.23 & 0.018 & 0.23 & 0.043 & 0.24 & 0.036 \\
1.413 & 0.27 & 0.23 & 0.016 & 0.22 & 0.042 & 0.23 & 0.045 \\
1.496 & 0.27 & 0.23 & 0.016 & 0.2 & 0.041 & 0.3 & 0.088 \\
1.585 & 0.27 & 0.25 & 0.016 & 0.22 & 0.04 & 0.24 & 0.042 \\
1.679 & 0.27 & 0.24 & 0.016 & 0.22 & 0.039 & 0.25 & 0.043 \\
1.778 & 0.27 & 0.22 & 0.015 & 0.2 & 0.038 & 0.25 & 0.051 \\
1.884 & 0.27 & 0.24 & 0.014 & 0.18 & 0.036 & 0.2 & 0.039 \\
1.995 & 0.27 & 0.26 & 0.014 & 0.22 & 0.035 & 0.28 & 0.045 \\
2.113 & 0.28 & 0.25 & 0.013 & 0.18 & 0.034 & 0.28 & 0.05 \\
2.239 & 0.27 & 0.24 & 0.013 & 0.22 & 0.033 & 0.29 & 0.04 \\
2.371 & 0.28 & 0.24 & 0.013 & 0.2 & 0.032 & 0.24 & 0.044 \\
2.512 & 0.28 & 0.25 & 0.013 & 0.24 & 0.031 & 0.23 & 0.034 \\
2.661 & 0.28 & 0.21 & 0.012 & 0.21 & 0.029 & 0.24 & 0.042 \\
2.818 & 0.29 & 0.22 & 0.012 & 0.25 & 0.028 & 0.25 & 0.051 \\
2.985 & 0.29 & 0.23 & 0.012 & 0.22 & 0.027 & 0.28 & 0.043 \\
3.162 & 0.29 & 0.21 & 0.012 & 0.2 & 0.026 & 0.28 & 0.044 \\
3.35 & 0.3 & 0.26 & 0.012 & 0.19 & 0.025 & 0.26 & 0.041 \\
3.548 & 0.3 & 0.23 & 0.012 & 0.24 & 0.023 & 0.3 & 0.049 \\
3.758 & 0.3 & 0.23 & 0.012 & 0.19 & 0.022 & 0.26 & 0.046 \\
3.981 & 0.31 & 0.23 & 0.012 & 0.24 & 0.021 & 0.27 & 0.045 \\
4.217 & 0.32 & 0.22 & 0.012 & 0.24 & 0.02 & 0.24 & 0.043 \\
4.467 & 0.32 & 0.25 & 0.012 & 0.23 & 0.02 & 0.29 & 0.04 \\
4.732 & 0.33 & 0.23 & 0.012 & 0.18 & 0.02 & 0.31 & 0.039 \\
5.012 & 0.34 & 0.22 & 0.013 & 0.21 & 0.021 & 0.32 & 0.043 \\
5.309 & 0.35 & 0.2 & 0.012 & 0.23 & 0.022 & 0.38 & 0.038 \\
5.623 & 0.36 & 0.23 & 0.013 & 0.22 & 0.023 & 0.3 & 0.04 \\
5.957 & 0.35 & 0.22 & 0.012 & 0.23 & 0.022 & 0.3 & 0.044 \\
6.31 & 0.34 & 0.2 & 0.0099 & 0.23 & 0.02 & 0.27 & 0.042 \\
6.683 & 0.33 & 0.22 & 0.0086 & 0.22 & 0.018 & 0.35 & 0.049 \\
7.079 & 0.33 & 0.28 & 0.008 & 0.22 & 0.017 & 0.29 & 0.058 \\
7.499 & 0.33 & 0.23 & 0.0076 & 0.21 & 0.016 & 0.29 & 0.053 \\
7.943 & 0.34 & 0.26 & 0.0076 & 0.17 & 0.015 & 0.27 & 0.062 \\
8.414 & 0.35 & 0.2 & 0.0077 & 0.2 & 0.015 & 0.32 & 0.057 \\
8.913 & 0.36 & 0.24 & 0.0079 & 0.2 & 0.014 & 0.29 & 0.055 \\
9.441 & 0.37 & 0.21 & 0.0082 & 0.22 & 0.015 & 0.31 & 0.053 \\
10 & 0.38 & 0.24 & 0.0087 & 0.21 & 0.016 & 0.31 & 0.054 \\
10.59 & 0.4 & 0.26 & 0.0089 & 0.23 & 0.018 & 0.31 & 0.056 \\
11.22 & 0.42 & 0.23 & 0.0095 & 0.2 & 0.022 & 0.39 & 0.059 \\
11.89 & 0.43 & 0.2 & 0.0093 & 0.19 & 0.026 & 0.41 & 0.074 \\
12.59 & 0.44 & 0.22 & 0.0092 & 0.2 & 0.027 & 0.34 & 0.069 \\
13.34 & 0.44 & 0.24 & 0.0087 & 0.19 & 0.029 & 0.35 & 0.065 \\
14.13 & 0.45 & 0.25 & 0.0087 & 0.23 & 0.031 & 0.4 & 0.072 \\
14.96 & 0.45 & 0.32 & 0.0085 & 0.21 & 0.031 & 0.42 & 0.065 \\
15.85 & 0.47 & 0.23 & 0.0083 & 0.21 & 0.031 & 0.42 & 0.071 \\
16.79 & 0.48 & 0.22 & 0.0084 & 0.19 & 0.029 & 0.35 & 0.066 \\
17.78 & 0.48 & 0.22 & 0.0081 & 0.24 & 0.023 & 0.32 & 0.08 \\
18.84 & 0.49 & 0.32 & 0.0088 & 0.21 & 0.016 & 0.39 & 0.092 \\
19.95 & 0.5 & 0.22 & 0.0084 & 0.2 & 0.016 & 0.39 & 0.084 \\
21.13 & 0.5 & 0.26 & 0.0083 & 0.24 & 0.016 & 0.4 & 0.079 \\
22.39 & 0.5 & 0.25 & 0.0082 & 0.2 & 0.016 & 0.45 & 0.087 \\
23.71 & 0.5 & 0.26 & 0.0079 & 0.19 & 0.016 & 0.38 & 0.08 \\
25.12 & 0.51 & 0.21 & 0.0075 & 0.22 & 0.014 & 0.35 & 0.084 \\
26.61 & 0.51 & 0.22 & 0.0075 & 0.23 & 0.011 & 0.42 & 0.08 \\
28.18 & 0.51 & 0.24 & 0.0071 & 0.2 & 0.007 & 0.43 & 0.083 \\
29.85 & 0.51 & 0.2 & 0.0073 & 0.21 & 0.0056 & 0.44 & 0.099 \\
31.62 & 0.52 & 0.26 & 0.0069 & 0.22 & 0.0056 & 0.41 & 0.093 \\
33.5 & 0.52 & 0.21 & 0.0066 & 0.24 & 0.0056 & 0.33 & 0.095 \\
35.48 & 0.52 & 0.18 & 0.0067 & 0.22 & 0.0056 & 0.44 & 0.087 \\
37.58 & 0.53 & 0.29 & 0.0065 & 0.22 & 0.0056 & 0.38 & 0.096 \\
39.81 & 0.54 & 0.24 & 0.0066 & 0.22 & 0.0056 & 0.43 & 0.096 \\
42.17 & 0.54 & 0.22 & 0.0064 & 0.21 & 0.0056 & 0.38 & 0.084 \\
44.67 & 0.54 & 0.24 & 0.0062 & 0.23 & 0.0056 & 0.45 & 0.092 \\
47.32 & 0.56 & 0.24 & 0.0062 & 0.21 & 0.0056 & 0.37 & 0.12 \\
50.12 & 0.57 & 0.36 & 0.006 & 0.22 & 0.0056 & 0.48 & 0.11 \\
53.09 & 0.57 & 0.24 & 0.0061 & 0.22 & 0.0056 & 0.41 & 0.11 \\
56.23 & 0.58 & 0.24 & 0.006 & 0.19 & 0.0056 & 0.46 & 0.095 \\
59.57 & 0.58 & 0.21 & 0.0064 & 0.2 & 0.0056 & 0.39 & 0.1 \\
63.1 & 0.59 & 0.21 & 0.0066 & 0.21 & 0.0056 & 0.48 & 0.12 \\
66.83 & 0.61 & 0.26 & 0.0064 & 0.22 & 0.0056 & 0.45 & 0.1 \\
70.79 & 0.63 & 0.25 & 0.0064 & 0.25 & 0.0056 & 0.42 & 0.11 \\
74.99 & 0.65 & 0.28 & 0.0068 & 0.21 & 0.0056 & 0.44 & 0.1 \\
79.43 & 0.66 & 0.24 & 0.007 & 0.21 & 0.0056 & 0.62 & 0.1 \\
84.14 & 0.68 & 0.27 & 0.0071 & 0.2 & 0.0056 & 0.43 & 0.1 \\
89.13 & 0.69 & 0.25 & 0.0073 & 0.21 & 0.0056 & 0.49 & 0.1 \\
94.41 & 0.74 & 0.28 & 0.0076 & 0.22 & 0.0056 & 0.53 & 0.11 \\
\end{longtable}